\pdfoutput=1
\documentclass[prologue, xcdraw,sigconf,table,screen]{acmart}
\usepackage{graphicx}
\usepackage{subcaption}
\usepackage{tikz}
\usepackage{hyperref}
\usepackage{pgfplots}
\usetikzlibrary{patterns}
\usepackage{rotating}
\usepackage{amsmath}

\usepackage{algorithm, algpseudocode}
\usepackage{placeins}
\usepackage{svg}
\usepackage{rotating}
\usepackage{diagbox}
\usetikzlibrary{decorations.markings}
\usetikzlibrary{decorations.pathreplacing,positioning}
\usepackage{enumitem}
\usepackage{caption}
\usepackage{hyperref}
\usepackage{pifont}

\usepackage{listings}
\usepackage{color}

\definecolor{dkgreen}{rgb}{0,0.6,0}
\definecolor{gray}{rgb}{0.5,0.5,0.5}
\definecolor{mauve}{rgb}{0.58,0,0.82}

\definecolor{dkgreen}{rgb}{0,0.6,0}
\definecolor{gray}{rgb}{0.5,0.5,0.5}
\definecolor{mauve}{rgb}{0.58,0,0.82}

\AtBeginDocument{%
  \providecommand\BibTeX{{%
    \normalfont B\kern-0.5em{\scshape i\kern-0.25em b}\kern-0.8em\TeX}}}

\setcopyright{none} 

\acmConference[]{}{}{}

\setcopyright{none}
\renewcommand\footnotetextcopyrightpermission[1]{} 
\usepackage{booktabs}

\begin{document}

\author{Yehonatan Fridman}
\affiliation{%
  \institution{Ben-Gurion University, NRCN, Israel}
  \country{}
}
\email{fridyeh@post.bgu.ac.il}

\author{Suprasad Mutalik Desai, Navneet Singh}
\affiliation{%
  \institution{Intel, India}
  \country{}
  }
\email{{suprasad.desai,navneet.singh}@intel.com}

\author{Thomas Willhalm}
\affiliation{%
  \institution{Intel, Germany}
  \country{}
  }
\email{thomas.willhalm@intel.com}

\author{Gal Oren}
\affiliation{%
 \institution{Technion, NRCN, Israel}
 \country{}
 }
\email{galoren@cs.technion.ac.il}

\title[CXL Memory as Persistent Memory for Disaggregated HPC: A Practical Approach]{CXL Memory as Persistent Memory for Disaggregated HPC:\\A Practical Approach}

\begin{abstract}

In the landscape of High-Performance Computing (HPC), the quest for efficient and scalable memory solutions remains paramount. The advent of Compute Express Link (CXL) introduces a promising avenue with its potential to function as a Persistent Memory (PMem) solution in the context of disaggregated HPC systems. This paper presents a comprehensive exploration of CXL memory's viability as a candidate for PMem, supported by physical experiments conducted on cutting-edge multi-NUMA nodes equipped with CXL-attached memory prototypes. Our study not only benchmarks the performance of CXL memory but also illustrates the seamless transition from traditional PMem programming models to CXL, reinforcing its practicality.

To substantiate our claims, we establish a tangible CXL prototype using an FPGA card embodying CXL 1.1/2.0 compliant endpoint designs (Intel FPGA CXL IP). Performance evaluations, executed through the STREAM and STREAM-PMem benchmarks, showcase CXL memory's ability to mirror PMem characteristics in \textit{App-Direct} and \textit{Memory Mode} while achieving impressive bandwidth metrics with Intel 4th generation Xeon (Sapphire Rapids) processors.

The results elucidate the feasibility of CXL memory as a persistent memory solution, outperforming previously established benchmarks. In contrast to published DCPMM results, our CXL-DDR4 memory module offers comparable bandwidth to local DDR4 memory configurations, albeit with a moderate decrease in performance. The modified STREAM-PMem application underscores the ease of transitioning programming models from PMem to CXL, thus underscoring the practicality of adopting CXL memory.


The sources of this work are available at: \textcolor{blue}{\url{https://github.com/Scientific-Computing-Lab-NRCN/STREAMer}}.

\end{abstract}

\settopmatter{printfolios=true}
\maketitle

\section{Introduction}
\subsection{Current     HPC Memory Solutions Limitations}
As the era of Exa-Scale computing unfolds, the demand for analyzing, manipulating, and storing massive amounts of data intensifies~\cite{reed2022reinventing}. Exascale systems are designed to meet these demands and enable the execution of a broad spectrum of computations, ranging from loosely to tightly coupled tasks, including CFD simulations and deep learning optimizations~\cite{evans2022survey}. Memory and storage resources play a crucial role in the performance and scalability of these computations~\cite{kogge2022frontier}. Memory factors such as capacity, latency, and bandwidth are responsible for successfully handling extensive tasks and delivering data to processing units promptly~\cite{mckee2004reflections}. In scientific computing, storage devices hold significance for preserving diagnostics throughout computations~\cite{lockwood2023storage}. Notably, the growing frequency of failures in exascale machines emphasizes the significance of storing vast data volumes to support recovery and bolster fault tolerance~\cite{bustos2023response, fridman2022recovery}.

However, the traditional memory and storage hierarchy in HPC systems reveals notable gaps that impose critical constraints on scientific computations~\cite{kogge2022frontier}. From the vantage point of memory architecture, DRAM has inherent limitations of bandwidth and capacity that impact performance and prevent the processing of large-scale problems~\cite{radulovic2015another, peng2021holistic}. From the storage perspective, traditional devices (such as HDDs and SSDs) provide large capacities but exhibit very slow access times, leading to significant overheads for I/O-bound applications and fault tolerance mechanisms~\cite{lockwood2023storage}. These gaps and limitations of traditional hardware highlight the ongoing endeavors to expand the memory-storage hierarchy and develop novel memory architectures and solutions. A notable example is Non-Volatile RAM~\cite{wang2023survey, rai2023nonvolatile} (on which we elaborate in \autoref{pmem_hpc}).

While High-Bandwidth Memory (HBM)~\cite{jun2017hbm} has been introduced as a solution to enhance memory performance, it doesn't entirely alleviate the problem~\cite{jun2016high}. HBM memory modules are stacked vertically, allowing for higher memory bandwidth due to their increased parallelism. However, even with HBM, the memory capacity remains limited compared to conventional DDR (Double Data Rate) memory modules~\cite{shipman2022early}. This limitation can still lead to constraints in memory-intensive applications that require larger memory spaces~\cite{McCalpin2023}. Moreover, while HBM addresses the bandwidth issue to some extent, it doesn't eliminate the underlying problem of memory hierarchy~\cite{McCalpin2023}. The processor still needs to access different memory levels, and the latency of transferring data between these levels can impact performance~\cite{McCalpin2023}. HBM improves bandwidth between the processor and certain memory modules, but the need to access different levels of memory introduces latency that can affect the execution of various tasks~\cite{McCalpin2023}.

In general, it is possible to proclaim that the conventional approach of locating memory modules directly on the board poses significant challenges in the context of HPC systems~\cite{ding2023evaluating}. This arrangement restricts memory bandwidth due to the limited number of connections between the processor and these modules~\cite{ding2023evaluating}. As a result, the data transfer rate between the processor and board-mounted memory becomes a bottleneck, hindering the overall performance of the system~\cite{ding2023evaluating}.

For an increase of memory capacity outside of the node, advanced communication technologies such as the Remote Direct Memory Access (RDMA) based Message Passing Interface (MPI) optimize inter-node communication~\cite{liu2003high}. However, these sophisticated frameworks are not devoid of challenges~\cite{gropp2012mpi}: MPI, a cornerstone for distributed computing communication, contends with latency and overhead issues during message transmission, disproportionately affecting efficiency for applications requiring frequent communication. Furthermore, the management complexity escalates with the cluster's scale due to heightened contention for network resources among a larger node count~\cite{bernholdt2020survey}. 

\subsection{Persistent Memory in HPC}
\label{pmem_hpc}
\begin{table*}[]
\centering
\begin{tabular}{|p{1.5cm}|p{6cm}|p{6cm}|}
\hline
\textbf{Property}    & \textbf{As a main memory extension} & \textbf{As a direct access to persistent memory}                                               \\ \hline
Volatility  & Volatile in memory extension mode & Non-volatile in direct access mode           \\ \hline
Access      & Cache-coherent memory expansion & Transactional byte-addressable object store    \\ \hline
Capacity    & Higher than main memory volume & Lower than storage volume     \\ \hline
Cost        & Cheaper than the main memory & More expansive than storage                       \\ \hline
Performance & Several factors below main memory bandwidth & High bandwidth compared to storage \\ \hline
\end{tabular}%
\caption{Properties of PMem modules, either as a memory extension (\textit{Memory Mode)} or as a direct access PMem (\textit{App-Direct}).}
\label{tab:pmemprop}
\end{table*}
A proposed solution aimed at bridging the gap between memory and storage is Persistent Memory (PMem)~\cite{lee2010phase,mutlu2013memory}. PMem implementations such as BBU (battery backed up) DIMM or Non-Volatile RAM (NVRAM) aim to deliver rapid byte-addressable data access alongside persistent data retention across power cycles. PMem technologies establish a new tier within the memory-storage hierarchy by combining memory and storage characteristics~\cite{wang2023survey, rai2023nonvolatile}. Basic solutions include battery-backed DRAM and have been accessible from diverse vendors over a significant timeframe, representing an established concept~\cite{narayanan2012whole, sainio2016nvdimm, kateja2017viyojit, malladi2016drampersist}. However, these solutions face challenges due to limited scalability and potential data loss risks. The reliance on batteries introduces concerns regarding power failures, leading to potential data corruption or loss if batteries deplete. Moreover, the approach's scalability is hampered by the need for individual batteries for each module, impacting cost-effectiveness and overall system performance. 

Yet, in recent years new PMem technologies have emerged, with 3D-Xpoint~\cite{hady2017platform} being the main technology and Intel Optane DCPMM~\cite{weiland2019early,hirofuchi2020prompt} the prominent product on the market. These modern PMem technologies offer byte-addressable memory in larger capacities compared to DRAM while maintaining comparable access times~\cite{izraelevitz2019basic}. Moreover, as these technologies are non-volatile in nature, they enable data retrieval even in instances of power failures. Moreover, PMem offers two configuration options based on these characteristics: (1) It can be utilized as main memory expansion, providing additional volatile memory, and (2) it can serve as a persistent memory pool that can be accessed by applications via a PMem-aware file system~\cite{wang2023survey} or be managed and accessed directly by applications~\cite{izraelevitz2019basic}. To simplify and streamline PMem programming and management, the Persistent Memory Development Kit (PMDK) was created~\cite{scargall2020programming}. 

During recent years, PMem has gained significant traction in HPC applications~\cite{patil2019performance, mironov2019performance, fridman2021assessing,rudoff2017persistent}, with two direct use cases of PMem for scientific applications that require no (or minimal) changes to applications. The first use-case involves PMem as memory expansion to support the execution of large scientific problems~\cite{mironov2019performance}. The second use case involves leveraging PMem as a fast storage device accessed by a PMem-aware file system (mainly based on the POSIX API), primarily for application diagnostics and checkpoint restart (C/R) mechanisms~\cite{logan2023evaluation}, but also for increasing the performance and inherent fault tolerance of scientific applications~\cite{fridman2022recovery}. 

In addition to the direct use cases of PMem in scientific applications, various frameworks and algorithms were developed to access and manage data structures on PMem ~\cite{baldassin2021persistent}. Among these are primary methods that are built on top of the PMDK library~\cite{khan2020persistent,fridman2022recovery}. For example, persistent memory object storage frameworks such as MOSIQS~\cite{khan2020persistent} and the NVM-ESR recovery model for exact state reconstruction of linear iterative solvers using PMem~\cite{fridman2022recovery}. 

Nevertheless, as HPC workloads advance, computing units evolve, and onboard processing elements increase, the demand for heightened memory bandwidth becomes essential~\cite{kogge2022frontier}. Existing PMem solutions demonstrate notable shortcomings in meeting these requirements, showing limitations in scalability beyond a certain threshold~\cite{fridman2021assessing}. Specifically, PMem devices exhibit limited bandwidth. For instance, the bandwidth of Optane DCPMM for reading and writing is multi-factor lower than that of DRAM~\cite{izraelevitz2019basic}. This, in part, is connected with the hybrid and in-between properties of a PMem module~\cite{gugnani2020understanding}, as schematically described in \autoref{tab:pmemprop}. 

Adding to these challenges, a significant limitation arises from the physical attachment of most PMem devices, like Optane DCPMM, to the CPU board through memory DIMMs. This configuration restricts the potential for memory expansion, as PMem contends for DIMM slots alongside conventional DRAM cards, presenting a bottleneck to achieving optimal memory configurations~\cite{tristian2019analyzing, tristian2021analyzing}. The HPC community as a whole --- both the super and cloud computing~\cite{ruan2023persistent} --- recognizes the drawbacks associated with tight integrating memory and compute resources, particularly in relation to capacity, bandwidth, elasticity, and overall system utilization~\cite{peng2020memory,ding2023evaluating}. PMem technologies that are tightly coupled with the CPU inherit these limitations. Now, as prominent PMem technologies are phased out (Optane DCPMM, for example, as announced in 2022~\cite{intel-pmem-cxl-tech, handy2023optane}), there is an active and prominent pursuit for the adoption of novel memory solutions in particular, and a strive to achieve more disaggregated computing in general~\cite{liu2023fabric}.

\subsection{Dissagregated Memory with CXL}

The emergence of discrete memory nodes housing DRAM and network interface controllers (NICs) is anticipated to revolutionize conventional memory paradigms, facilitating distributed and shared memory access and reshaping HPC landscapes~\cite{ding2023evaluating}. This shift aligns with the concept of disaggregation, where compute resources and memory units are decoupled for optimized resource utilization, scalability, and adaptability. 

The concept of memory disaggregation has been facilitated recently by the development of advanced interconnect technologies, exemplified by Compute Express Link (CXL)~\cite{sharma2023introduction}.
CXL is an open standard to support cache-coherent interconnect between a variety of devices~\cite{sharma2023introduction}.
After its introduction in 2019, the standard has evolved and continues to be enhanced.
CXL 1.1 defines the protocol for three major device types~\cite{sharma2023introduction}: 
Accelerators with cache-only (type 1), cache with attached memory (type 2), and memory expansion (type 3).
CXL 2.0 expands the specification -- among other capabilities -- to memory pools using CXL switches on a device level.
CXL 3.0 introduces fabric capabilities and management, improved memory sharing and pooling with dynamic capacity capability, enhanced coherency, and peer-to-peer communication. Bandwidth-wise, CXL 1.1 and 2.0 employ PCIe 5.0, achieving 32 GT/s for transfers up to 64 GB/s in each direction via a 16-lane link. On the other hand, CXL 3.0 utilizes PCIe 6.0, doubling the speed to 64 GT/s, supporting 128 GB/s bi-directional communication via an x16 link.

Since the market of CXL memory modules is emerging, several vendors have announced products using the CXL protocol.
For example, Samsung~\cite{samsung} and SK Hynix~\cite{hynix} introduce CXL DDR5 modules,
AsteraLabs~\cite{asteralabs} announced a CXL memory accelerator, and Montage Technology~\cite{montage} will offer a CXL memory expander controller.

\begin{figure}[htp!]
\centerline{\includegraphics[width=0.5\textwidth]{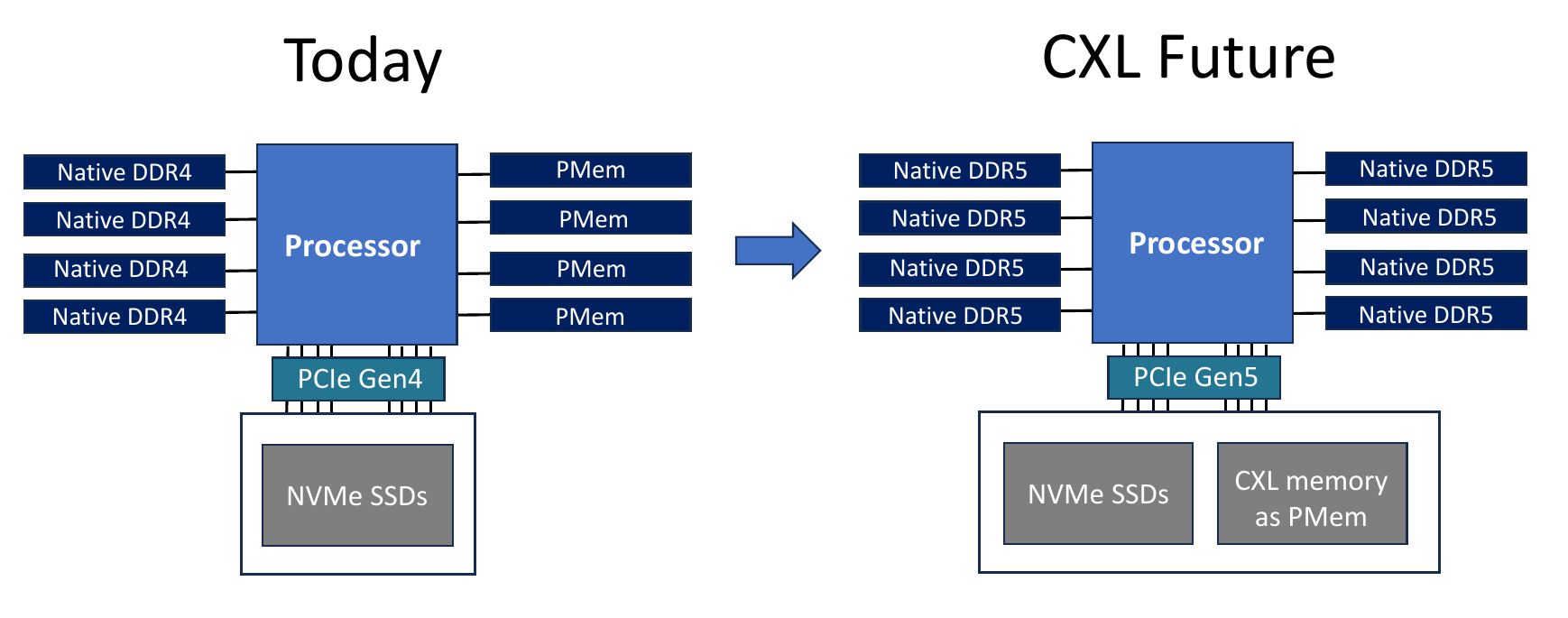}}
\caption{The migration from PMem as hardware to CXL memory as PMem in future systems.}
\label{fig:optane_to_cxl}
\end{figure}

Leveraging CXL, memory nodes will be interconnected through high-speed links, enabling adaptive memory provisioning to compute nodes in real time~\cite{wahlgren2022evaluating}. The practice of intra-rack disaggregation holds the potential to effectively address the memory demands of applications while concurrently ensuring an adequate supply of efficient remote memory bandwidth~\cite{michelogiannakis2022case, michelogiannakis2023efficient}. \autoref{fig:optane_to_cxl} demonstrates the expected phase change from the processor's point of view, from previous years' DDR4+PMem memory access, equipped with NVMe SSDs via the PCIe Gen4, to the upcoming future of DDR5 local memory equipped with local or remote NVMe SSDs and CXL memory for memory expansion or persistency over the new generations of PCIe.

Nevertheless, while the concept of memory disaggregation with technologies like CXL holds significant promise, it is important to acknowledge that there are still challenges and considerations that need to be addressed~\cite{al2023memory, geyer2023working}; challenges and considerations that resemble the ones of persistent memory integration in HPC~\cite{benson2023we}. For example, software and programming models need to evolve to take advantage of disaggregated memory fully; Applications and algorithms must be designed or adapted to work seamlessly across distributed memory nodes; and efficient data placement and movement strategies are crucial to minimize the impact of network latencies and ensure that data-intensive workloads can effectively utilize CXL-based disaggregated memory resources, especially when cache-coherence or direct access is enabled. Notwithstanding, when comparing CXL memory aspects to the ones of PMem as non-volatile RAM (NVRAM), in general, it can be observed (\autoref{table:comp_cxl_nvram}) that from the disaggregated HPC usage perspective, there should be a prevalence to CXL over NVRAM considering bandwidth, data transfer, and scalability, but also considering memory coherency, integration, pooling and sharing.

\begin{table*}[h]
\centering
\begin{tabular}{|p{2.5cm}|p{7cm}|p{7cm}|}
\hline
\textbf{Aspect} & \textbf{CXL Memory} & \textbf{NVRAM} \\
\hline
Bandwidth \& \newline Data Transfer & Significantly higher bandwidth enabling fast data transfers between processors and memory devices. & Non-volatile storage with potential data transfer rate limitations due to underlying interface and technology. \\
\hline
Memory\newline Coherency & Provides memory-coherent links, ensuring consistent data across different memory tiers. & Requires additional mechanisms for memory coherency, except with local RAM, when integrated with other memory technologies. \\
\hline
Heterogeneous Memory\newline Integration & Allows seamless integration of various memory technologies within a unified architecture. & Effective for extending memory capacity, but integration may require additional considerations due to unique characteristics. \\
\hline
Memory Pooling\newline and Sharing & Facilitates memory pooling and sharing, enabling efficient resource utilization and dynamic allocation based on workload requirements. & Extends memory capacity, but inherent flexibility in memory sharing and pooling may be limited. \\
\hline
Industry\newline Standardization & Open industry standard supported by major technology players, ensuring compatibility, interoperability, and broader adoption. & Solutions may vary, potentially leading to compatibility challenges and limited integration options. \\
\hline
Scalability & Architecture designed for scalability with multiple lanes and protocols, catering to evolving data center needs. & Scalability may be constrained by underlying technology characteristics, such as DIMM count and RAM/NVRAM tradeoff. \\
\hline
Relevance to HPC & Higher bandwidth, memory coherency, and memory pooling capabilities enhance HPC workload performance. Standardization compatibility in heterogeneous environments and scalability cater to evolving demands. & Offers non-volatility but is constrained by limitations in bandwidth, coherency management, and scalability, affecting its applicability to complex HPC memory needs. \\
\hline
\end{tabular}
\caption{General comparison between common aspects of CXL memory and NVRAM for disaggregated HPC.}
\label{table:comp_cxl_nvram}
\end{table*}

\subsection{Contribution}

In this work, based on actual physical experiments with multi-NUMA nodes and multi-core high-performance SOTA hardware (\autoref{hpchardware}) and CXL-remote memory (\autoref{cxlhardware}), we claim that it is not only possible to exemplify most persistent memory modules characteristics (as described in \autoref{tab:pmemprop}) with CXL memory fully but also that in terms of performance, we can achieve much better bandwidth than previously published Optane DCPMM ones (such in \cite{izraelevitz2016linearizability}, which, for a single Optane DCPMM, discovers that its max read bandwidth is 6.6 GB/s, whereas its max write bandwidth is 2.3 GB/s). In fact, we show (\autoref{resultsandanalysis}) that by approaching our CXL-DDR4 memory module -- much cheaper than DDR5 -- we achieve comparable results to the local DDR4 module and exhibit performance degradation of only about 60\% in bandwidth in comparison to local DDR5 module access (noting that DDR4 has about 50\% bandwidth of DDR5). Our tests were made in multiple configurations (\autoref{confs}) in relation to the memory distance from the working threads using the well-known STREAM benchmark (\autoref{streaming}). 

In order to demonstrate the non-volatile properties of the memory as PMem, the CXL memory was located outside of the node, in an FPGA device (\autoref{cxlhardware}), potentially backed by battery, like previous battery-backed DIMMs. 
As many nodes can approach the device, the battery-backed consideration is no longer considered by us as a major overhead since it will be applied only once for the memory modules and not in each compute node. 

Moreover, besides the cache-coherent performance benchmarks with STREAM~\cite{McCalpin1995,McCalpin2007}, we retested the memory bandwidth in an equivalent of the \textit{App-Direct} approach with a modified STREAM application, named STREAM-PMem~\cite{fagerheim2021benchmarking} when all of the main arrays were allocated as a PMDK's \textit{pmemobj} and manipulated accordingly \cite{scargall2020libpmemobj}. \textit{pmemobj} provides an assurance that the condition of objects will remain internally consistent regardless of when the program concludes. Additionally, it offers a \textit{transaction} function that can encompass various modifications made to persistent objects. This function ensures that either all of the modifications are successfully applied or none of them take effect. 

We stress that as our CXL memory module is located outside of the node and can be backed by a battery, the ability to transactionally and directly access the memory, exactly as previously done with Optane DCPMM, while achieving even better performances, is a key to our practical approach, which consider CXL memory as a persistent memory for the future of disaggregated HPC. 

Finally, we open-sourced the entire benchmarking methodology as an easy-to-use and automated tool named STREAMer for future CXL memory device evaluations for HPC purposes.


\section{Physical Experimental Setup}
\subsection{HPC hardware}
\label{hpchardware}
\begin{figure*}[htp!]
  \begin{minipage}{0.47\textwidth}
    \centering
    \includegraphics[width=\columnwidth]{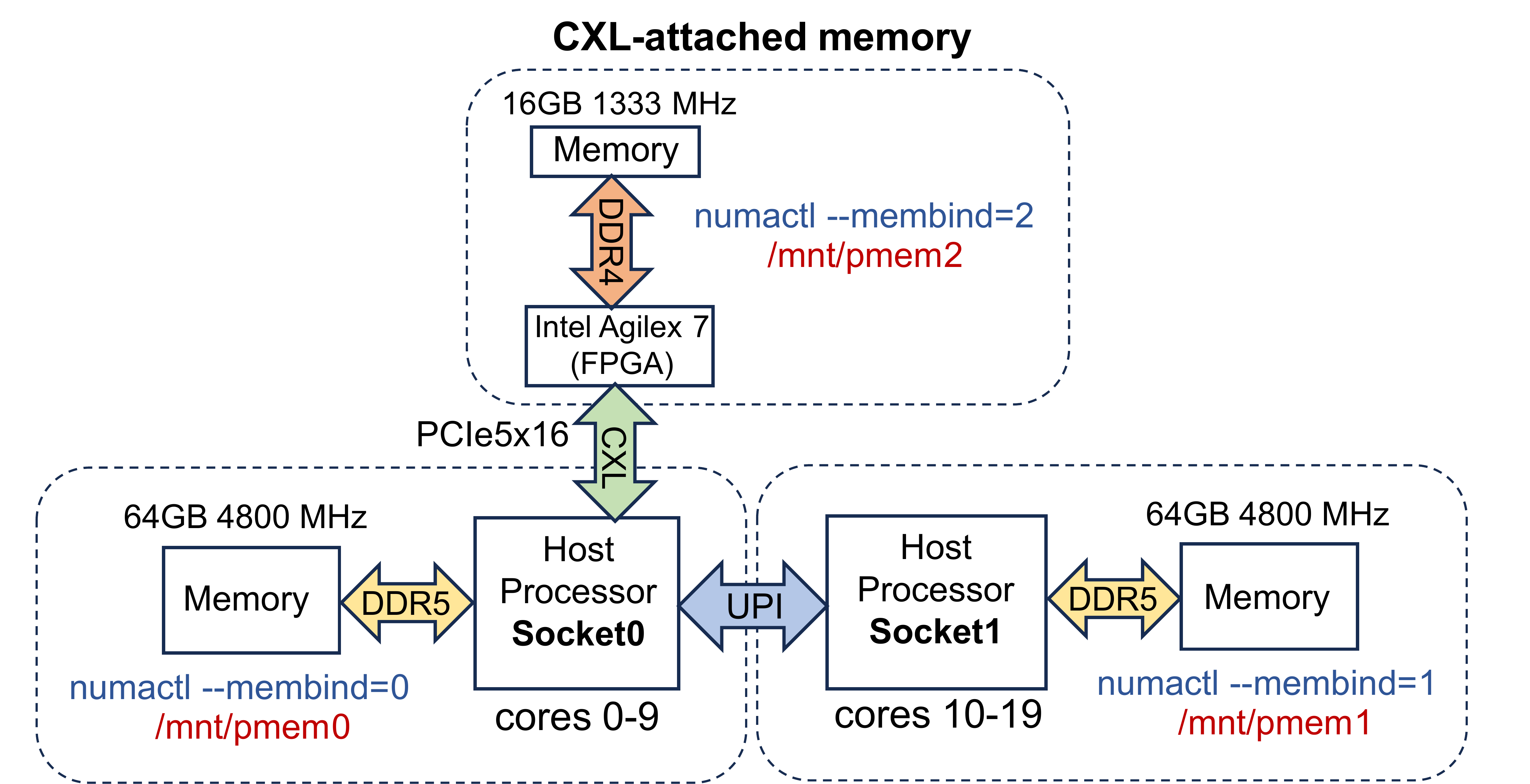}
    \caption{Setup \#1 with DDR5 on-node memory and DDR4 CXL-attached memory.}
    \label{fig:cxl_diagram}
        \includegraphics[width=\columnwidth]{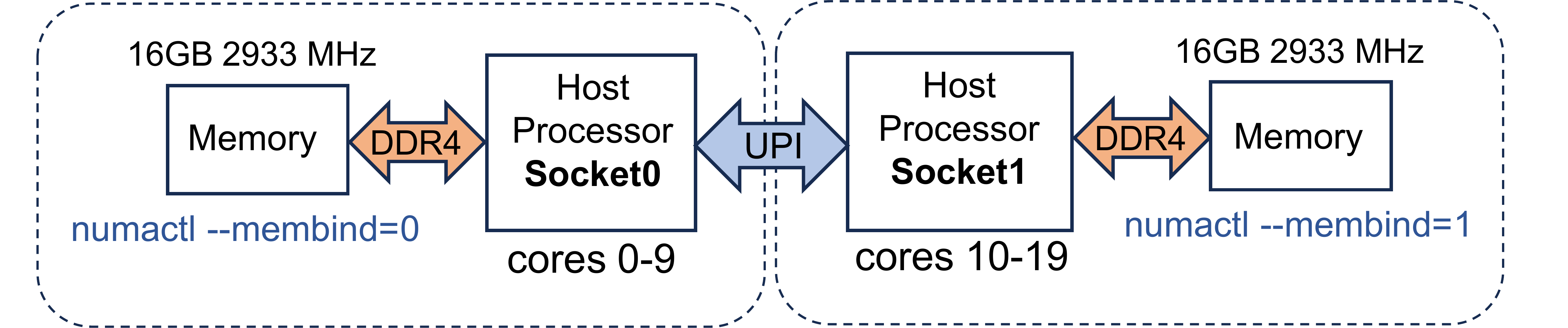}
    \caption{Setup \#2 with DDR4 on-node memory.}
    \label{fig:ddr4_node_diagram}
  \end{minipage}\hspace{0.04\textwidth}%
  \begin{minipage}{0.47\textwidth}
    \centering
\centerline{\includegraphics[width=\columnwidth]{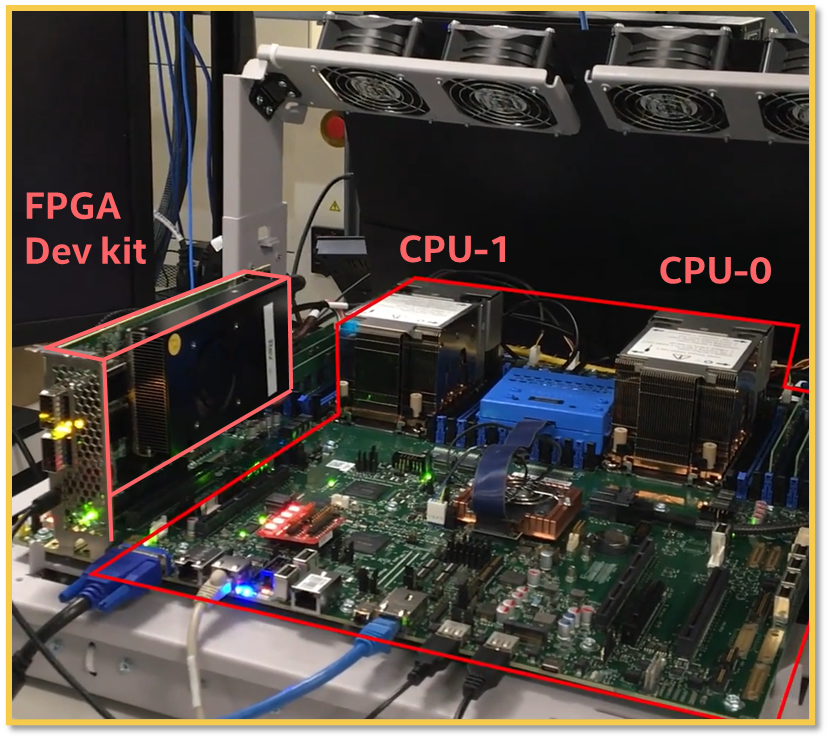}}
\caption{Overview of CXL IP for Intel® Agilex® 7 I-Series FPGA~\cite{cxl-fpga-intel}, demonstrated in \autoref{fig:cxl_diagram} (setup \#1).}
\label{fig:system}
  \end{minipage}
\end{figure*}

Our HPC hardware experimental environment is based on 2 setups:
\begin{enumerate}[wide, labelwidth=!, labelindent=0pt]
    \item Node equipped with two Intel $4^{th}$ generation Xeon (Sapphire Rapids) processors with a base frequency of 2.1GHz and 48 cores each, plus Hyper-Threading. BIOS was updated to support only 10 cores per socket. Each processor has one memory DIMM (64GB DDR5 4800MHz DIMM). The system is equipped with a CXL prototype device, implemented as DDR4 memory on a PCIe-attached FPGA (see Figure \ref{fig:cxl_diagram}).    
    \item Node equipped with two Intel Xeon Gold 5215 processors with a base frequency of 2.5GHz and 10 cores each, plus Hyper-Threading. Each processor has total 96GB DRAM in 6 channels, 16GB DDR4 2666MHz DIMM per channel.  (see Figure \ref{fig:ddr4_node_diagram}).
\end{enumerate}



  
  



\subsection{CXL prototype}
\label{cxlhardware}
We provide an in-depth overview of our CXL prototype's implementation on an FPGA card~\cite{fpga}. \autoref{fig:cxl_diagram}
and \autoref{fig:system} grant a more detailed view into the implementation of our CXL memory pool on the FPGA card (while \autoref{fig:ddr4_node_diagram} show the reference system, without any CXL attachment, with DDR4 main memory). The prototype aims to harness the capabilities of the R-Tile Intel FPGA IP for CXL, encompassing critical functionalities for CXL link establishment and transaction layer management. This comprehensive solution facilitates the construction of FPGA-based CXL 1.1/2.0 compliant endpoint designs, including Type 1, Type 2, and Type 3 configurations. It's built upon a previously proven prototype, with necessary slight modifications for PMem activity~\cite{lee2023elastic}.

The architecture of our CXL implementation revolves around a synergistic pairing of protocol Soft IP within the FPGA main fabric die and the Hard IP counterpart, the R-Tile. This cohesive arrangement ensures effective management of CXL link functions, which are pivotal for seamless communication. Specifically, the R-Tile interfaces with a CPU host via a PCIe Gen5x16 connection, delivering a theoretical bandwidth of up to 64GB/s. As a key facet of our implementation, the FPGA device is duly enumerated as a CXL endpoint within the host system.

Complementing this link management, the Soft IP assumes the mantle of transaction layer functions, vital for the successful execution of different CXL endpoint types. For Type 3 configurations, the CXL.mem transaction layer adeptly handles incoming CXL.mem requests originating from the CPU host. It orchestrates the generation of host-managed device memory (HDM) requests directed toward an HDM subsystem. Simultaneously, the CXL.io transaction layer undertakes the responsibility of processing CXL.io requests. These requests encompass both configuration and memory space inquiries initiated from the CPU host, seamlessly forwarding them to their designated control and status registers. A noteworthy augmentation is the User Streaming Interface, offering a conduit for custom CXL.io features that can be seamlessly integrated into the user design.

Integral to our FPGA card is the inclusion of two onboard DDR4 memory modules, each boasting a capacity of 8GB and operating at a clock frequency of 1333 MHz. These modules are accessible from the host system as conventional memory resources. It is imperative to highlight a distinctive attribute of this prototype configuration: the CXL link facilitates access to an identical memory volume. In essence, this means that the same far memory segment can be made available to two distinct NUMA nodes, eliminating any concerns of address overlap. However, due to the absence of a unified cache-coherent domain, the onus of maintaining coherency between the two NUMA nodes assigned to the shared far memory rests with the applications leveraging this configuration.

Notably, the bandwidth attainable from this prototype configuration is subject to current implementation constraints and does not reflect an intrinsic limitation of the CXL standard. Potential avenues for enhancing bandwidth include several considerations. First, transitioning to a higher-speed FPGA, supporting DDR4 speeds of 3200 Mbps or even embracing the capabilities of DDR5 at 5600 Mbps, could appreciably enhance throughput. Additionally, scaling the resources allocated to the CXL IP by increasing the number of slices is a viable strategy. Furthermore, expanding the FPGA's capacity to accommodate multiple independent DDR channels, possibly transitioning from one channel to four, holds promise in augmenting the prototype's bandwidth potential.


In our discussion, the fact that the CXL memory device is DDR4 and not DDR5 is key, as usually, PMem is slower and cheaper than the main memory. By using DDR4 CXL memory and not DDR5, while main memory is DDR5, we keep on this important relation.

\section{Performance Evaluation}

\subsection{STREAM and STREAM-PMem Benchmarks}
\label{streaming}
The STREAM benchmark~\cite{stream} is a synthetic benchmark program that measures sustainable memory bandwidth for simple vector kernels in high-performance computers. STREAM was developed as a proxy for the basic computational kernels in scientific computations~\cite{mccalpin1995stream} and includes Copy, Scale, Sum, and Triad kernels. STREAM has a dedicated version to benchmark PMem modules by allocating and accessing PMem via PMDK (STREAM-PMem~\cite{fagerheim2021benchmarking}). 

The excerpt presented in \autoref{lst:orgStreamDram} constitutes a portion of the initial codebase that has since been extracted from the current version of the code.
\vspace{-0.2cm}
\begin{lstlisting}[caption={Original STREAM benchmark code at line 175-181.}, label={lst:orgStreamDram}]
#ifndef STREAM_TYPE
#define STREAM_TYPE double
#endif
static STREAM_TYPE  a[STREAM_ARRAY_SIZE+OFFSET],
                    b[STREAM_ARRAY_SIZE+OFFSET],
                    c[STREAM_ARRAY_SIZE+OFFSET];
\end{lstlisting}
The content represented in \autoref{lst:orgStreamDram} has been substituted in STREAM-PMem~\cite{fagerheim2021benchmarking} with the code demonstrated in \autoref{lst:orgStreamNvdimm}. The code commences by accessing the memory pool. Furthermore, a function named \textit{initiate} is employed to initialize the three arrays. Following this initialization, the code proceeds to execute the remaining segments of the STREAM benchmark code, mirroring the structure of the original STREAM benchmark code.
\vspace{-0.2cm}
\begin{lstlisting}[caption={Code that has replaced original code.}, label={lst:orgStreamNvdimm}]
PMEMobjpool *pop;
POBJ_LAYOUT_BEGIN(array);
POBJ_LAYOUT_TOID(array, double);
POBJ_LAYOUT_END(array); //Declearing the arrays
TOID(double) a, b, c;
void initiate() { //Initiating the arrays.
	POBJ_ALLOC(pop, &a, double, (STREAM_ARRAY_SIZE+OFFSET)*sizeof(STREAM_TYPE), NULL, NULL); //Same for b and c.
int main(){
	const char path[] = ".../pool.obj";
	pop = pmemobj_create(path, LAYOUT_NAME, 10737418240, 0666);
	if (pop == NULL)
		pop = pmemobj_open(path, LAYOUT_NAME);
	if (pop == NULL) {
		perror(path);
		exit(1);	}
	initiate();
	//The rest of the STREAM benchmark after this.
 }
\end{lstlisting}

In this work, we employ STREAM in those two versions to showcase the shift from PMem to CXL. Throughout this demonstration, we illustrate how programs designed for PMem can seamlessly operate on CXL-enabled devices. Furthermore, we provide performance assessments to anticipate the impact of CXL on performance in relation to local RAM (DDR4 and DDR5) and local PMem-like devices (emulation of remote sockets either for memory expansion or as a direct access device, as done in~\cite{foyer2023survey, bergstrom2011measuring}). 

In contrast to previous research that primarily emphasizes demonstrating the use of CXL memory for in-memory database queries or file system operations~\cite{ahn2022enabling, lee2023elastic}, STREAM memory access involves accessing and manipulating large arrays, making it particularly applicable and significant for scientific computations in HPC systems. Moreover, STREAM is implemented with OpenMP threads, which is the common shared-memory paradigm in scientific computing for parallelism~\cite{dagum1998openmp}. 

\subsection{Test Configurations}
\label{confs}
The methodology of this work is to employ STREAM and STREAM-PMem in various CPU and memory configurations, taking into account the availability of DRAM and CXL memory available on the HPC setups, as will be described next. 
The results presented in \autoref{Scale_results}, \autoref{Add_results}, \autoref{Copy_results}, \autoref{Triad_results} refer to STREAM executions with 100M array elements for Scale, Add, Copy, and Triad operations correspondingly. For each STREAM method, the results of our tests are presented in 2 classes (and a total of 5 groups), divided conceptually for unique comparisons. We sub-divide those 5 groups into two classes. The first class (Class 1, \ref{item:group_a}-\ref{item:group_c}) refers to the equivalent of the \textit{App-Direct} mode in PMem in which we directly access the local or remote memory (either in the alternative socket or in the CXL memory), and the second class (Class 2, \ref{item:group_d}-\ref{item:group_e}) refers to the \textit{Memory Mode} in PMem, in which we increase the available memory using other CC-NUMA nodes:

\noindent\textbf{Class 1 --- \textit{App-Direct}}:
\begin{enumerate}[label=(\alph*),wide, labelwidth=!, labelindent=0pt]

    \item\label{item:group_a} 
    \textbf{Local memory access as PMem:} Configurations within this group involve accessing local memory (on-socket memory) in \textit{App-Direct} mode (thus benchmarking STREAM-PMem). 

    \item\label{item:group_b} \textbf{Remote memory access as PMem:} Configurations within this group involve computing cores on a single socket that access remote memory  in \textit{App-Direct} mode (thus benchmarking STREAM-PMem). The term "remote memory" in this context encompasses both CXL-attached memory and on-node memory accessed from the alternative CPU socket (i.e., memory accessed through the UPI).  
    \item\label{item:group_c} 
    \textbf{ Remote memory as PMem (thread affinity):} Configurations within this group involve computing cores in both sockets that access remote memory  in \textit{App-Direct} mode (thus benchmarking STREAM-PMem) using two distinct thread affinity methods: \textit{close} and \textit{spread}. The \textit{close} method populates an entire socket first and then adds cores from the second socket. The \textit{spread} method, on the opposite, adds cores alternately from both sockets.

\end{enumerate}

\noindent\textbf{Class 2 --- \textit{Memory Mode}}:

\begin{enumerate}[label=(\alph*),wide, labelwidth=!, labelindent=0pt]
    
    \item\label{item:group_d} \textbf{Remote CC-NUMA:} Configurations within this group involve computing cores on a single socket that access remote memory as CC-NUMA.
    
    \item\label{item:group_e}
     \textbf{Remote CC-NUMA (all cores):} Configurations within this group involve cores on both CPU sockets accessing remote memory as CC-NUMA. This includes configurations where both sockets operate and access memory on one of them since these workloads include remote accesses.  
    
\end{enumerate}

For better clarity, the data flow for each test configuration is demonstrated in Figure~\ref{dataflows}. Each row in Figure~\ref{dataflows} contains the data flow examinations of the test groups of the two classes. Thus, in each of our test groups, for each of the STREAM operations (Figures ~\ref{Scale_results}, \ref{Add_results}, \ref{Copy_results}, \ref{Triad_results}), the way to understand each trend, and its correspondence to the relevant dataflow, is given in the trend itself by a combination of three: symbol, color and memory annotation. The symbol is used to distinguish between accessing on-node DDR4 ({\tiny\ding{115}}), on-node DDR5 ({\tiny\ding{108}}) or CXL-attached DDR4 ({\textbf{$\times$}}). The color implies the active compute cores –-- either in socket0, socket1, or both. The annotations \textit{pmem$\#\{0,1,2\}$} or \textit{numa$\#\{0,1,2\}$} accompanying each trend provide an explanation of the accesses memory location: 0 for socket0; 1 for socket1; and 2 for CXL memory. \textit{numa} signifies STREAM accessing memory as NUMA memory expansion, while \textit{pmem} represents STREAM-PMem accessing memory using PMDK. 

\section{Results and Analysis}

\begin{figure*}
  \begin{subfigure}{0.32\textwidth}
    \centering
    \includegraphics[width=\linewidth]{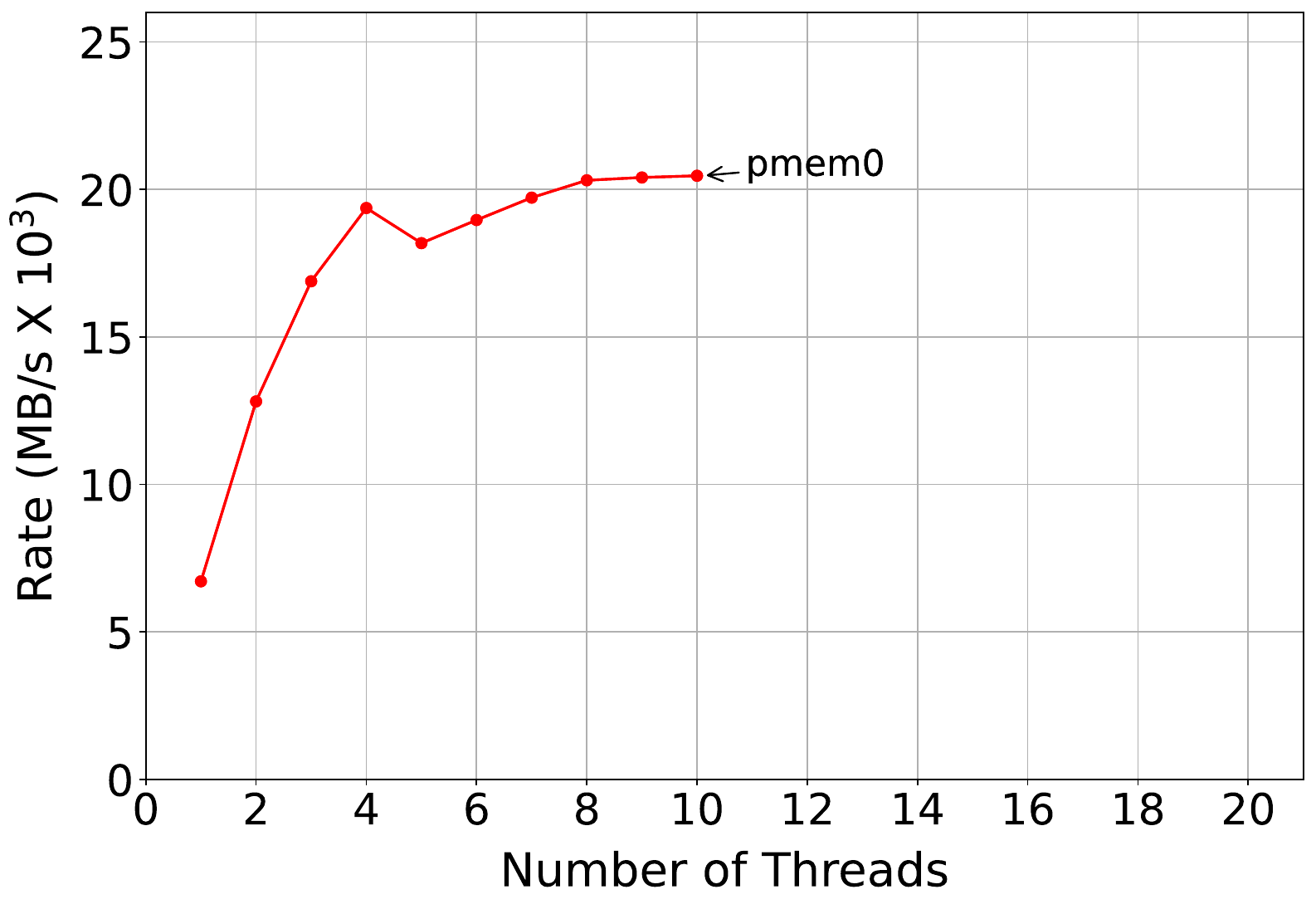}
    \caption{Class 1.a: Local memory access as PMem} 
    \label{fig:Scale_d}
  \end{subfigure}\hfill
  \begin{subfigure}{0.32\textwidth}
    \centering
     \includegraphics[width=\linewidth]
    {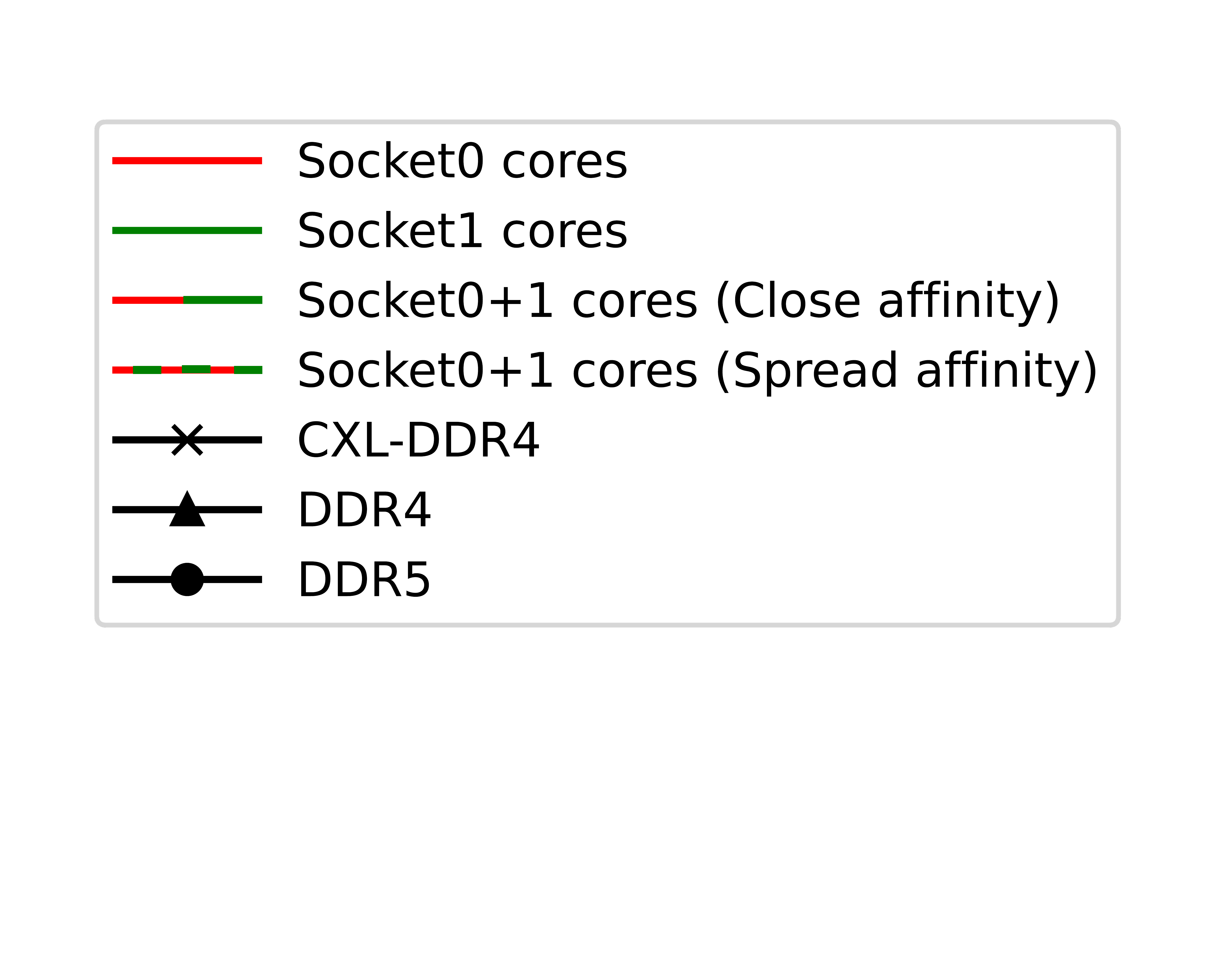}
  \end{subfigure}\hfill
  \begin{subfigure}{0.32\textwidth}
    \centering
     \includegraphics[width=\linewidth]{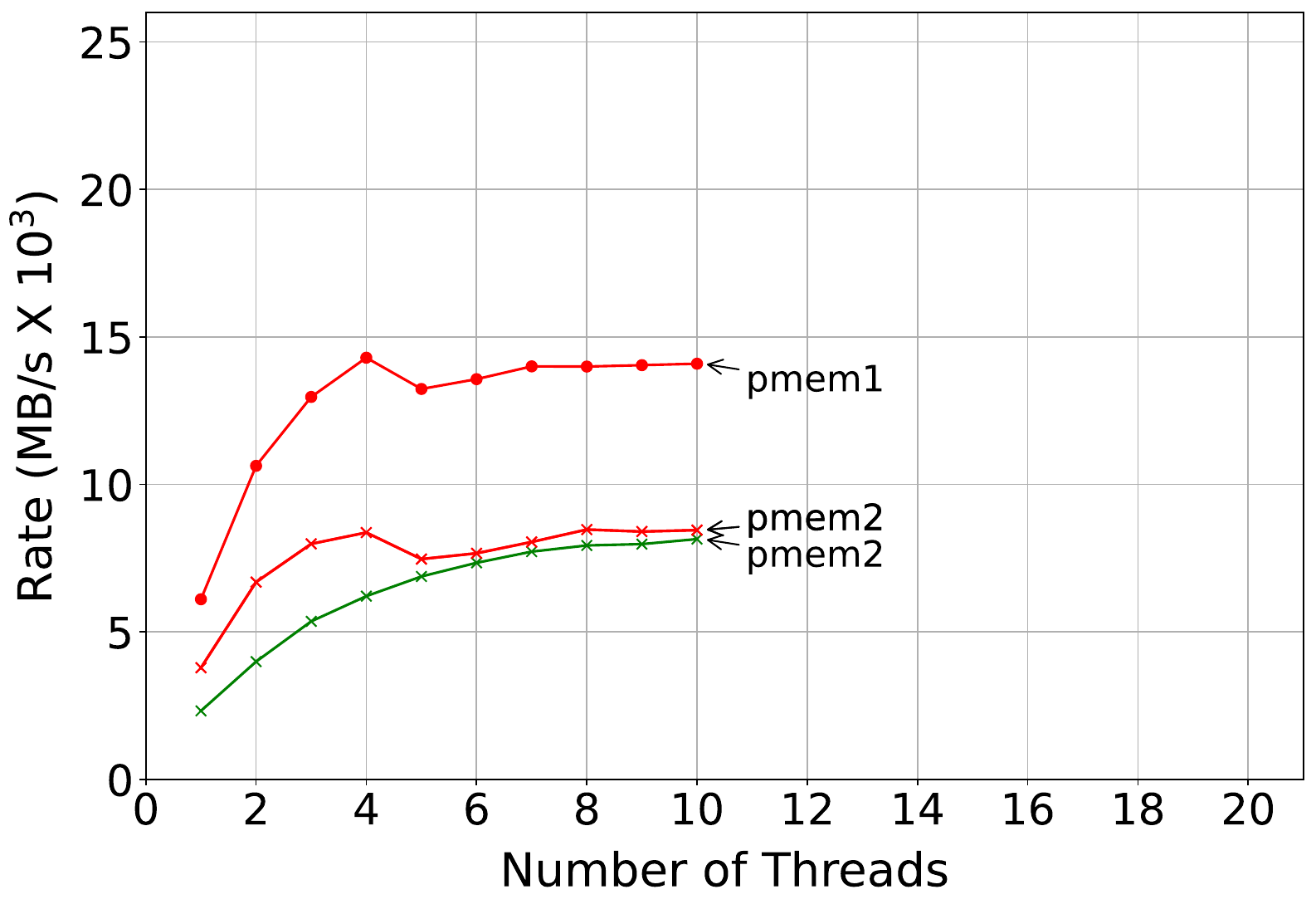}
    \caption{Class 1.b: Remote memory access as PMem} 
    \label{fig:Scale_a}
  \end{subfigure}

  \medskip

  \begin{subfigure}{0.32\textwidth}
    \centering
     \includegraphics[width=\linewidth]{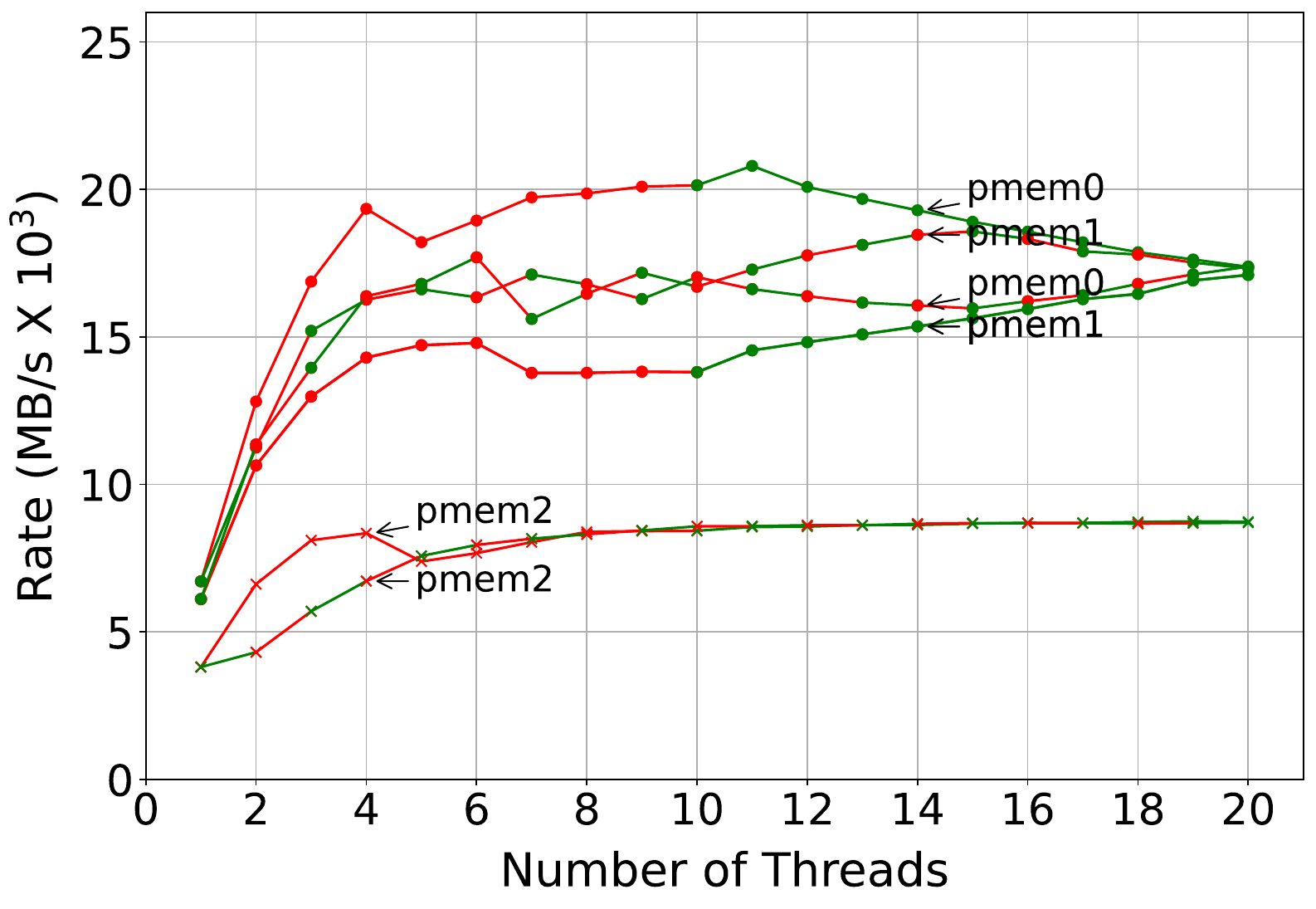}
    \caption{Class 1.c: Remote memory as PMem (thread affinity)} 
    \label{fig:Scale_e}
  \end{subfigure}\hfill
  \begin{subfigure}{0.32\textwidth}
    \centering
     \includegraphics[width=\linewidth]{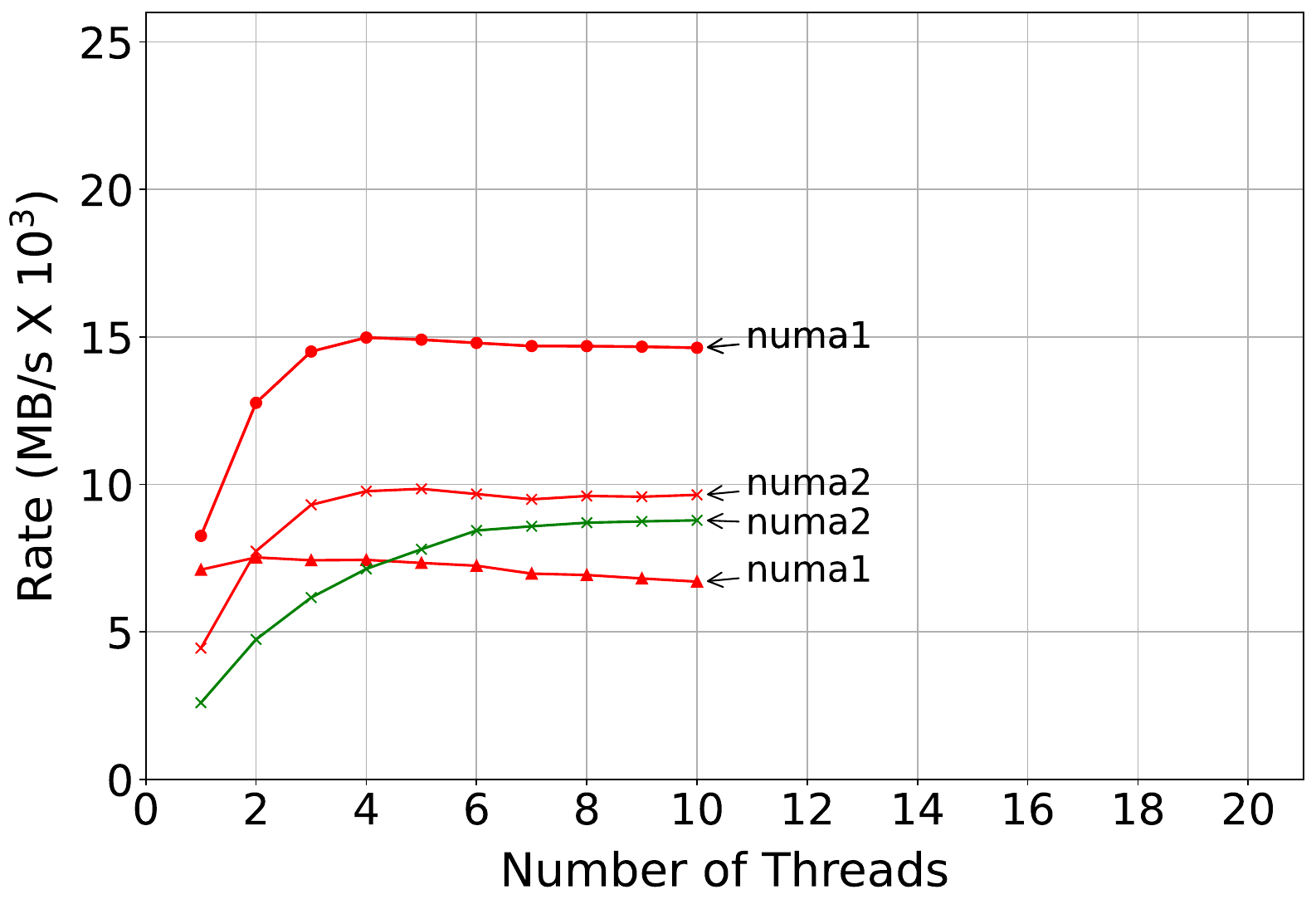}
    \caption{Class 2.a: Remote CC-NUMA \vspace{10pt}} %
    \label{fig:Scale_b}
  \end{subfigure}\hfill
  \begin{subfigure}{0.32\textwidth}
    \centering
     \includegraphics[width=\linewidth]{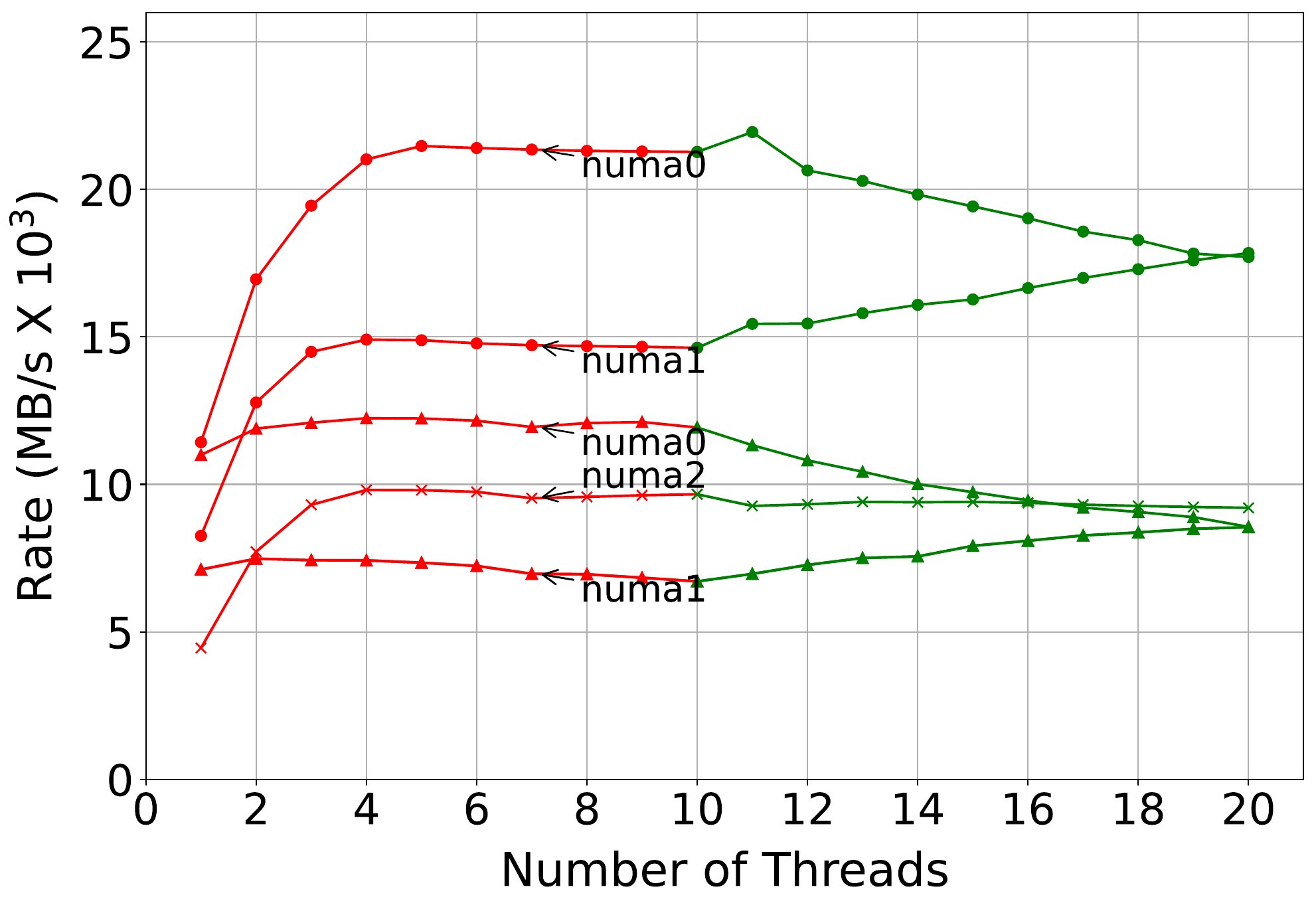}
        \caption{Class 2.b: Remote CC-NUMA (all cores) \vspace{10pt}}
    \label{fig:Scale_c}
  \end{subfigure}
  \caption{\textbf{SCALE ---} Various STREAM test configurations. Refer to Section~\ref{confs} for definition of test groups 1.\ref{item:group_a}, 1.\ref{item:group_b}, 1.\ref{item:group_c}, 2.\ref{item:group_d}, 2.\ref{item:group_e} and legend clarifications.}
        \captionsetup{format=default} 
 \rule{\linewidth}{0.4pt}
\label{Scale_results}
\end{figure*}

\begin{figure*}
  \begin{subfigure}{0.32\textwidth}
    \centering
    \includegraphics[width=\linewidth]{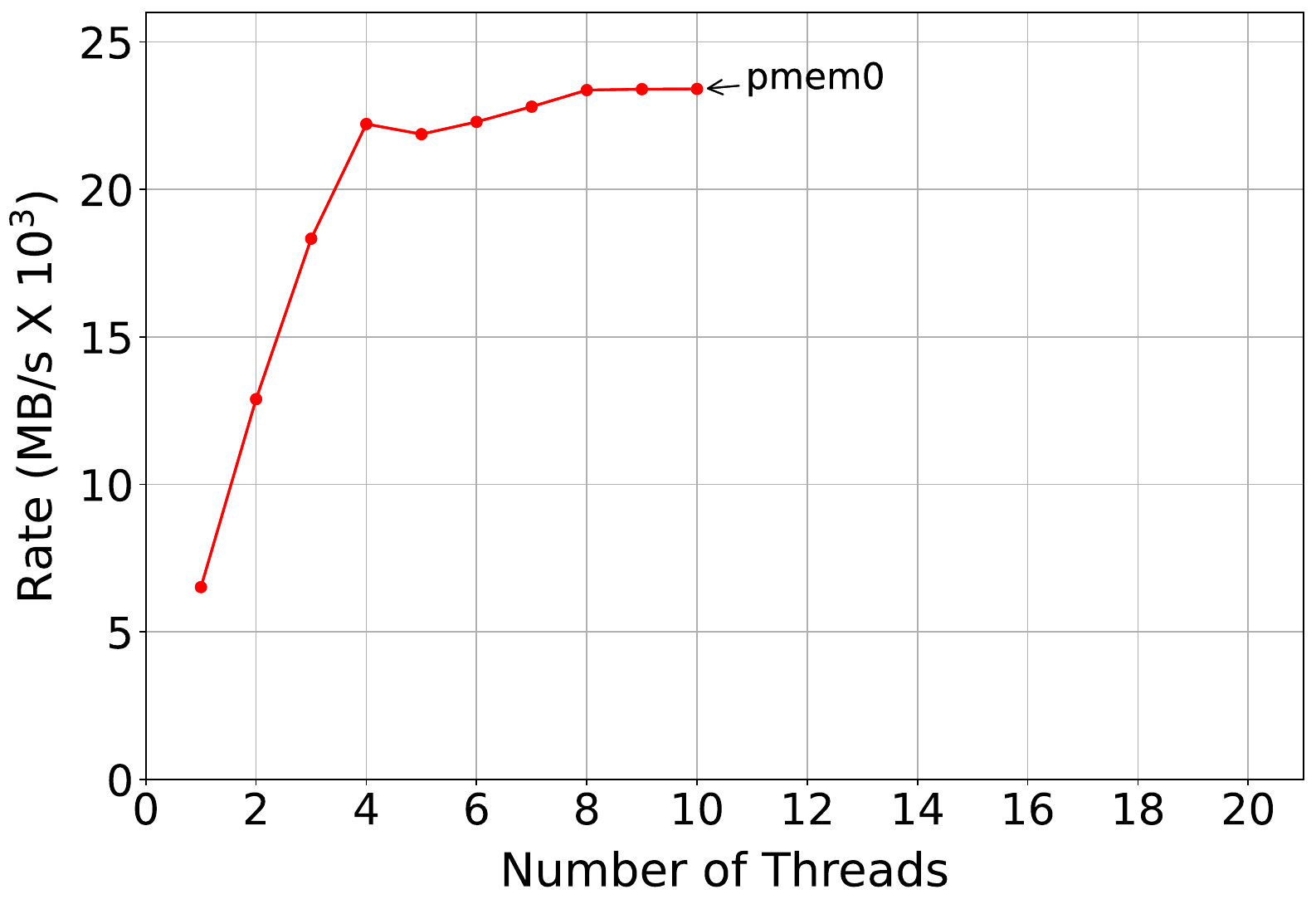}
    \caption{Class 1.a: Local memory access as PMem} 
    \label{fig:Add_d}
  \end{subfigure}\hfill
  \begin{subfigure}{0.32\textwidth}
    \centering
    \includegraphics[width=\linewidth]
    {figures/results/legend_fix.pdf}
  \end{subfigure}\hfill
  \begin{subfigure}{0.32\textwidth}
    \centering
    \includegraphics[width=\linewidth]{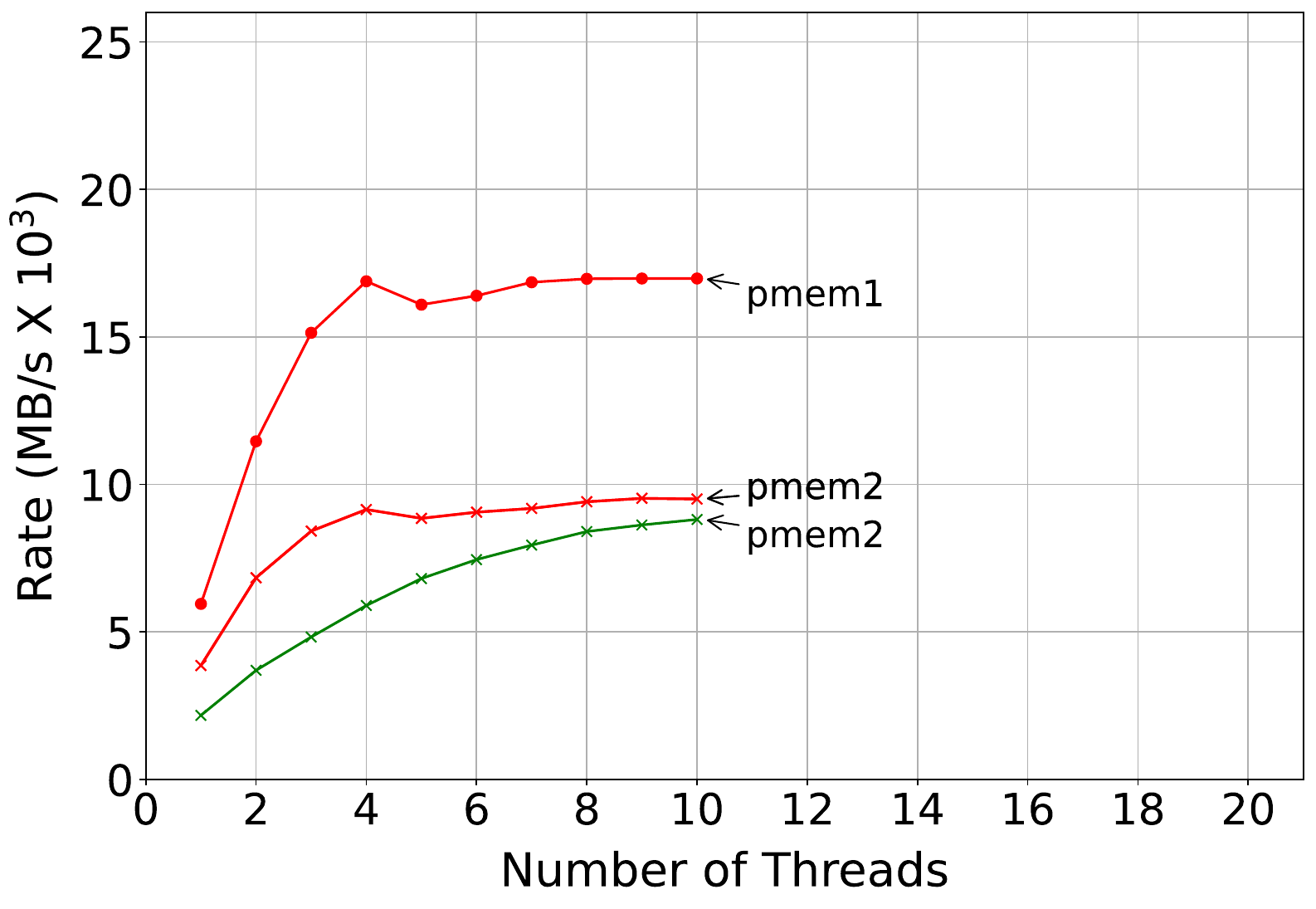}
    \caption{Class 1.b: Remote memory access as PMem} 
    \label{fig:Add_a}
  \end{subfigure}

  \medskip

  \begin{subfigure}{0.32\textwidth}
    \centering
    \includegraphics[width=\linewidth]{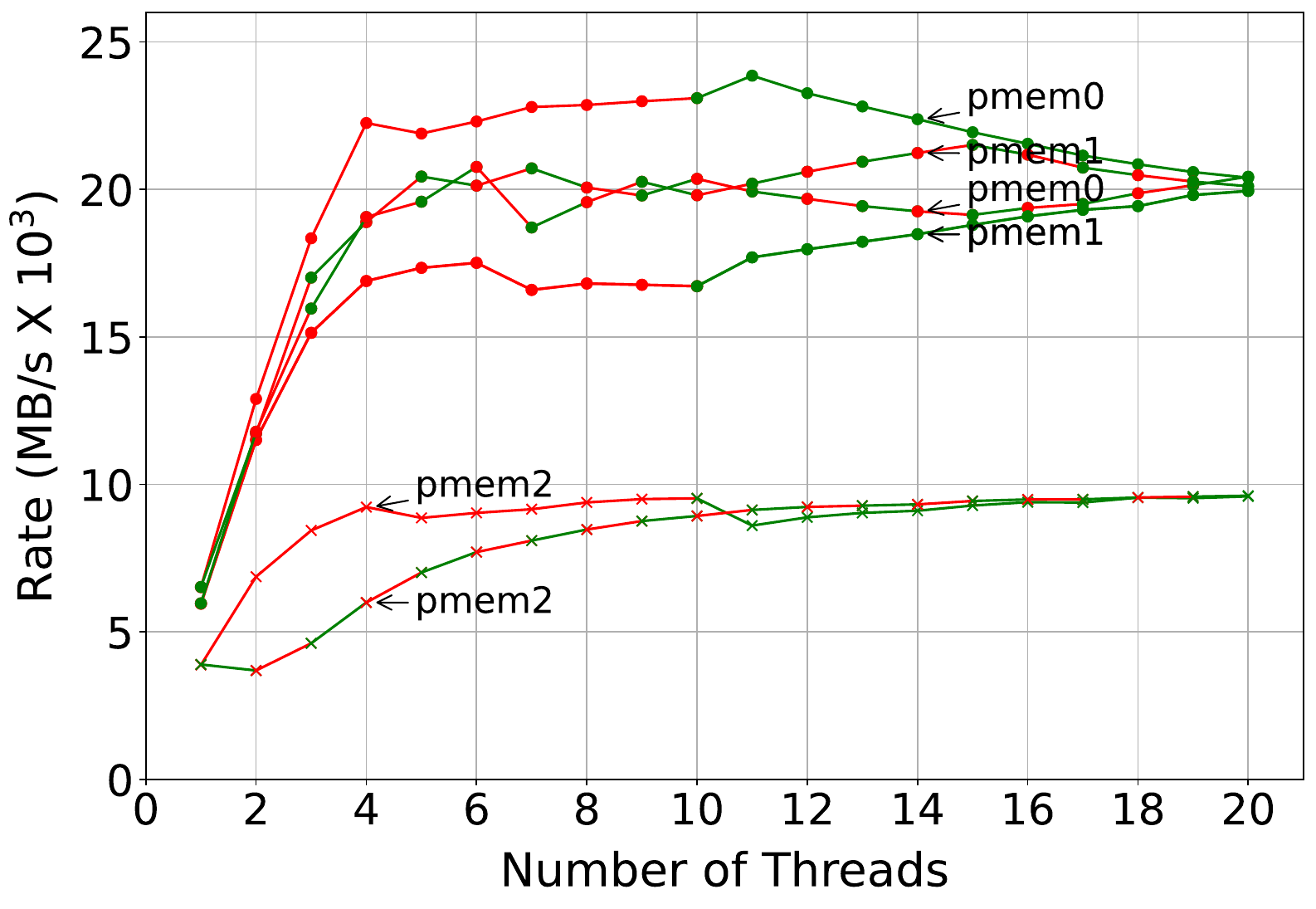}
    \caption{Class 1.c: Remote memory as PMem (thread affinity)} 
    \label{fig:Add_e}
  \end{subfigure}\hfill
  \begin{subfigure}{0.32\textwidth}
    \centering
    \includegraphics[width=\linewidth]{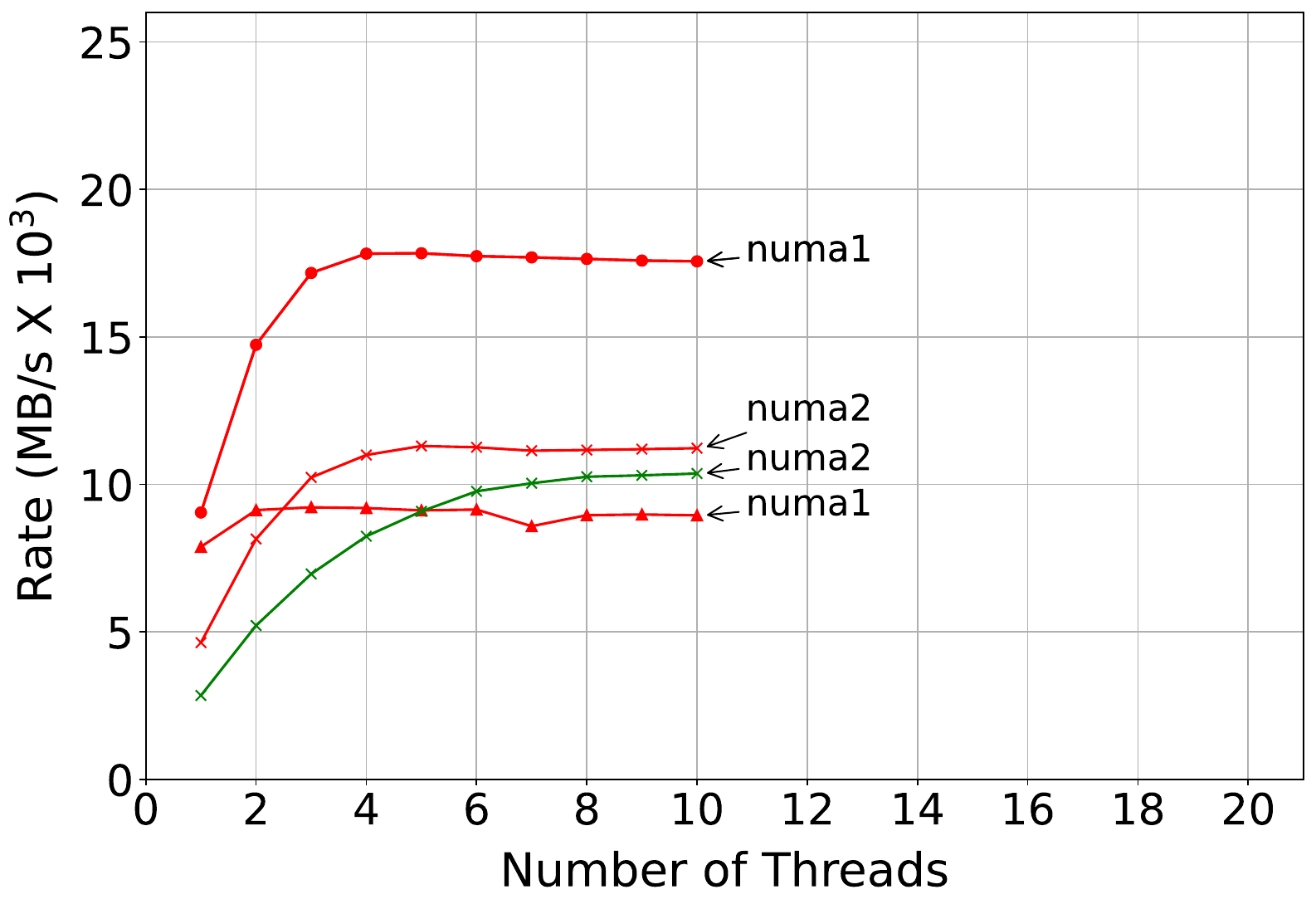}
    \caption{Class 2.a: Remote CC-NUMA \vspace{10pt}} %
    \label{fig:Add_b}
  \end{subfigure}\hfill
  \begin{subfigure}{0.32\textwidth}
    \centering
    \includegraphics[width=\linewidth]{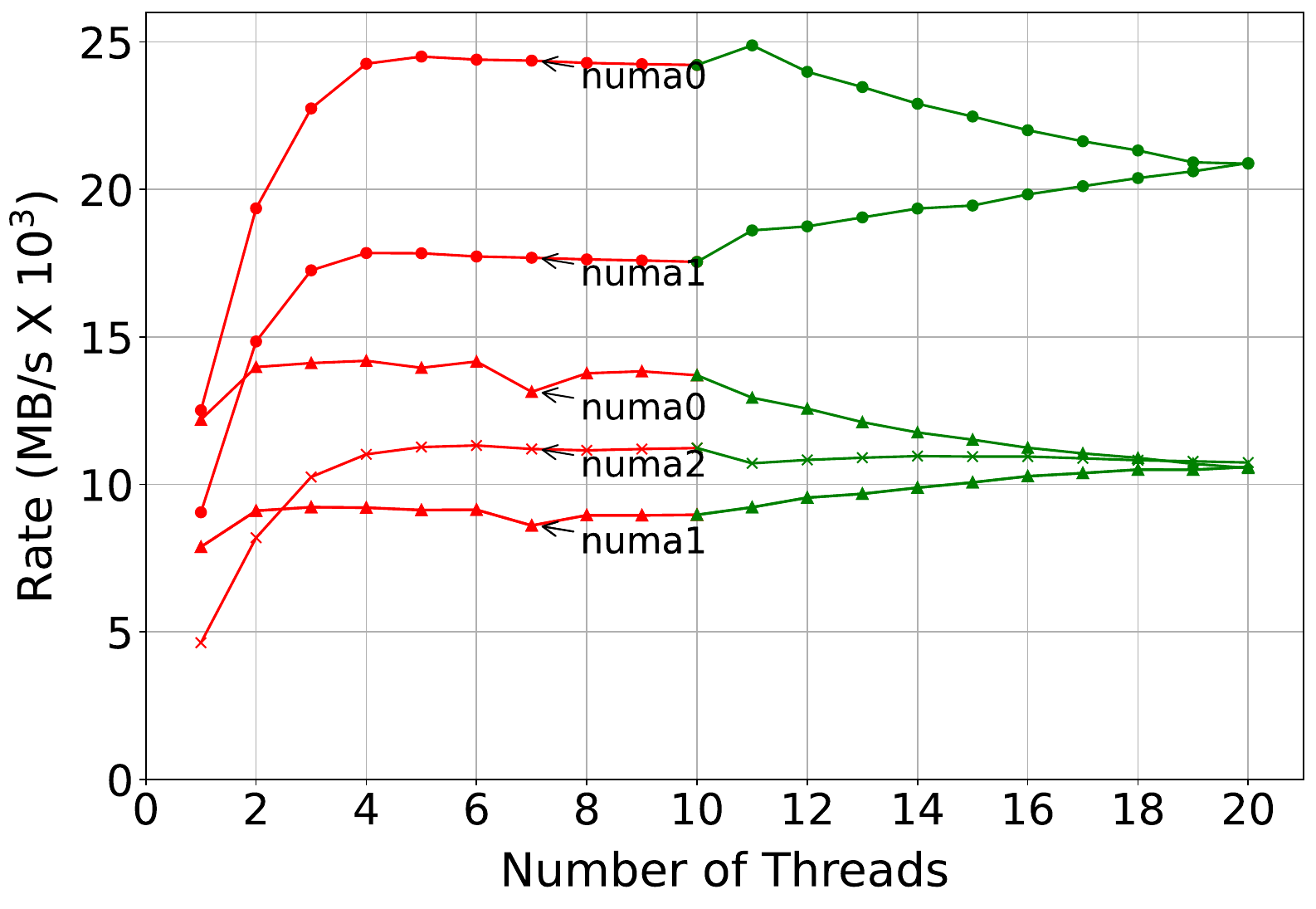}
        \caption{Class 2.b: Remote CC-NUMA (all cores) \vspace{10pt}}
    \label{fig:Add_c}
  \end{subfigure}
  \caption{\textbf{ADD ---} Various STREAM test configurations. Refer to Section~\ref{confs} for definition of test groups 1.\ref{item:group_a}, 1.\ref{item:group_b}, 1.\ref{item:group_c}, 2.\ref{item:group_d}, 2.\ref{item:group_e} and legend clarifications.}
        \captionsetup{format=default} 
 
\label{Add_results}
\end{figure*}

\begin{figure*}
  \begin{subfigure}{0.32\textwidth}
    \centering
    \includegraphics[width=\linewidth]{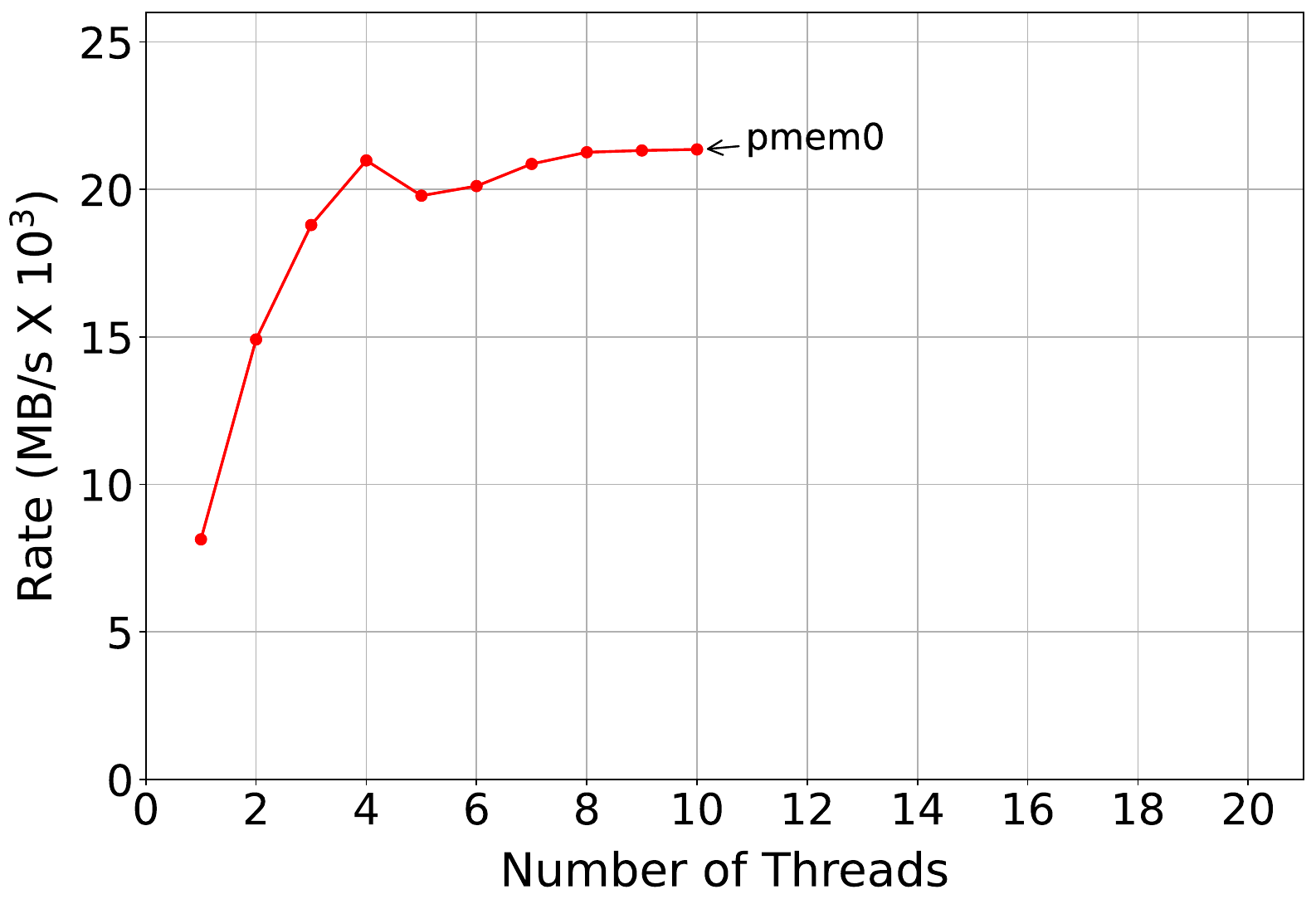}
    \caption{Class 1.a: Local memory access as PMem} 
    \label{fig:Copy_d}
  \end{subfigure}\hfill
  \begin{subfigure}{0.32\textwidth}
    \centering
    \includegraphics[width=\linewidth]
    {figures/results/legend_fix.pdf}
  \end{subfigure}\hfill
  \begin{subfigure}{0.32\textwidth}
    \centering
    \includegraphics[width=\linewidth]{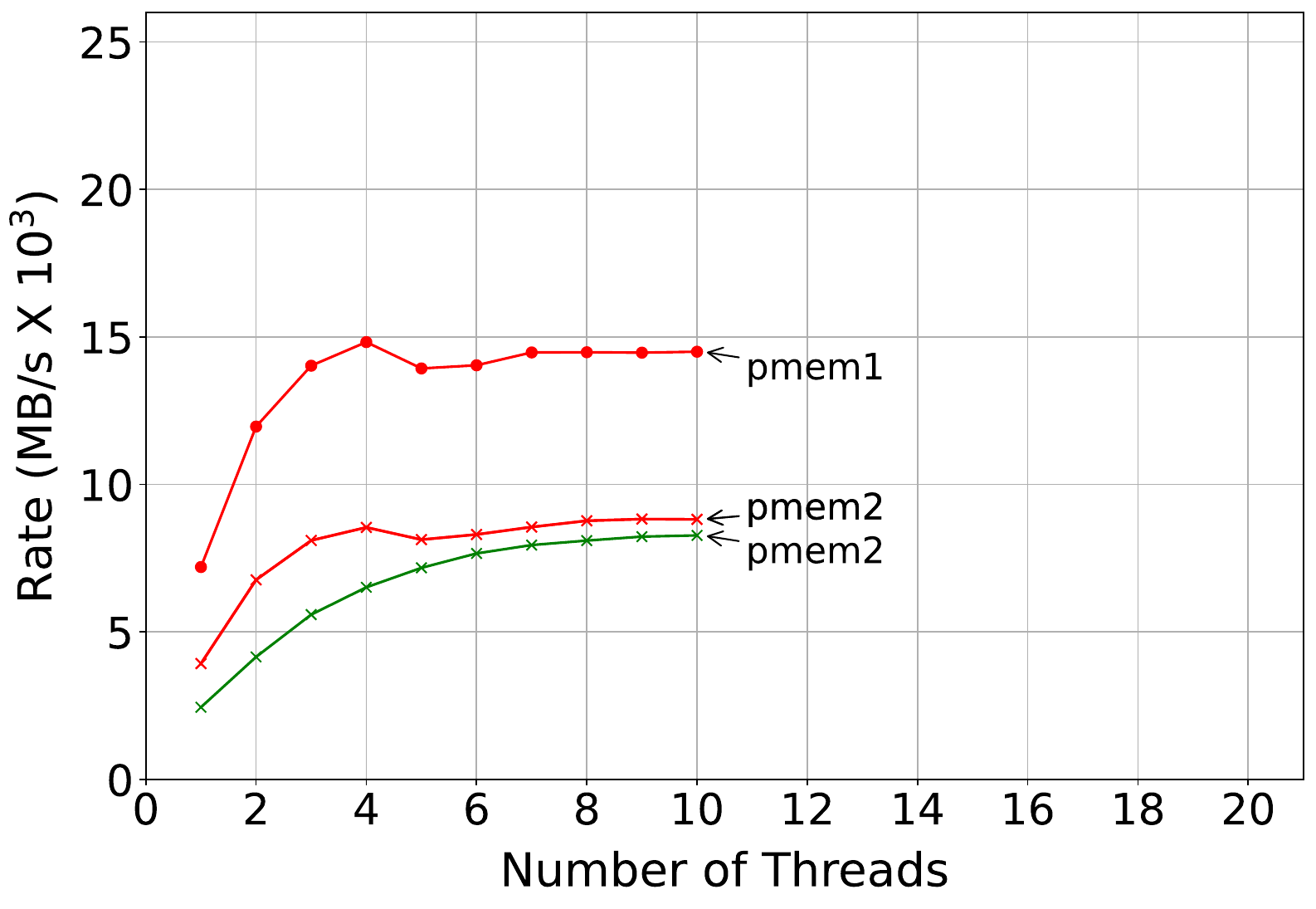}
    \caption{Class 1.b: Remote memory access as PMem} 
    \label{fig:Copy_a}
  \end{subfigure}

  \medskip

  \begin{subfigure}{0.32\textwidth}
    \centering
    \includegraphics[width=\linewidth]{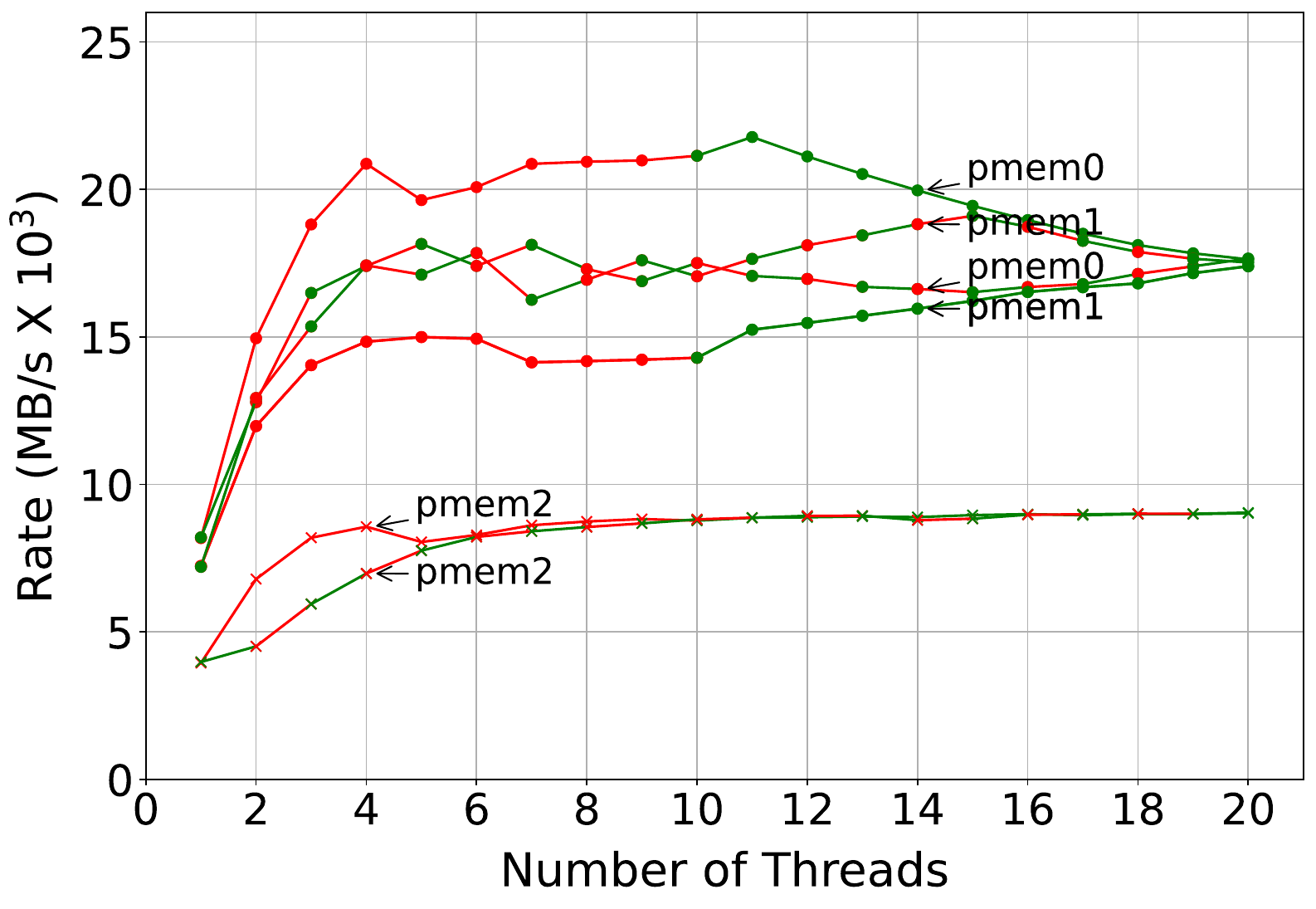}
    \caption{Class 1.c: Remote memory as PMem (thread affinity)} 
    \label{fig:Copy_e}
  \end{subfigure}\hfill
  \begin{subfigure}{0.32\textwidth}
    \centering
    \includegraphics[width=\linewidth]{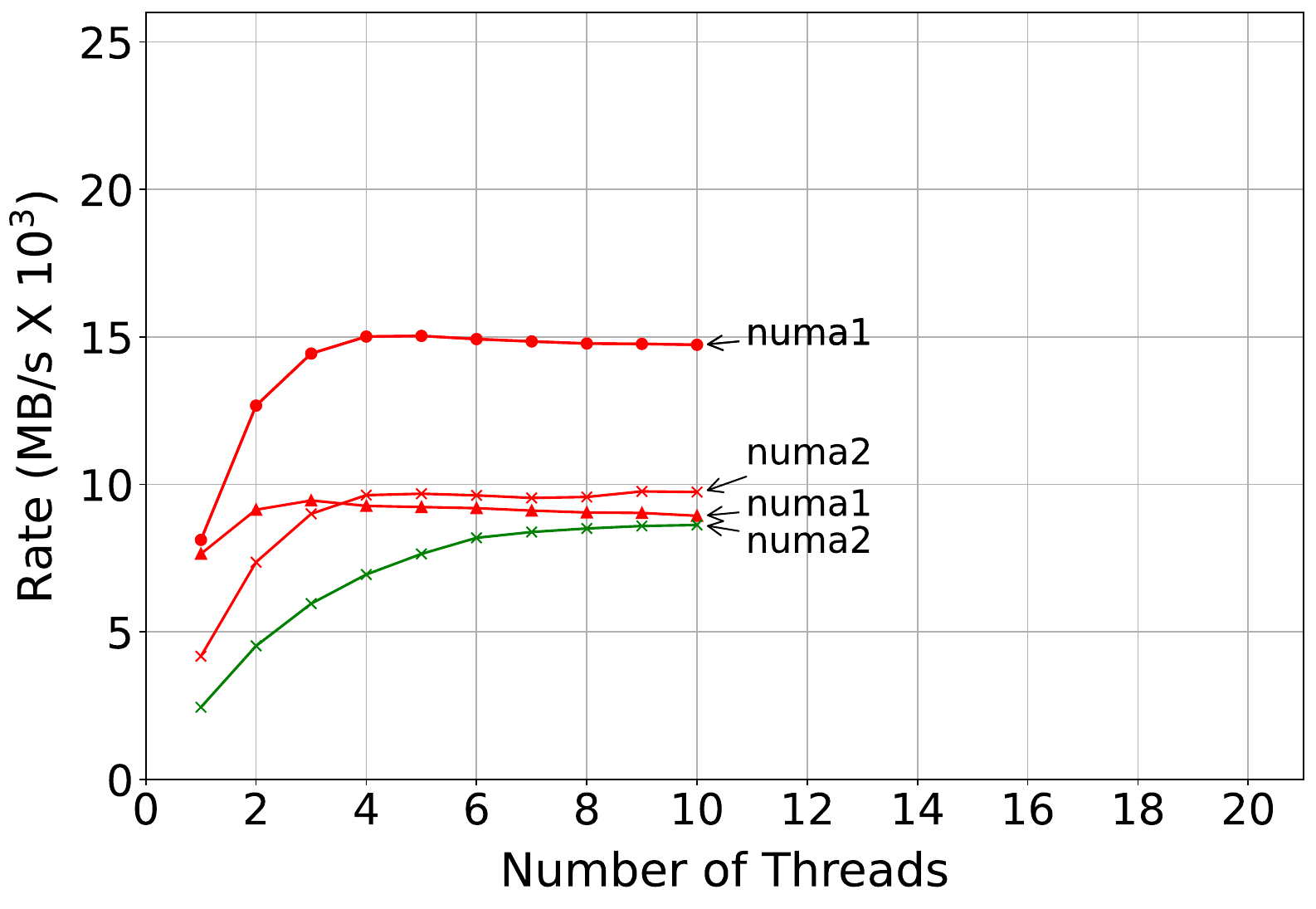}
    \caption{Class 2.a: Remote CC-NUMA \vspace{10pt}} %
    \label{fig:Copy_b}
  \end{subfigure}\hfill
  \begin{subfigure}{0.32\textwidth}
    \centering
    \includegraphics[width=\linewidth]{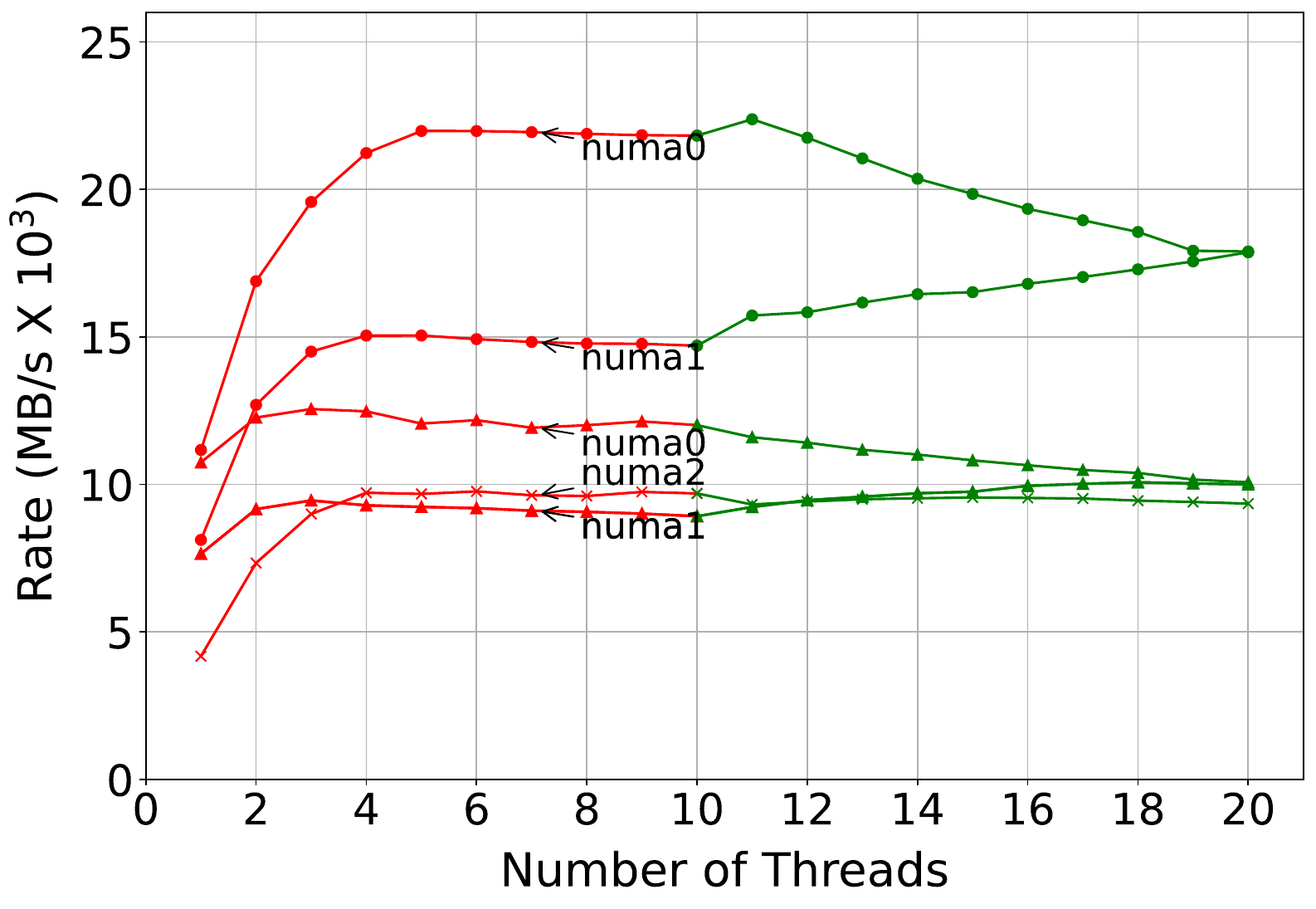}
        \caption{Class 2.b: Remote CC-NUMA (all cores) \vspace{10pt}}
    \label{fig:Copy_c}
  \end{subfigure}
  \caption{\textbf{COPY ---} Various STREAM test configurations. Refer to Section~\ref{confs} for definition of test groups 1.\ref{item:group_a}, 1.\ref{item:group_b}, 1.\ref{item:group_c}, 2.\ref{item:group_d}, 2.\ref{item:group_e} and legend clarifications.}
        \captionsetup{format=default} 
 \rule{\linewidth}{0.4pt}
\label{Copy_results}
\end{figure*}

\begin{figure*}
  \begin{subfigure}{0.32\textwidth}
    \centering
    \includegraphics[width=\linewidth]{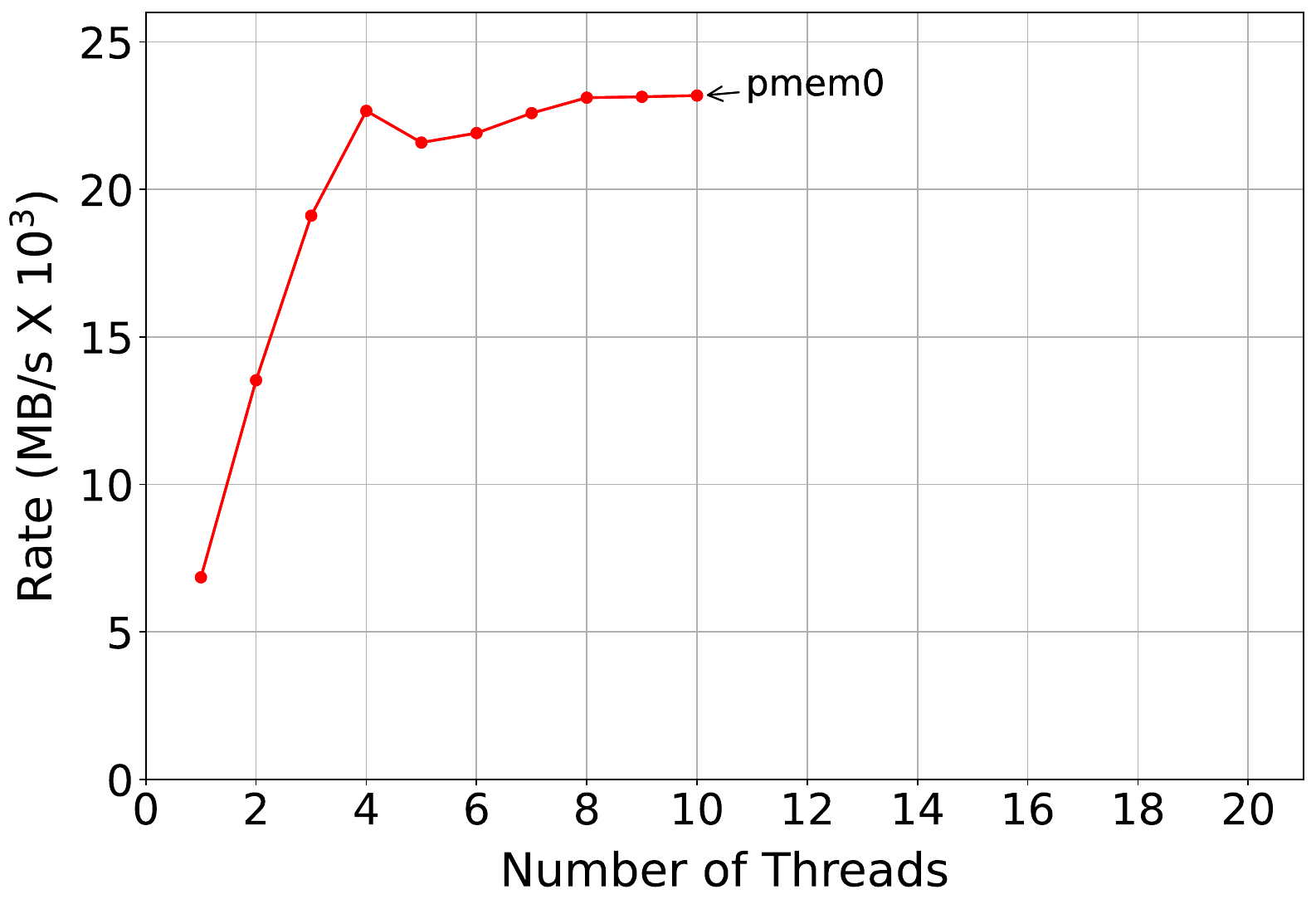}
    \caption{Class 1.a: Local memory access as PMem} 
    \label{fig:Triad_d}
  \end{subfigure}\hfill
  \begin{subfigure}{0.32\textwidth}
    \centering
    \includegraphics[width=\linewidth]
    {figures/results/legend_fix.pdf}
  \end{subfigure}\hfill
  \begin{subfigure}{0.32\textwidth}
    \centering
    \includegraphics[width=\linewidth]{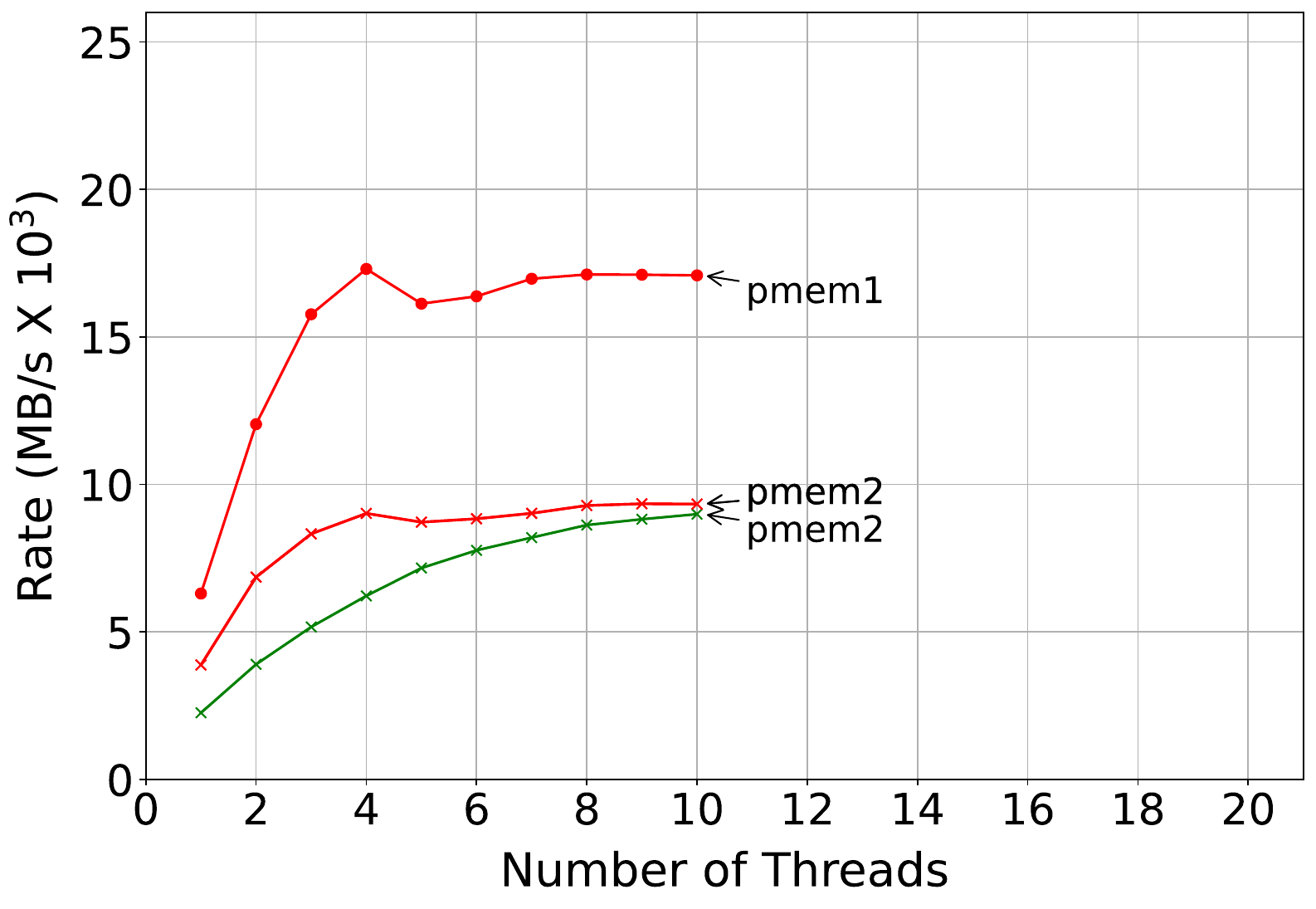}
    \caption{Class 1.b: Remote memory access as PMem} 
    \label{fig:Triad_a}
  \end{subfigure}

  \medskip

  \begin{subfigure}{0.32\textwidth}
    \centering
    \includegraphics[width=\linewidth]{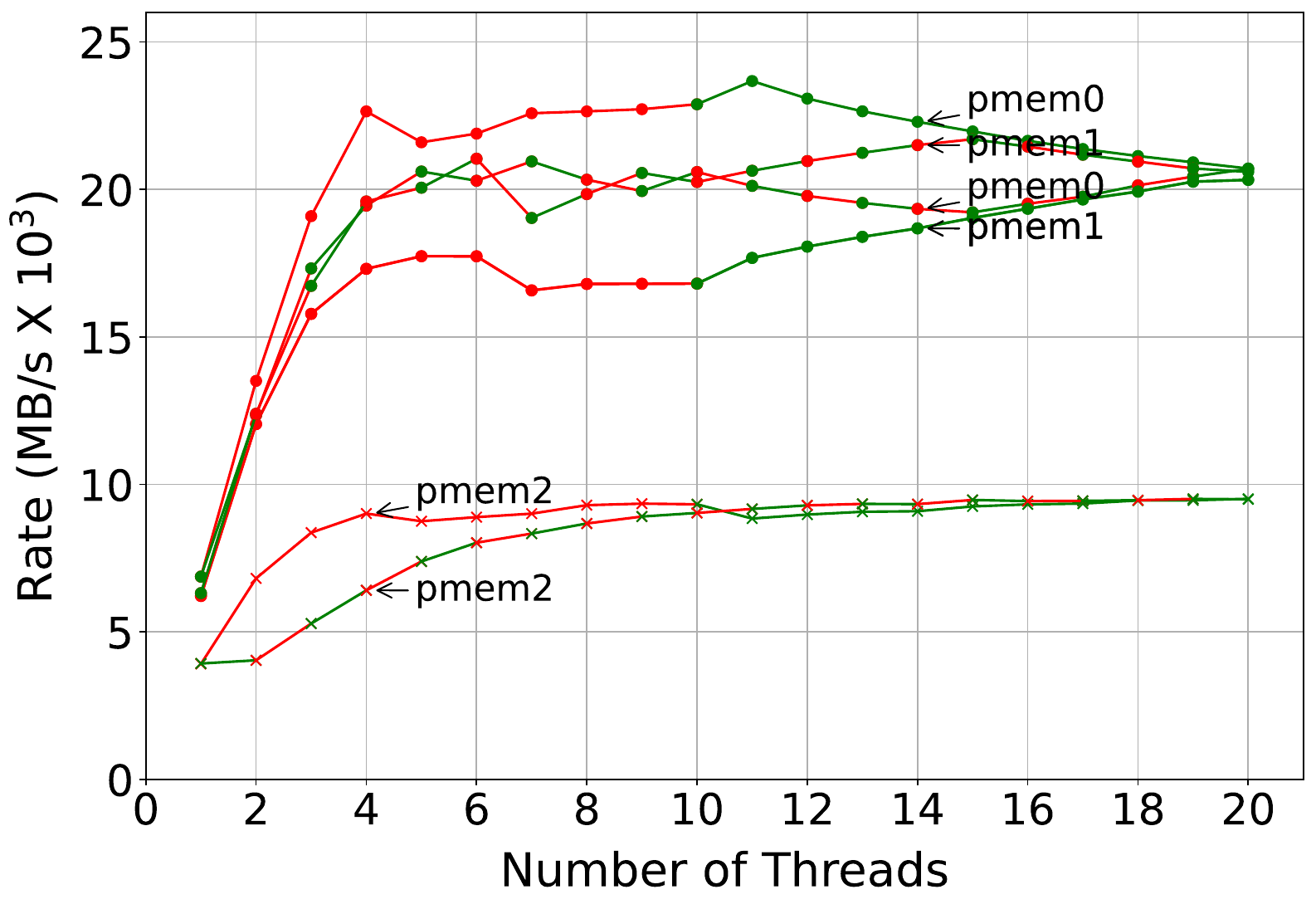}
    \caption{Class 1.c: Remote memory as PMem (thread affinity)} 
    \label{fig:Triad_e}
  \end{subfigure}\hfill
  \begin{subfigure}{0.32\textwidth}
    \centering
    \includegraphics[width=\linewidth]{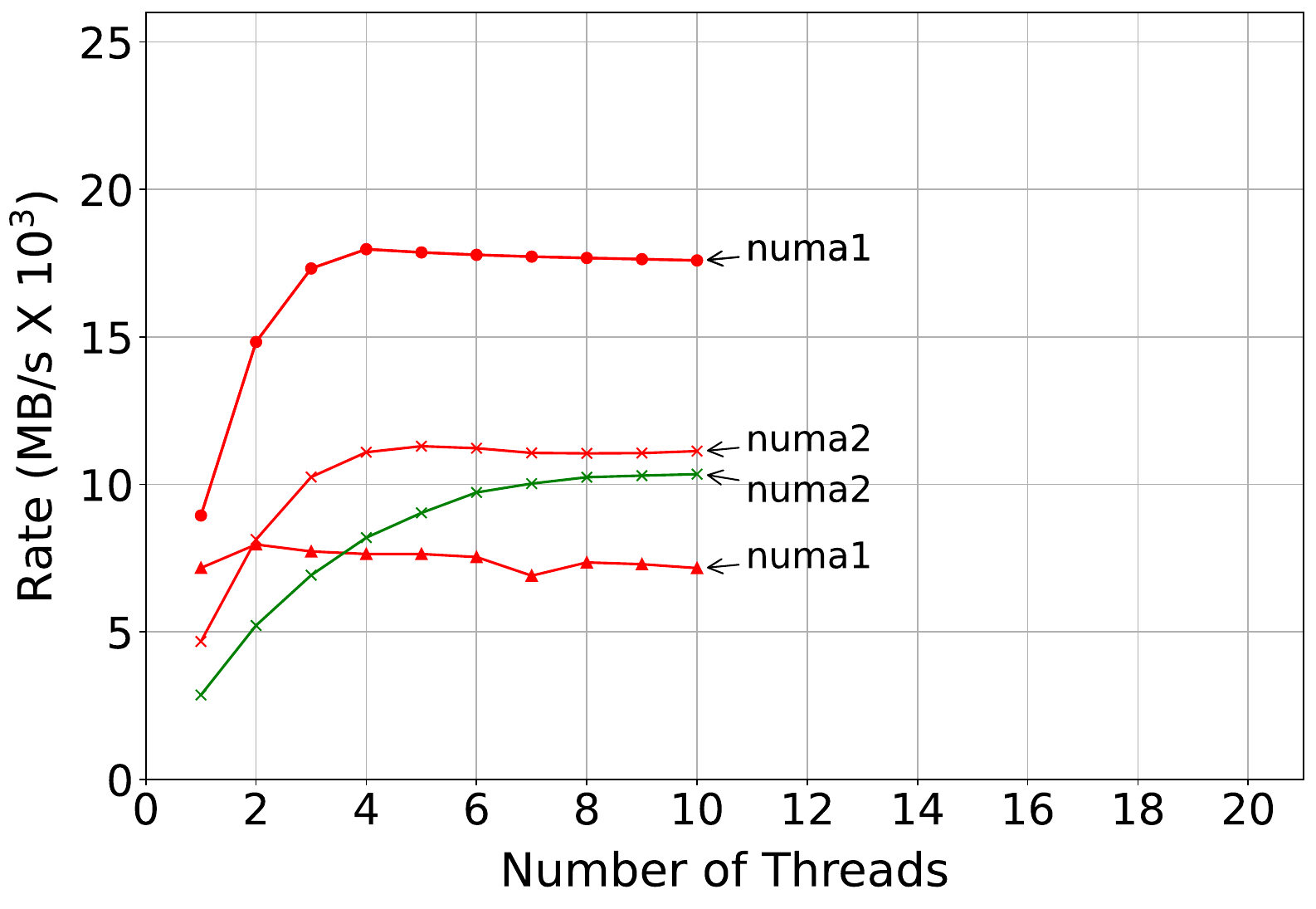}
    \caption{Class 2.a: Remote CC-NUMA \vspace{10pt}} %
    \label{fig:Triad_b}
  \end{subfigure}\hfill
  \begin{subfigure}{0.32\textwidth}
    \centering
    \includegraphics[width=\linewidth]{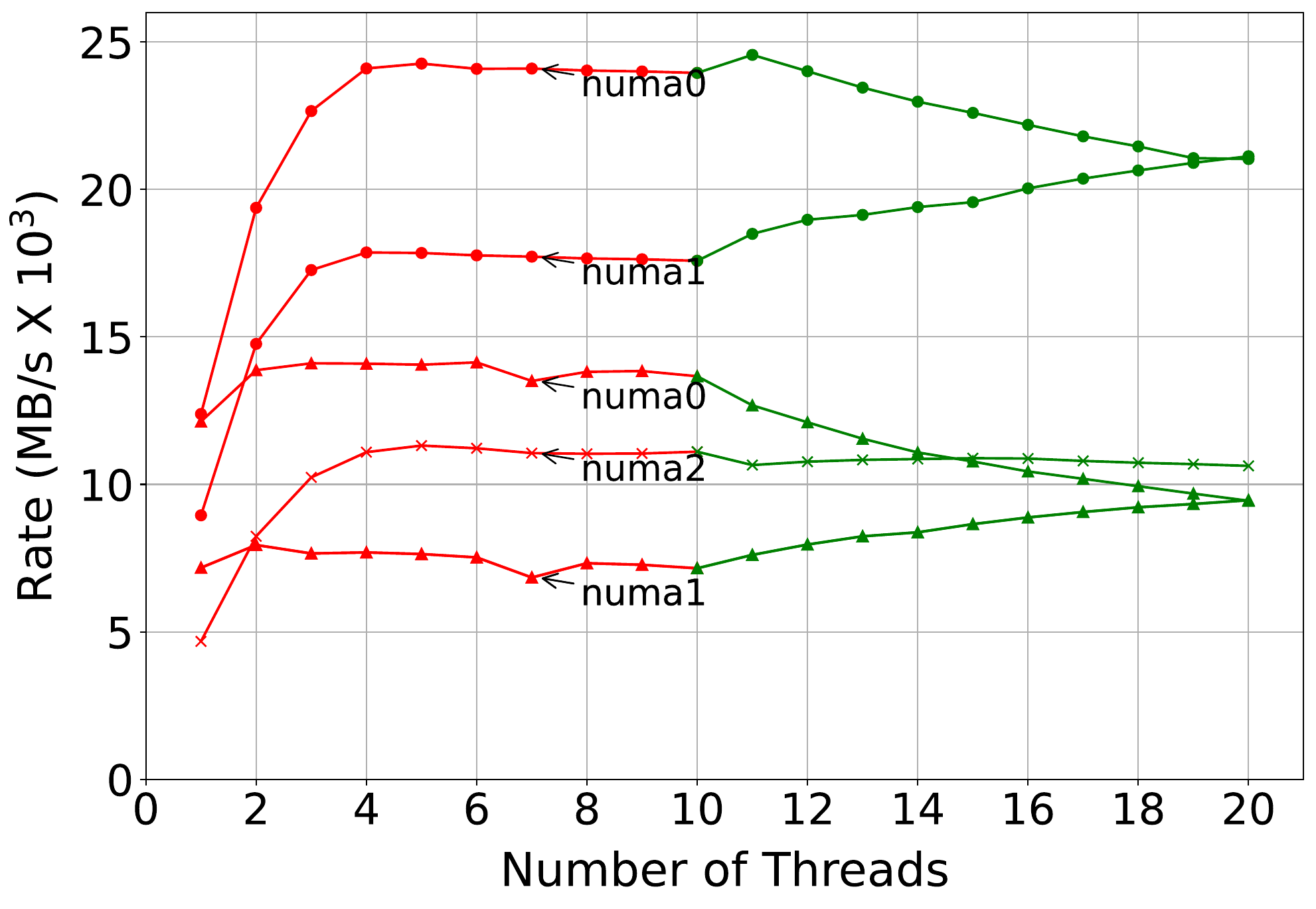}
        \caption{Class 2.b: Remote CC-NUMA (all cores) \vspace{10pt}}
    \label{fig:Triad_c}
  \end{subfigure}
  \caption{\textbf{TRIAD ---} Various STREAM test configurations. Refer to Section~\ref{confs} for definition of test groups 1.\ref{item:group_a}, 1.\ref{item:group_b}, 1.\ref{item:group_c}, 2.\ref{item:group_d}, 2.\ref{item:group_e} and legend clarifications.}
        \captionsetup{format=default} 

\label{Triad_results}
\end{figure*}

\begin{figure*}
  \centering
  \begin{sideways}
    \begin{minipage}{\textheight}
      \centering
      
      \textbf{(Class 1.a)}\hspace{1em} 
      \begin{subfigure}{0.26\textwidth}
        \centering
        \includegraphics[width=\columnwidth]{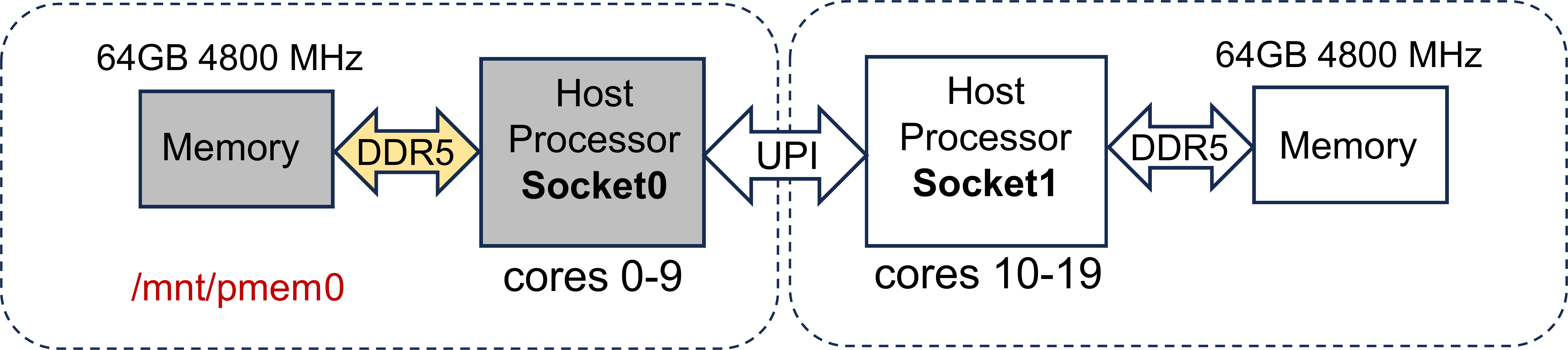}
      \end{subfigure}
        \captionsetup{format=default} 
 \rule{\linewidth}{0.4pt}
      \medskip

      \textbf{(Class 1.b)}\hspace{1em}
      \begin{subfigure}{0.26\textwidth}
        \centering
        \includegraphics[width=\columnwidth]{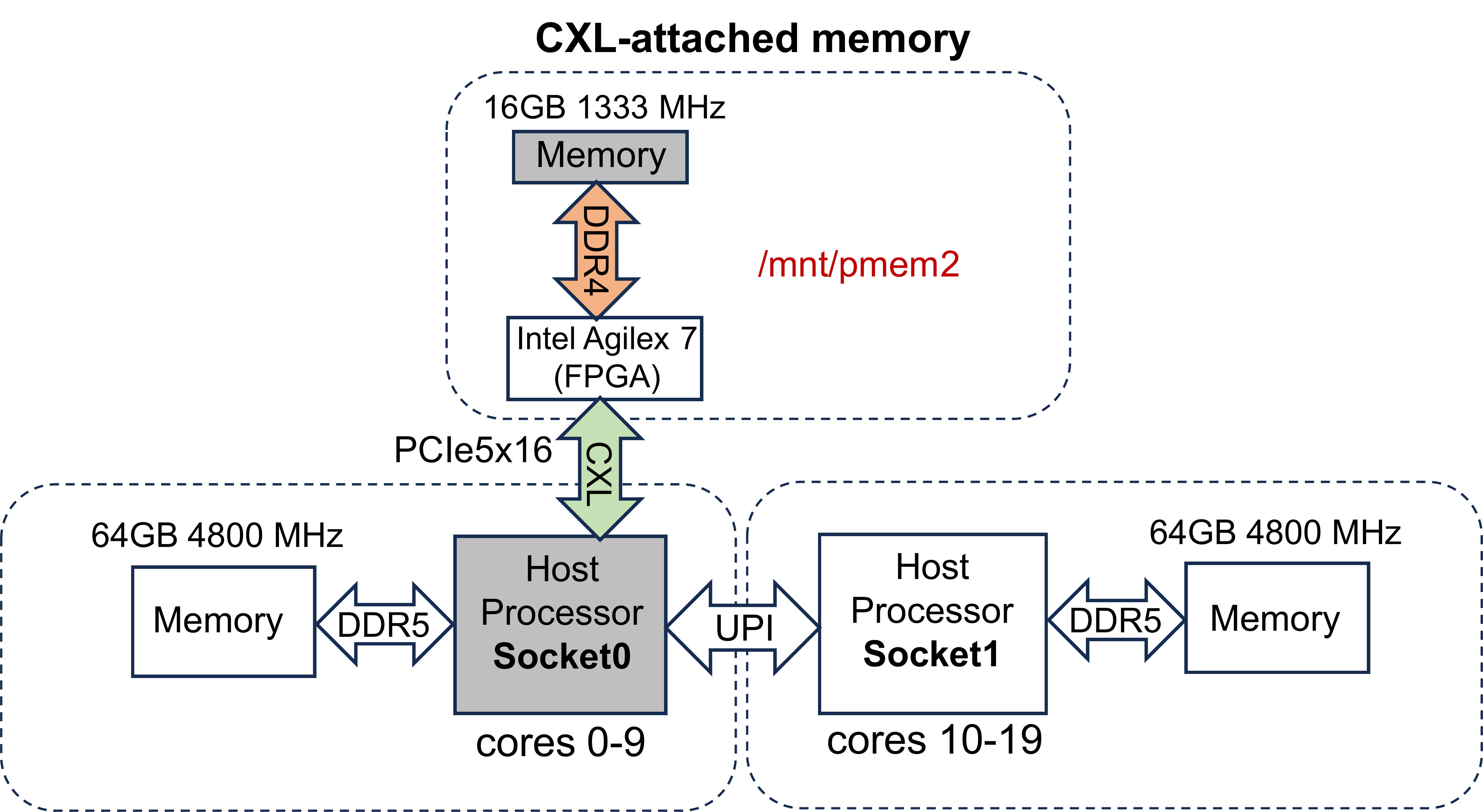}
      \end{subfigure}%
      \hspace{0.02\textwidth}
      \begin{subfigure}{0.26\textwidth}
        \centering
        \includegraphics[width=\columnwidth]{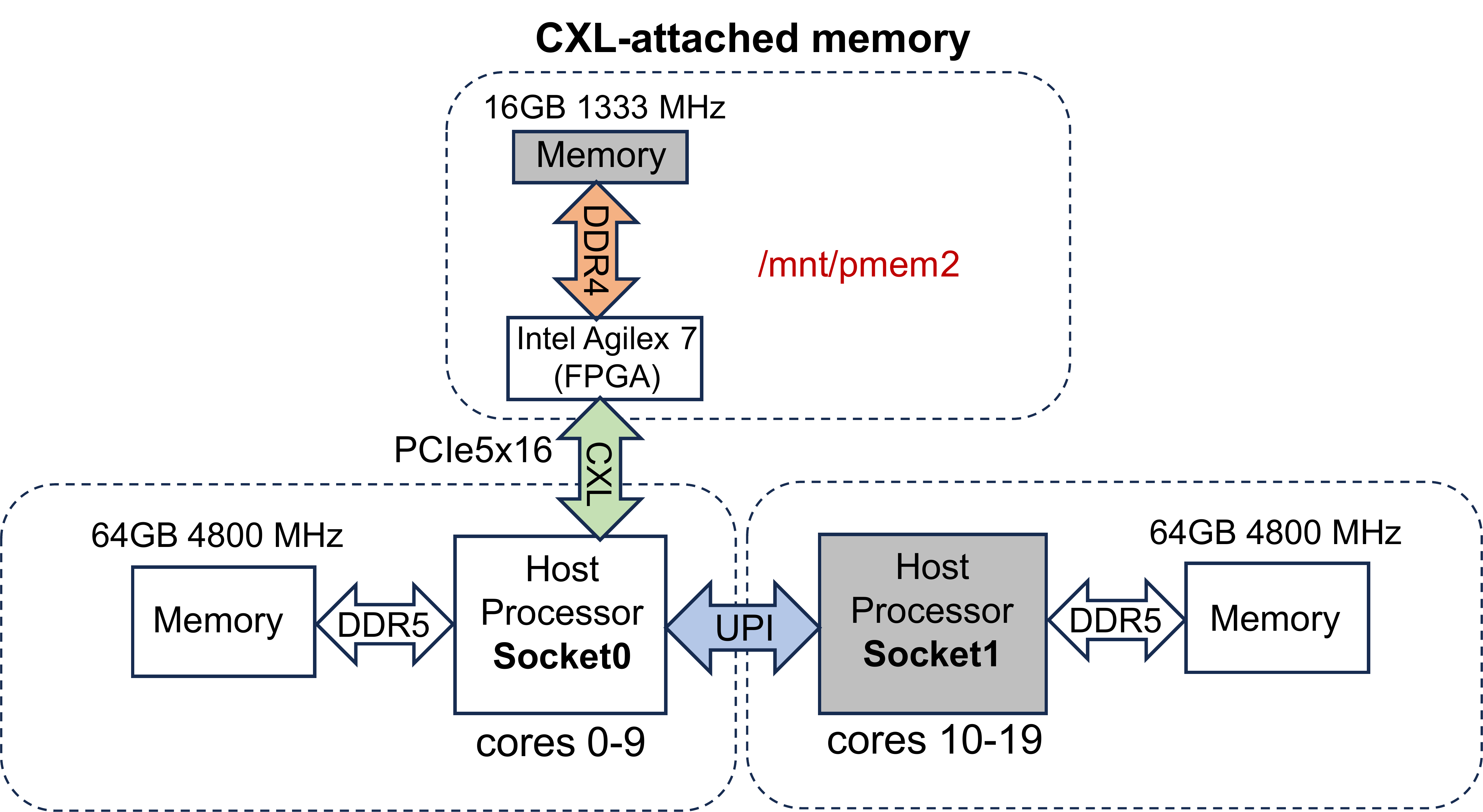}
      \end{subfigure}%
      \hspace{0.02\textwidth}
      \begin{subfigure}{0.26\textwidth}
        \centering
        \includegraphics[width=\columnwidth]{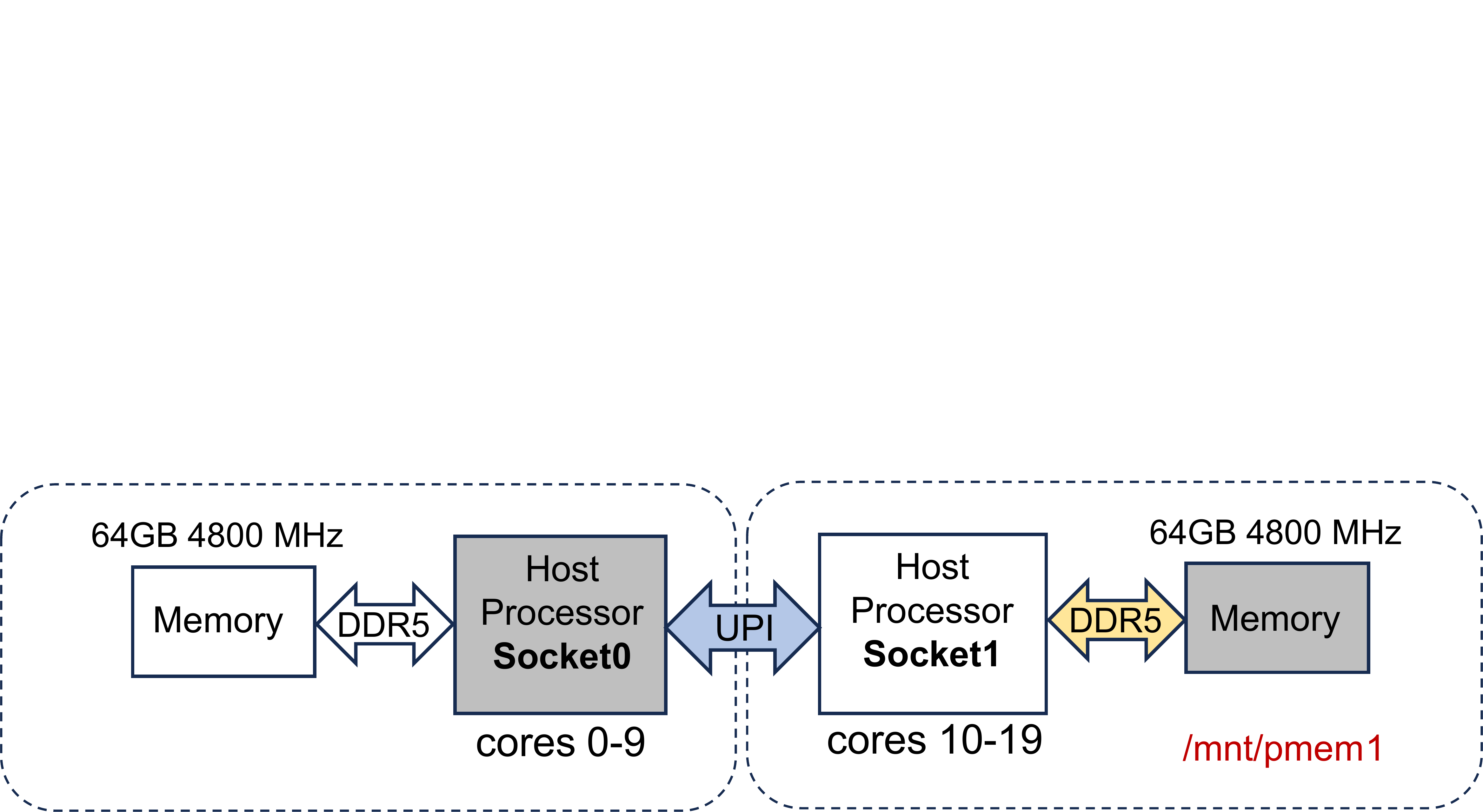}
      \end{subfigure}
        \captionsetup{format=default} 
 \rule{\linewidth}{0.4pt}
      \medskip

      \textbf{(Class 1.c)}\hspace{1em}
      \begin{subfigure}{0.26\textwidth}
        \centering
        \includegraphics[width=\columnwidth]{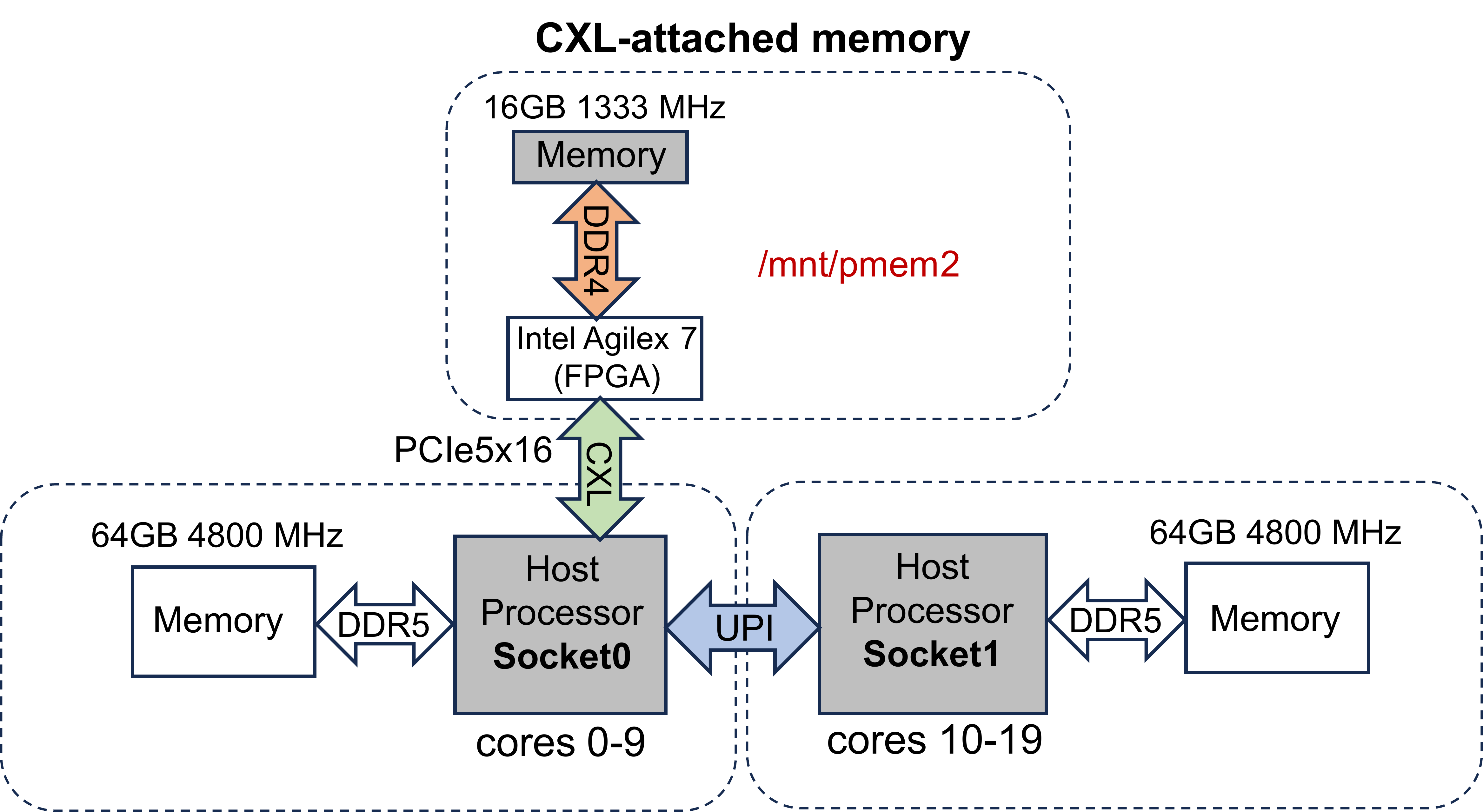}
      \end{subfigure}%
      \hspace{0.02\textwidth}
      \begin{subfigure}{0.26\textwidth}
        \centering
        \includegraphics[width=\columnwidth]{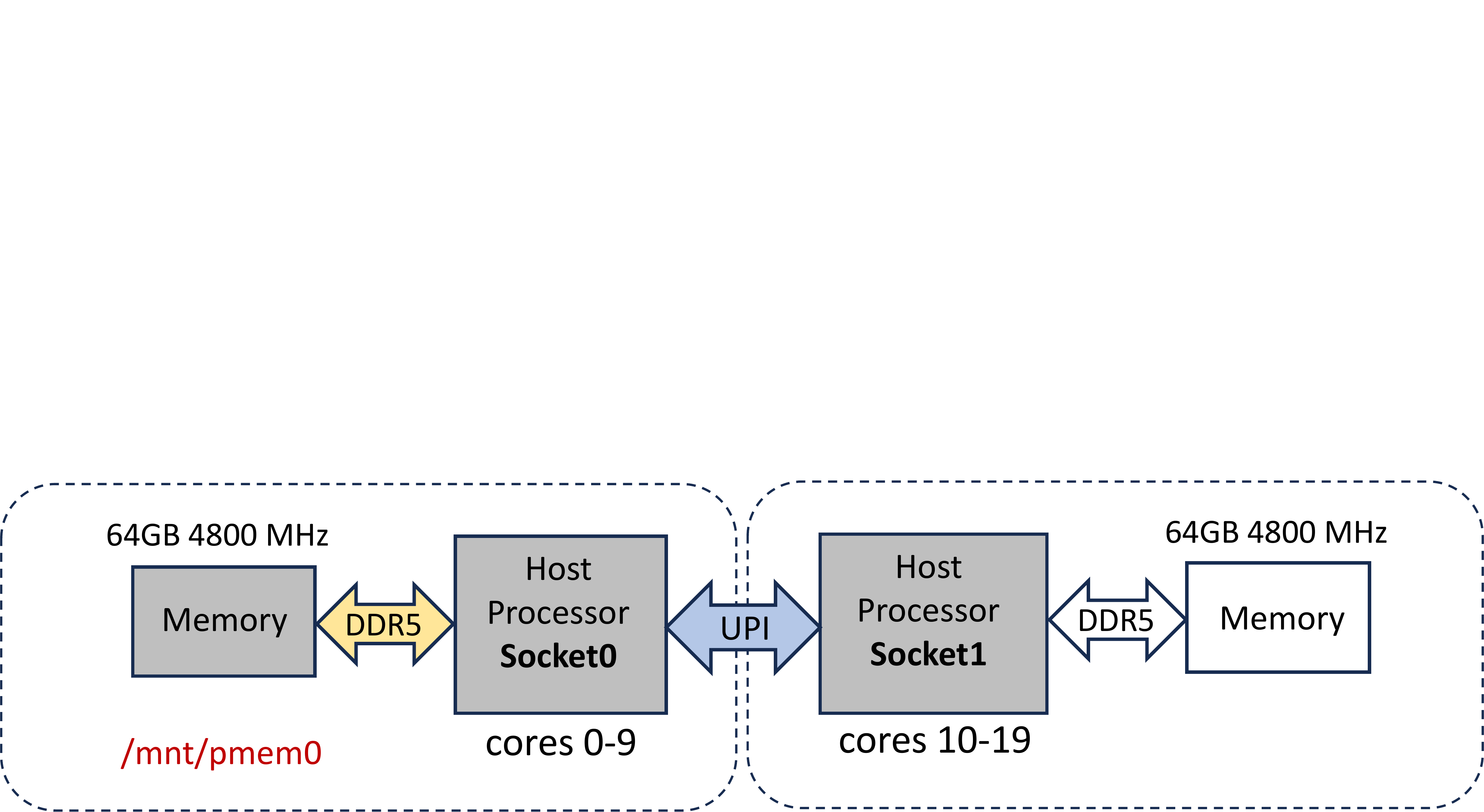}
      \end{subfigure}%
      \hspace{0.02\textwidth}
      \begin{subfigure}{0.26\textwidth}
        \centering
        \includegraphics[width=\columnwidth]{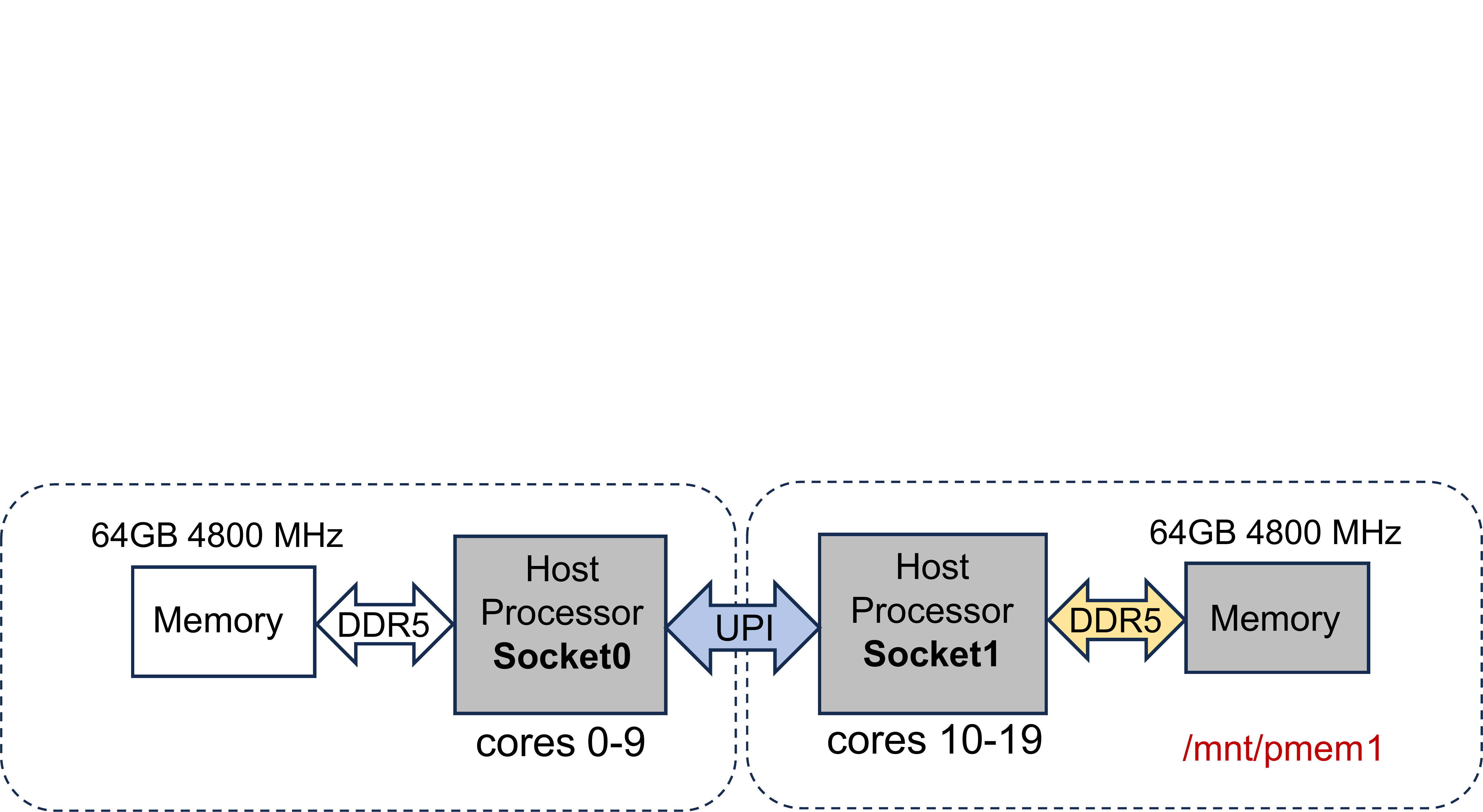}
      \end{subfigure}
    \captionsetup{format=default} 
        \rule{\linewidth}{0.4pt}
      
      \textbf{(Class 2.a)}\hspace{1em}
      \begin{subfigure}{0.26\textwidth}
        \centering
        \includegraphics[width=\columnwidth]{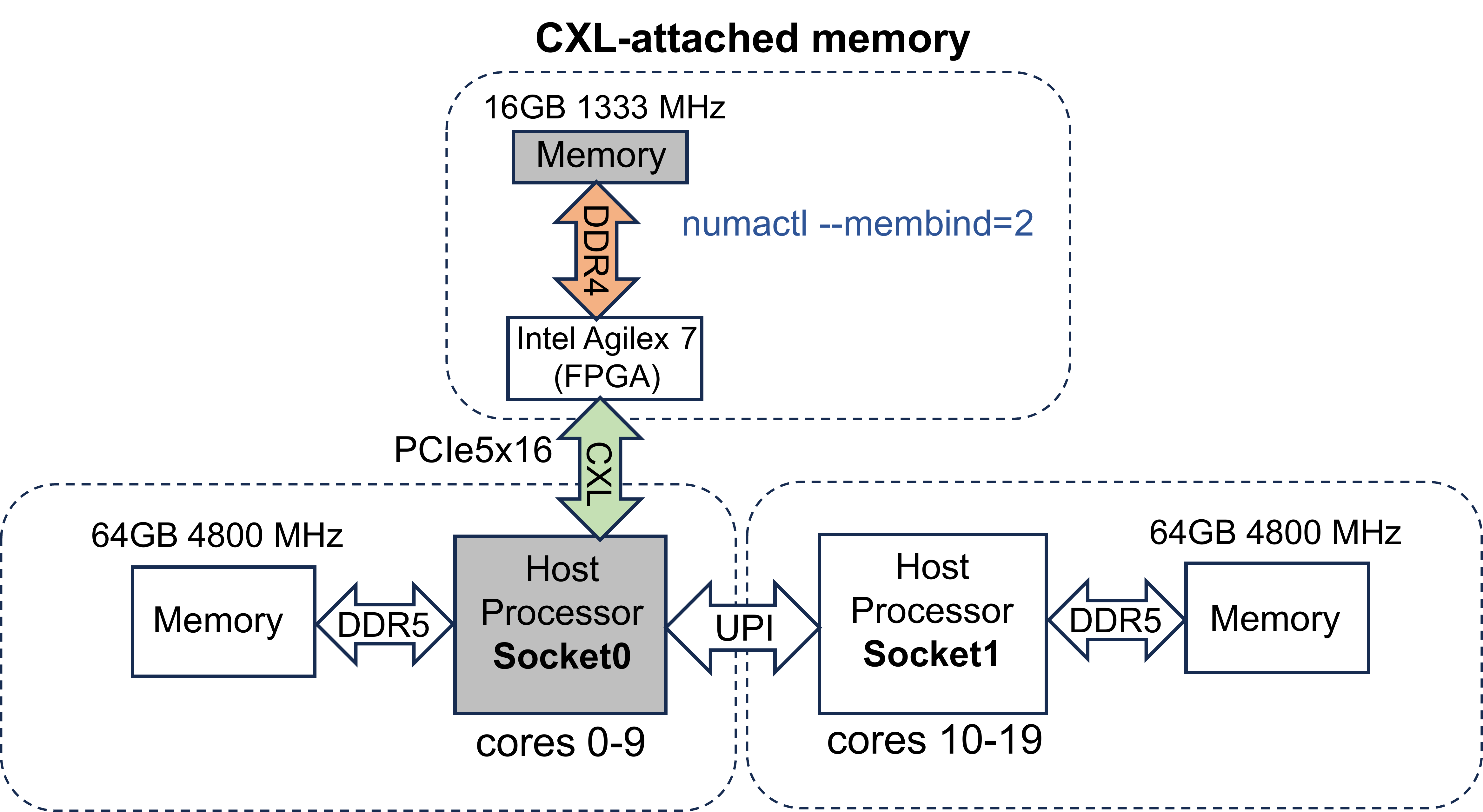}
      \end{subfigure}%
      \hspace{0.02\textwidth}
      \begin{subfigure}{0.26\textwidth}
        \centering
        \includegraphics[width=\columnwidth]{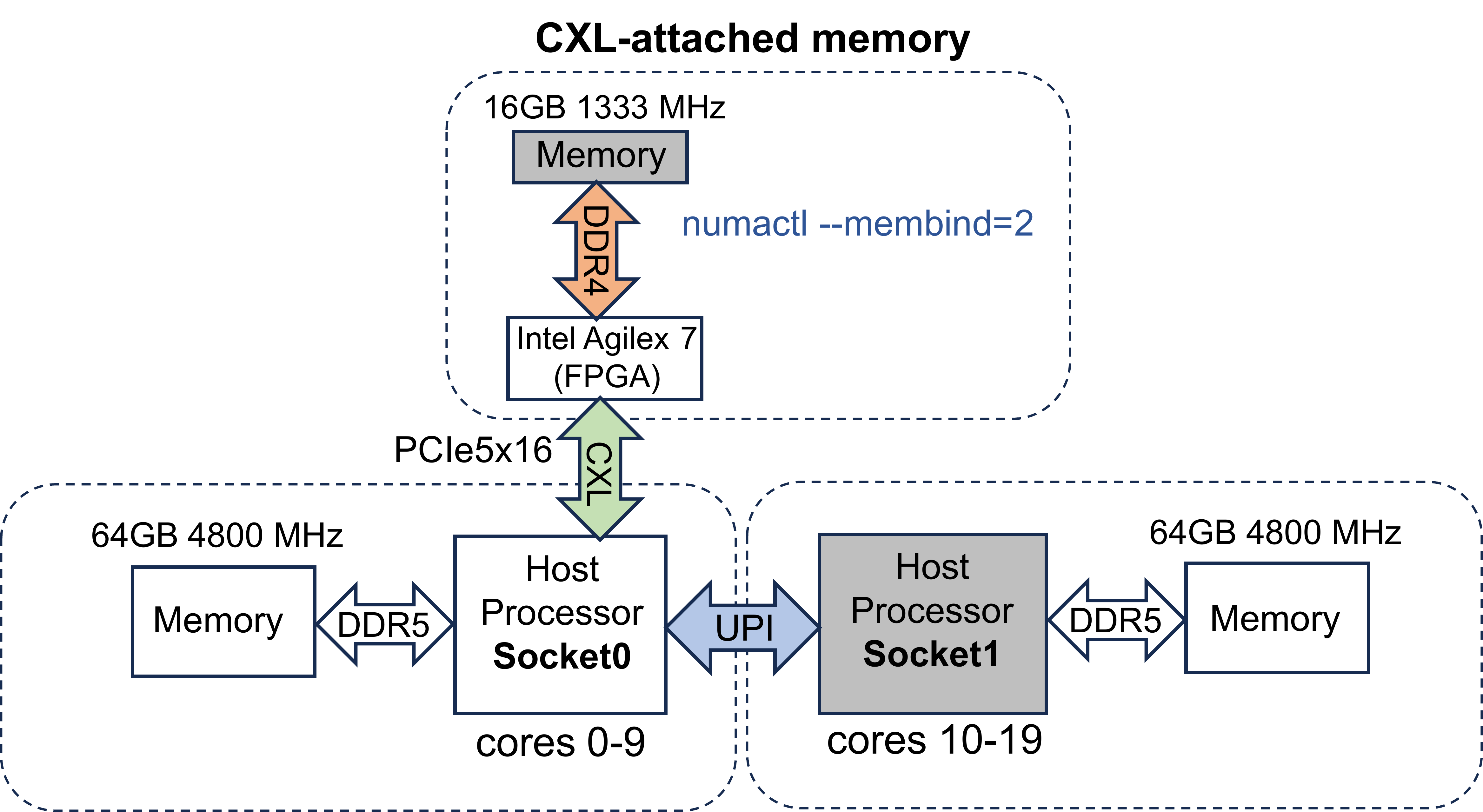}
      \end{subfigure}%
      \hspace{0.02\textwidth}
      \begin{subfigure}{0.26\textwidth}
        \centering
        \includegraphics[width=\columnwidth]{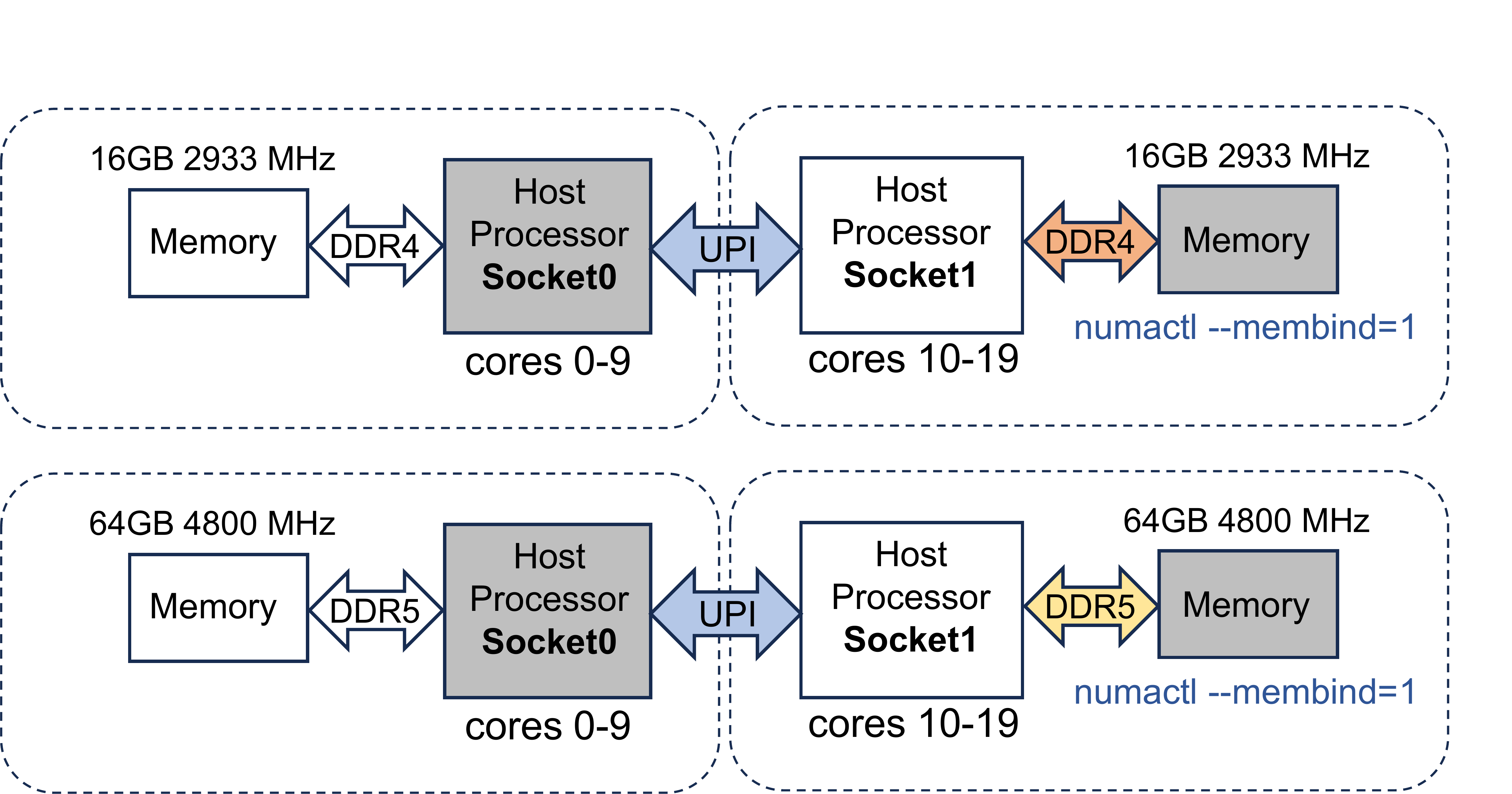}
      \end{subfigure}
        \captionsetup{format=default} 
 \rule{\linewidth}{0.4pt}
      \medskip

      \textbf{(Class 2.b)}\hspace{1em}
      \begin{subfigure}{0.26\textwidth}
        \centering
        \includegraphics[width=\columnwidth]{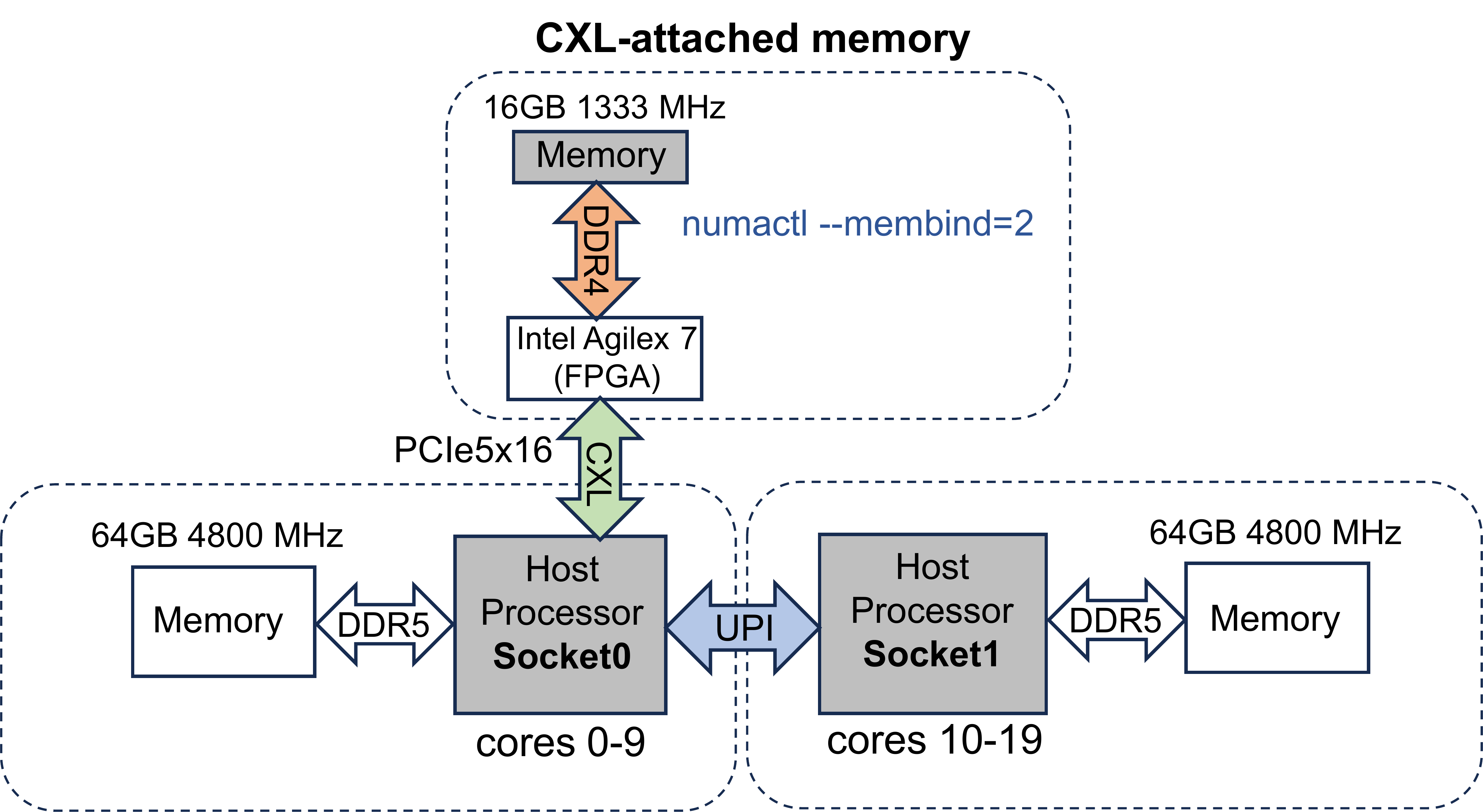}
      \end{subfigure}%
      \hspace{0.02\textwidth}
      \begin{subfigure}{0.26\textwidth}
        \centering
        \includegraphics[width=\columnwidth]{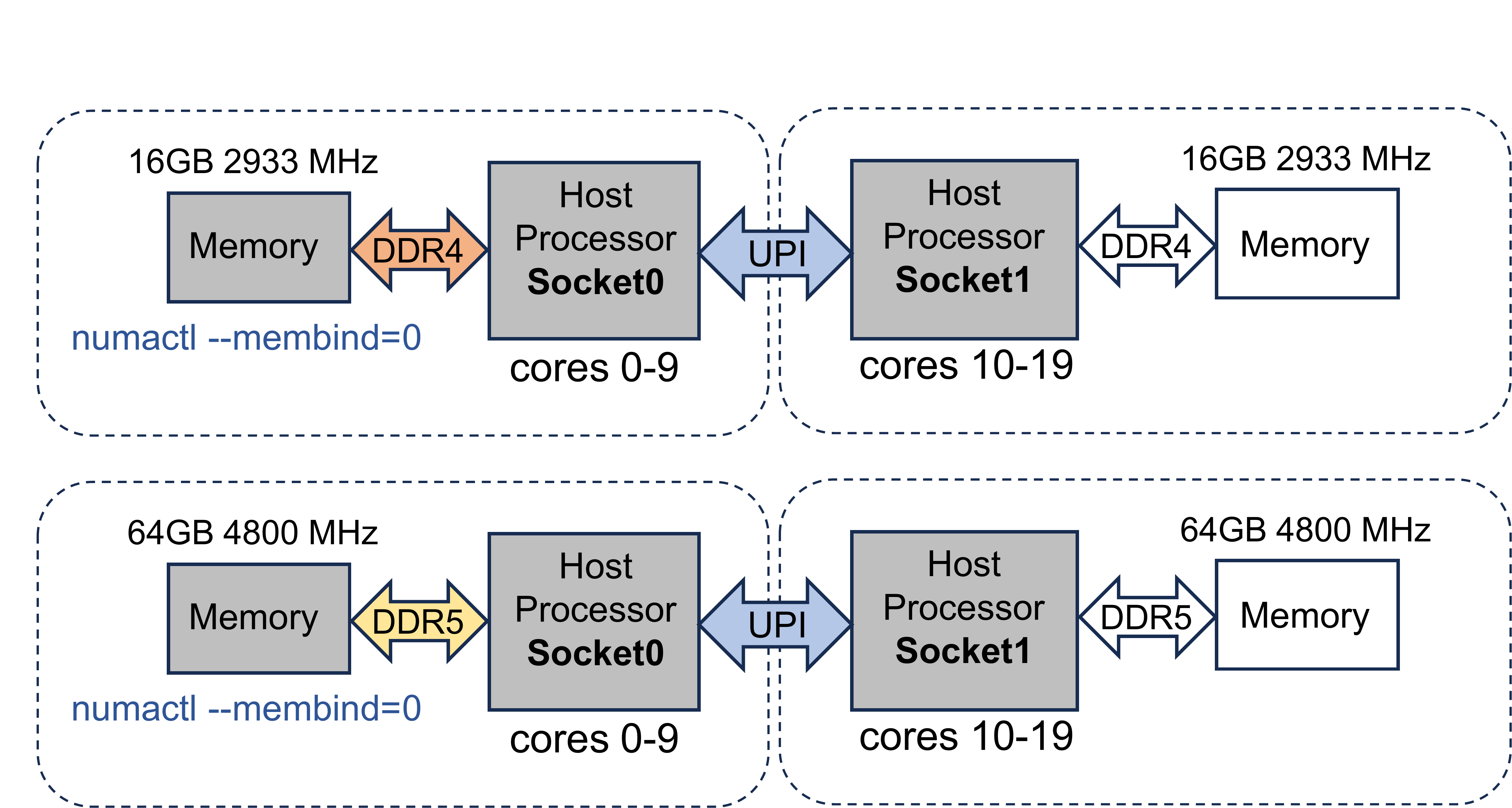}
      \end{subfigure}%
      \hspace{0.02\textwidth}
      \begin{subfigure}{0.26\textwidth}
        \centering
        \includegraphics[width=\columnwidth]{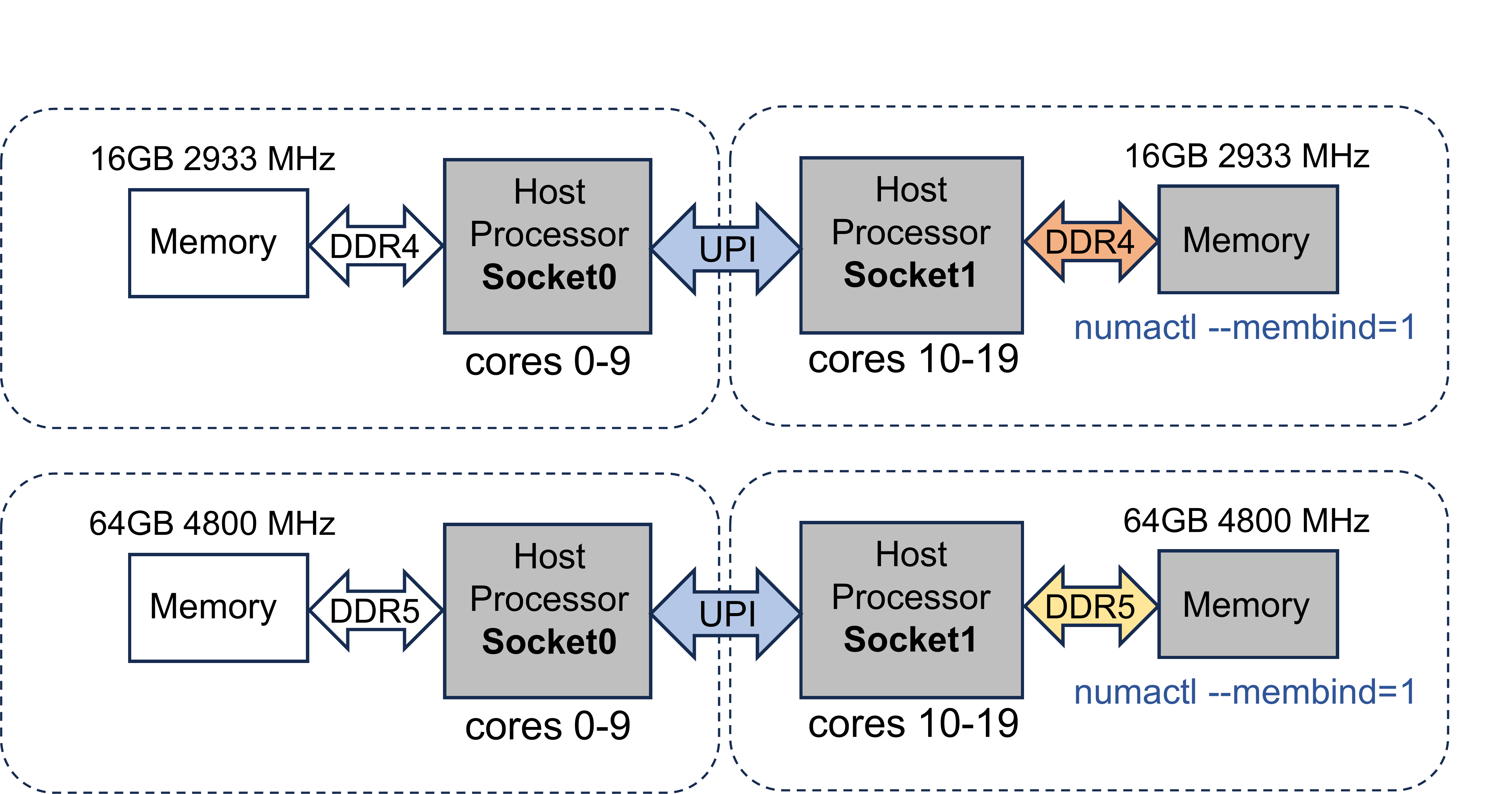}
      \end{subfigure}
      \medskip

    \end{minipage}
  \end{sideways}
  \caption{Data flow demonstrations for the two classes (\textit{App Direct} and \textit{Memory Mode}). Each test group is evaluated in corresponding to subfigures of~\autoref{Scale_results}, \autoref{Add_results}, \autoref{Copy_results}, \autoref{Triad_results}. Each row corresponds to a test group.}
  \label{dataflows}
\end{figure*}

\label{resultsandanalysis}
\autoref{Scale_results}, \autoref{Add_results}, \autoref{Copy_results} and \autoref{Triad_results} present STREAM results for the Scale, Add, Copy, and Triad operations correspondingly, and for the test configurations defined in \autoref{confs} as will be described next. \autoref{fig:Scale_d}, \autoref{fig:Add_d}, \autoref{fig:Copy_d} and \autoref{fig:Triad_d} through \autoref{fig:Scale_c}, \autoref{fig:Add_c}, \autoref{fig:Copy_c} and \autoref{fig:Triad_c} present results for Class 1.\ref{item:group_a} group though Class 2.\ref{item:group_e} group correspondingly.

The results explain the costs associated with memory access across varied configurations distinguished by parameters such as memory type (on-node or CXL-attached), memory placement (local to the socket, on the alternate CPU socket, or the CXL-attached memory), access mode (\textit{App-Direct} vs. \textit{Memory Mode}), and thread affinity (Close or Spread).



Next, we will examine and analyze the achieved results in relation to the configuration classes and groups presented in \autoref{confs}:

\noindent\textbf{Class 1 --- \textit{App-Direct}}:
\begin{enumerate}[label=(\alph*),wide, labelwidth=!, labelindent=0pt]

    \item\label{item:group_a_result} 
    \textbf{Local memory access as PMem:} It is possible to observe that among all of the STREAM actions, the \textit{App-Direct} access using PMDK to the local DDR5 memory is saturated around 20-22 GB/s. This test is a reference for the remote access presented in the following group, either to a nearby remote socket or to the CXL memory (with PMDK). 

    \item\label{item:group_b_result} \textbf{Remote memory access as PMem:} \textit{App-Direct} access to the emulated remote PMem (DDR5 on the alternate socket) results in a decrease of 30\% ($\sim$15 GB/s) of performance on average for all STREAM operations, in comparison to local \textit{App-Direct} access. In the case of \textit{App-Direct} access to remote CXL memory (DDR4), we experience 50\% decrease in performance in comparison to the emulated PMem on DDR5. However, we note that DDR5 inherently has about 50\% higher bandwidth than DDR4, meaning that the rest of the overhead -- about 2-3 GB/s loss in bandwidth -- can be attributed to the CXL fabric. 
    
    \item\label{item:group_c_result} 
    \textbf{Remote memory as PMem (thread affinity):}
    As observed in previous groups, local \textit{App-Direct} accesses result in higher bandwidth than remote accesses. In the case of \textit{close} thread affinity, after populating the entire socket, adding remote accesses of compute cores to the workload negatively impacts the bandwidth, whereas adding local accesses contributes positively. With \textit{spread} affinity, the performance demonstrates an average between local and remote accesses due to the inclusion of alternating accesses. Eventually, when both sockets are operating with the entire core count, the results converge for on-node DDR5 and remote CXL memory, separately. Notably, accessing remote CXL memory (DDR4) leads to a 50\% observed degradation compared to on-node DDR5.
    

\end{enumerate}

\noindent\textbf{Class 2 --- \textit{Memory Mode}}:

\begin{enumerate}[label=(\alph*),wide, labelwidth=!, labelindent=0pt]
    
    \item\label{item:group_d_result} \textbf{Remote CC-NUMA:} Evaluating DDR4 CC-NUMA, whether on the remote socket or CXL-attached memory, yields comparable figures (with average gaps of up to 2-5 GB/s). However, following a small number of threads, a slight advantage is observed for accessing CXL memory. This advantage can be attributed to the larger caches in Setup \#1  utilizing CXL (Shappire Rapids), as opposed to Setup \#2 (Xeon Gold) with on-node DDR4 (\autoref{hpchardware}). This indicates that the CXL fabric overhead is constrained by the performance reduction when transitioning back from Sapphire Rapids to Xeon Gold.  
    Moreover, the gap between the \textit{CC-NUMA} to DDR5 and DDR4 (on-node or CXL-attached) stands on a factor of two, as already observed in 1.\ref{item:group_b_result} and 1.\ref{item:group_c_result}. In addition, in comparison to the results of the \textit{App-Direct} tests in 1.\ref{item:group_b_result}, it is observed that PMDK overheads over CC-NUMA are 10\%-15\% (in all STREAM methods).
        
    \item\label{item:group_e_result}
     \textbf{Remote CC-NUMA (all cores):} 
     The observed gap between DDR4 and DDR5 repeats here. Moreover, accessing on-node DDR4 using all cores converges to the same results as accessing DDR4 CXL memory. 
         
\end{enumerate}

To conclude, the analysis reveals that direct access to local DDR5 memory using PMDK saturates at 20-22 GB/s, while direct remote access to emulated PMem and CXL memory results in 30\% and 50\% performance decreases, respectively, with about 2-3 GB/s bandwidth loss attributed to CXL fabric. In terms of memory expansion, accessing remote DDR4 CC-NUMA and DDR4 CXL-attached memory exhibit similar performance gaps of 2-3 GB/s, while DDR5 CC-NUMA maintains an advantage gap of a factor of 1.5 compared to DDR4, and on-node DDR4 access converges with off-node DDR4 access under varying thread affinities.

\section{Conclusions}

In this study, we embarked on a comprehensive exploration of the potential of CXL memory as a promising candidate for serving as a persistent memory solution in the context of disaggregated HPC systems. By conducting physical experiments on state-of-the-art multi-NUMA nodes equipped with high-performance processors and CXL-attached memory prototypes, we have provided empirical evidence that supports the feasibility of using CXL memory to exhibit all the characteristics of persistent memory modules while achieving impressive performance metrics.

Our findings demonstrate that CXL memory has the capability to outperform previously published benchmarks for Optane DCPMM in terms of bandwidth. Specifically, by employing a CXL-DDR4 memory module, which is a cost-effective alternative to DDR5 memory, we achieved bandwidth results comparable to local DDR4 memory configurations, with only a marginal decrease of around 50\% when compared to local DDR5 memory configurations. These results, attained across various memory distances from the working threads, were assessed through the well-established STREAM benchmark underscoring the reliability and versatility of CXL memory in the HPC landscape.

The shift from PMem to CXL was not only demonstrated through performance evaluations but was also highlighted through the modification of the STREAM application into STREAM-PMem. We showcased the seamless transition of programming models from PMem to CXL, leveraging the PMDK's \textit{pmemobj} to ensure transactional integrity and consistency of operations on persistent objects. Furthermore, the ability to access CXL memory directly and transactionally, akin to Optane DCPMM, was underscored as a key advantage for practical implementation.

Our study extends beyond theoretical considerations by implementing a practical CXL prototype on an FPGA card. This prototype embodies CXL 1.1/2.0 compliant endpoint designs, demonstrating effective link establishment and transaction layer management through a combination of Soft and Hard IP components. The prototype's performance, while constrained by current implementation limitations, stands as a testament to the extensibility of this solution and offers a blueprint for potential enhancements, including higher-speed FPGAs and increased resources.

\section{Future Work}

While this study provides valuable insights into the feasibility and potential benefits of using CXL-enabled memory in HPC systems, several avenues for future research and exploration remain:

\begin{itemize}[wide, labelwidth=!, labelindent=0pt]
    \item \textbf{Scalability and Performance Optimization}: Further investigation is warranted to explore the scalability of CXL-enabled memory in larger HPC clusters, with more than one node accessing the CXL memory. Optimizing communication protocols and memory access patterns can help maximize memory disaggregation benefits.
        
    \item \textbf{Hybrid Architectures}: Combining different memory technologies, such as DDR, PMem, and CXL memory, in a hybrid memory architecture could offer a balanced solution that leverages the strengths of each technology. Also, the CXL memory could also use DDR5 and even Optane DCPMM, and as such, revisiting the results with those CXL memories would be beneficial.
    
    \item \textbf{Real-World Applications}: Extending the evaluation to real-world HPC applications beyond benchmarks can provide a clearer understanding of how CXL memory performs in practical scenarios.
    
    \item \textbf{Fault Tolerance and Reliability}: Investigating fault tolerance mechanisms and data reliability in the context of CXL-enabled memory is crucial, especially in large-scale distributed environments. Specifically, code systems that have previously been built upon PMDK and Optane DCPMM presence in the HPC system.
        
\end{itemize}




\begin{acks}
This work was supported by Pazy grant 226/20, the Lynn and William Frankel Center for Computer Science, and Intel Corporation (oneAPI Center of Excellence program). Computational support was provided by the NegevHPC project~\cite{negevhpc} and Intel Developer Cloud~\cite{intel-developercloud}. The authors would like to thank Gabi Dadush, Israel Hen, and Emil Malka for their hardware support on NegevHPC. The authors also want to thank Jay Mahalingam and Guy Tamir of Intel for their great help in forming this collaboration.
\end{acks}
\bibliographystyle{ACM-Reference-Format}
\bibliography{sample-base}


\begin{thebibliography}{72}


\ifx \showCODEN    \undefined \def \showCODEN     #1{\unskip}     \fi
\ifx \showDOI      \undefined \def \showDOI       #1{#1}\fi
\ifx \showISBNx    \undefined \def \showISBNx     #1{\unskip}     \fi
\ifx \showISBNxiii \undefined \def \showISBNxiii  #1{\unskip}     \fi
\ifx \showISSN     \undefined \def \showISSN      #1{\unskip}     \fi
\ifx \showLCCN     \undefined \def \showLCCN      #1{\unskip}     \fi
\ifx \shownote     \undefined \def \shownote      #1{#1}          \fi
\ifx \showarticletitle \undefined \def \showarticletitle #1{#1}   \fi
\ifx \showURL      \undefined \def \showURL       {\relax}        \fi
\providecommand\bibfield[2]{#2}
\providecommand\bibinfo[2]{#2}
\providecommand\natexlab[1]{#1}
\providecommand\showeprint[2][]{arXiv:#2}

\bibitem[Ahn et~al\mbox{.}(2022)]%
        {ahn2022enabling}
\bibfield{author}{\bibinfo{person}{Minseon Ahn}, \bibinfo{person}{Andrew
  Chang}, \bibinfo{person}{Donghun Lee}, \bibinfo{person}{Jongmin Gim},
  \bibinfo{person}{Jungmin Kim}, \bibinfo{person}{Jaemin Jung},
  \bibinfo{person}{Oliver Rebholz}, \bibinfo{person}{Vincent Pham},
  \bibinfo{person}{Krishna Malladi}, {and} \bibinfo{person}{Yang~Seok Ki}.}
  \bibinfo{year}{2022}\natexlab{}.
\newblock \showarticletitle{Enabling CXL memory expansion for in-memory
  database management systems}. In \bibinfo{booktitle}{\emph{Proceedings of the
  18th International Workshop on Data Management on New Hardware}}.
  \bibinfo{pages}{1--5}.
\newblock


\bibitem[Al~Maruf and Chowdhury(2023)]%
        {al2023memory}
\bibfield{author}{\bibinfo{person}{Hasan Al~Maruf} {and}
  \bibinfo{person}{Mosharaf Chowdhury}.} \bibinfo{year}{2023}\natexlab{}.
\newblock \showarticletitle{Memory Disaggregation: Open Challenges in the Era
  of CXL}. In \bibinfo{booktitle}{\emph{Workshop on HotTopics in System
  Infrastructure}}, Vol.~\bibinfo{volume}{18}.
\newblock


\bibitem[AsteraLabs(2022)]%
        {asteralabs}
\bibfield{author}{\bibinfo{person}{AsteraLabs}.}
  \bibinfo{year}{2022}\natexlab{}.
\newblock \bibinfo{booktitle}{\emph{CXL Memory Accelerators}}.
\newblock
\urldef\tempurl%
\url{https://www.asteralabs.com/products/cxl-memory-platform/}
\showURL{%
\tempurl}


\bibitem[Baldassin et~al\mbox{.}(2021)]%
        {baldassin2021persistent}
\bibfield{author}{\bibinfo{person}{Alexandro Baldassin}, \bibinfo{person}{Joao
  Barreto}, \bibinfo{person}{Daniel Castro}, {and} \bibinfo{person}{Paolo
  Romano}.} \bibinfo{year}{2021}\natexlab{}.
\newblock \showarticletitle{Persistent memory: A survey of programming support
  and implementations}.
\newblock \bibinfo{journal}{\emph{ACM Computing Surveys (CSUR)}}
  \bibinfo{volume}{54}, \bibinfo{number}{7} (\bibinfo{year}{2021}),
  \bibinfo{pages}{1--37}.
\newblock


\bibitem[Benson et~al\mbox{.}(2023)]%
        {benson2023we}
\bibfield{author}{\bibinfo{person}{Lawrence Benson}, \bibinfo{person}{Marcel
  Weisgut}, {and} \bibinfo{person}{Tilmann Rabl}.}
  \bibinfo{year}{2023}\natexlab{}.
\newblock \showarticletitle{What We Can Learn from Persistent Memory for CXL}.
\newblock \bibinfo{journal}{\emph{BTW 2023}} (\bibinfo{year}{2023}).
\newblock


\bibitem[Bergstrom(2011)]%
        {bergstrom2011measuring}
\bibfield{author}{\bibinfo{person}{Lars Bergstrom}.}
  \bibinfo{year}{2011}\natexlab{}.
\newblock \showarticletitle{Measuring NUMA effects with the STREAM benchmark}.
\newblock \bibinfo{journal}{\emph{arXiv preprint arXiv:1103.3225}}
  (\bibinfo{year}{2011}).
\newblock


\bibitem[Bernholdt et~al\mbox{.}(2020)]%
        {bernholdt2020survey}
\bibfield{author}{\bibinfo{person}{David~E Bernholdt}, \bibinfo{person}{Swen
  Boehm}, \bibinfo{person}{George Bosilca}, \bibinfo{person}{Manjunath
  Gorentla~Venkata}, \bibinfo{person}{Ryan~E Grant}, \bibinfo{person}{Thomas
  Naughton}, \bibinfo{person}{Howard~P Pritchard}, \bibinfo{person}{Martin
  Schulz}, {and} \bibinfo{person}{Geoffroy~R Vallee}.}
  \bibinfo{year}{2020}\natexlab{}.
\newblock \showarticletitle{A survey of MPI usage in the US exascale computing
  project}.
\newblock \bibinfo{journal}{\emph{Concurrency and Computation: Practice and
  Experience}} \bibinfo{volume}{32}, \bibinfo{number}{3}
  (\bibinfo{year}{2020}), \bibinfo{pages}{e4851}.
\newblock


\bibitem[Bustos et~al\mbox{.}(2023)]%
        {bustos2023response}
\bibfield{author}{\bibinfo{person}{Andr{\'e}s Bustos},
  \bibinfo{person}{Antonio~Juan Rubio-Montero}, \bibinfo{person}{Roberto
  M{\'e}ndez}, \bibinfo{person}{Sergio Rivera}, \bibinfo{person}{Francisco
  Gonz{\'a}lez}, \bibinfo{person}{Xandra Campo}, \bibinfo{person}{Hern{\'a}n
  Asorey}, {and} \bibinfo{person}{Rafael Mayo-Garc{\'\i}a}.}
  \bibinfo{year}{2023}\natexlab{}.
\newblock \showarticletitle{Response of HPC hardware to neutron radiation at
  the dawn of exascale}.
\newblock \bibinfo{journal}{\emph{The Journal of Supercomputing}}
  (\bibinfo{year}{2023}), \bibinfo{pages}{1--22}.
\newblock


\bibitem[Dagum and Menon(1998)]%
        {dagum1998openmp}
\bibfield{author}{\bibinfo{person}{Leonardo Dagum} {and}
  \bibinfo{person}{Ramesh Menon}.} \bibinfo{year}{1998}\natexlab{}.
\newblock \showarticletitle{OpenMP: An industry-standard API for shared-memory
  programming}.
\newblock \bibinfo{journal}{\emph{Computing in Science \& Engineering}}
  \bibinfo{number}{1} (\bibinfo{year}{1998}), \bibinfo{pages}{46--55}.
\newblock


\bibitem[Ding et~al\mbox{.}(2023)]%
        {ding2023evaluating}
\bibfield{author}{\bibinfo{person}{Nan Ding}, \bibinfo{person}{Pieter Maris},
  \bibinfo{person}{Hai~Ah Nam}, \bibinfo{person}{Taylor Groves},
  \bibinfo{person}{Muaaz~Gul Awan}, \bibinfo{person}{LeAnn Lindsey},
  \bibinfo{person}{Christopher Daley}, \bibinfo{person}{Oguz Selvitopi},
  \bibinfo{person}{Leonid Oliker}, {and} \bibinfo{person}{Nicholas Wright}.}
  \bibinfo{year}{2023}\natexlab{}.
\newblock \showarticletitle{Evaluating the Potential of Disaggregated Memory
  Systems for HPC applications}.
\newblock \bibinfo{journal}{\emph{arXiv preprint arXiv:2306.04014}}
  (\bibinfo{year}{2023}).
\newblock


\bibitem[Evans et~al\mbox{.}(2022)]%
        {evans2022survey}
\bibfield{author}{\bibinfo{person}{Thomas~M Evans}, \bibinfo{person}{Andrew
  Siegel}, \bibinfo{person}{Erik~W Draeger}, \bibinfo{person}{Jack Deslippe},
  \bibinfo{person}{Marianne~M Francois}, \bibinfo{person}{Timothy~C Germann},
  \bibinfo{person}{William~E Hart}, {and} \bibinfo{person}{Daniel~F Martin}.}
  \bibinfo{year}{2022}\natexlab{}.
\newblock \showarticletitle{A survey of software implementations used by
  application codes in the Exascale Computing Project}.
\newblock \bibinfo{journal}{\emph{The International Journal of High Performance
  Computing Applications}} \bibinfo{volume}{36}, \bibinfo{number}{1}
  (\bibinfo{year}{2022}), \bibinfo{pages}{5--12}.
\newblock


\bibitem[Fagerheim(2021)]%
        {fagerheim2021benchmarking}
\bibfield{author}{\bibinfo{person}{Svein~Gunnar Fagerheim}.}
  \bibinfo{year}{2021}\natexlab{}.
\newblock \emph{\bibinfo{title}{Benchmarking Persistent Memory with Respect to
  Performance and Programmability}}.
\newblock \bibinfo{thesistype}{Master's\ thesis}.
\newblock


\bibitem[Foyer et~al\mbox{.}(2023)]%
        {foyer2023survey}
\bibfield{author}{\bibinfo{person}{Cl{\'e}ment Foyer}, \bibinfo{person}{Brice
  Goglin}, {and} \bibinfo{person}{Andr{\`e}s~Rubio Proa{\~n}o}.}
  \bibinfo{year}{2023}\natexlab{}.
\newblock \showarticletitle{A survey of software techniques to emulate
  heterogeneous memory systems in high-performance computing}.
\newblock \bibinfo{journal}{\emph{Parallel Comput.}} (\bibinfo{year}{2023}),
  \bibinfo{pages}{103023}.
\newblock


\bibitem[Fridman et~al\mbox{.}(2022)]%
        {fridman2022recovery}
\bibfield{author}{\bibinfo{person}{Yehonatan Fridman}, \bibinfo{person}{Yaniv
  Snir}, \bibinfo{person}{Harel Levin}, \bibinfo{person}{Danny Hendler},
  \bibinfo{person}{Hagit Attiya}, {and} \bibinfo{person}{Gal Oren}.}
  \bibinfo{year}{2022}\natexlab{}.
\newblock \showarticletitle{Recovery of Distributed Iterative Solvers for
  Linear Systems Using Non-Volatile RAM}. In \bibinfo{booktitle}{\emph{2022
  IEEE/ACM 12th Workshop on Fault Tolerance for HPC at eXtreme Scale (FTXS)}}.
  IEEE, \bibinfo{pages}{11--23}.
\newblock


\bibitem[Fridman et~al\mbox{.}(2021)]%
        {fridman2021assessing}
\bibfield{author}{\bibinfo{person}{Yehonatan Fridman}, \bibinfo{person}{Yaniv
  Snir}, \bibinfo{person}{Matan Rusanovsky}, \bibinfo{person}{Kfir Zvi},
  \bibinfo{person}{Harel Levin}, \bibinfo{person}{Danny Hendler},
  \bibinfo{person}{Hagit Attiya}, {and} \bibinfo{person}{Gal Oren}.}
  \bibinfo{year}{2021}\natexlab{}.
\newblock \showarticletitle{Assessing the use cases of persistent memory in
  high-performance scientific computing}. In \bibinfo{booktitle}{\emph{2021
  IEEE/ACM 11th Workshop on Fault Tolerance for HPC at eXtreme Scale (FTXS)}}.
  IEEE, \bibinfo{pages}{11--20}.
\newblock


\bibitem[Geyer et~al\mbox{.}(2023)]%
        {geyer2023working}
\bibfield{author}{\bibinfo{person}{Andreas Geyer}, \bibinfo{person}{Daniel
  Ritter}, \bibinfo{person}{Dong~Hun Lee}, \bibinfo{person}{Minseon Ahn},
  \bibinfo{person}{Johannes Pietrzyk}, \bibinfo{person}{Alexander Krause},
  \bibinfo{person}{Dirk Habich}, {and} \bibinfo{person}{Wolfgang Lehner}.}
  \bibinfo{year}{2023}\natexlab{}.
\newblock \showarticletitle{Working with Disaggregated Systems. What are the
  Challenges and Opportunities of RDMA and CXL?}
\newblock \bibinfo{journal}{\emph{BTW 2023}} (\bibinfo{year}{2023}).
\newblock


\bibitem[Gropp(2012)]%
        {gropp2012mpi}
\bibfield{author}{\bibinfo{person}{William Gropp}.}
  \bibinfo{year}{2012}\natexlab{}.
\newblock \showarticletitle{MPI 3 and beyond: why MPI is successful and what
  challenges it faces}. In \bibinfo{booktitle}{\emph{European MPI Users' Group
  Meeting}}. Springer, \bibinfo{pages}{1--9}.
\newblock


\bibitem[Gugnani et~al\mbox{.}(2020)]%
        {gugnani2020understanding}
\bibfield{author}{\bibinfo{person}{Shashank Gugnani}, \bibinfo{person}{Arjun
  Kashyap}, {and} \bibinfo{person}{Xiaoyi Lu}.}
  \bibinfo{year}{2020}\natexlab{}.
\newblock \showarticletitle{Understanding the idiosyncrasies of real persistent
  memory}.
\newblock \bibinfo{journal}{\emph{Proceedings of the VLDB Endowment}}
  \bibinfo{volume}{14}, \bibinfo{number}{4} (\bibinfo{year}{2020}),
  \bibinfo{pages}{626--639}.
\newblock


\bibitem[Hady et~al\mbox{.}(2017)]%
        {hady2017platform}
\bibfield{author}{\bibinfo{person}{Frank~T Hady}, \bibinfo{person}{Annie
  Foong}, \bibinfo{person}{Bryan Veal}, {and} \bibinfo{person}{Dan Williams}.}
  \bibinfo{year}{2017}\natexlab{}.
\newblock \showarticletitle{Platform storage performance with 3D XPoint
  technology}.
\newblock \bibinfo{journal}{\emph{Proc. IEEE}} \bibinfo{volume}{105},
  \bibinfo{number}{9} (\bibinfo{year}{2017}), \bibinfo{pages}{1822--1833}.
\newblock


\bibitem[Handy and Coughlin(2023)]%
        {handy2023optane}
\bibfield{author}{\bibinfo{person}{Jim Handy} {and} \bibinfo{person}{Tom
  Coughlin}.} \bibinfo{year}{2023}\natexlab{}.
\newblock \showarticletitle{Optane’s Dead: Now What?}
\newblock \bibinfo{journal}{\emph{Computer}} \bibinfo{volume}{56},
  \bibinfo{number}{3} (\bibinfo{year}{2023}), \bibinfo{pages}{125--130}.
\newblock


\bibitem[Hirofuchi and Takano(2020)]%
        {hirofuchi2020prompt}
\bibfield{author}{\bibinfo{person}{Takahiro Hirofuchi} {and}
  \bibinfo{person}{Ryousei Takano}.} \bibinfo{year}{2020}\natexlab{}.
\newblock \showarticletitle{A prompt report on the performance of intel optane
  dc persistent memory module}.
\newblock \bibinfo{journal}{\emph{IEICE TRANSACTIONS on Information and
  Systems}} \bibinfo{volume}{103}, \bibinfo{number}{5} (\bibinfo{year}{2020}),
  \bibinfo{pages}{1168--1172}.
\newblock


\bibitem[Intel(2022)]%
        {intel-pmem-cxl-tech}
\bibfield{author}{\bibinfo{person}{Intel}.} \bibinfo{year}{2022}\natexlab{}.
\newblock \bibinfo{title}{{Migration from Direct-Attached Intel Optane
  Persistent Memory to CXL-Attached Memory}}.
\newblock
  \bibinfo{howpublished}{\url{https://www.intel.com/content/dam/www/central-libraries/us/en/documents/2022-11/optane-pmem-to-cxl-tech-brief.pdf}}.
\newblock
\newblock
\shownote{[Online]}.


\bibitem[Intel(2023a)]%
        {intel-developercloud}
\bibfield{author}{\bibinfo{person}{Intel}.} \bibinfo{year}{2023}\natexlab{a}.
\newblock \bibinfo{title}{{Intel Developer Cloud}}.
\newblock
  \bibinfo{howpublished}{\url{https://www.intel.com/content/www/us/en/developer/tools/devcloud/overview.html}}.
\newblock
\newblock
\shownote{[Online]}.


\bibitem[Intel(2023b)]%
        {cxl-fpga-intel}
\bibfield{author}{\bibinfo{person}{Intel}.} \bibinfo{year}{2023}\natexlab{b}.
\newblock \bibinfo{title}{{Intel® FPGA Compute Express Link (CXL) IP}}.
\newblock
  \bibinfo{howpublished}{\url{https://www.intel.com/content/www/us/en/products/details/fpga/intellectual-property/interface-protocols/cxl-ip.html}}.
\newblock
\newblock
\shownote{[Online]}.


\bibitem[Intel(2023c)]%
        {fpga}
\bibfield{author}{\bibinfo{person}{Intel}.} \bibinfo{year}{2023}\natexlab{c}.
\newblock \bibinfo{title}{Intel® FPGA Compute Express Link (CXL) IP}.
\newblock
  \bibinfo{howpublished}{\url{https://www.intel.com/content/www/us/en/products/details/fpga/intellectual-property/interface-protocols/cxl-ip.html}}.
\newblock
\newblock
\shownote{[Online]}.


\bibitem[Izraelevitz et~al\mbox{.}(2016)]%
        {izraelevitz2016linearizability}
\bibfield{author}{\bibinfo{person}{Joseph Izraelevitz},
  \bibinfo{person}{Hammurabi Mendes}, {and} \bibinfo{person}{Michael~L Scott}.}
  \bibinfo{year}{2016}\natexlab{}.
\newblock \showarticletitle{Linearizability of persistent memory objects under
  a full-system-crash failure model}. In
  \bibinfo{booktitle}{\emph{International Symposium on Distributed Computing}}.
  Springer, \bibinfo{pages}{313--327}.
\newblock


\bibitem[Izraelevitz et~al\mbox{.}(2019)]%
        {izraelevitz2019basic}
\bibfield{author}{\bibinfo{person}{Joseph Izraelevitz}, \bibinfo{person}{Jian
  Yang}, \bibinfo{person}{Lu Zhang}, \bibinfo{person}{Juno Kim},
  \bibinfo{person}{Xiao Liu}, \bibinfo{person}{Amirsaman Memaripour},
  \bibinfo{person}{Yun~Joon Soh}, \bibinfo{person}{Zixuan Wang},
  \bibinfo{person}{Yi Xu}, \bibinfo{person}{Subramanya~R Dulloor},
  {et~al\mbox{.}}} \bibinfo{year}{2019}\natexlab{}.
\newblock \showarticletitle{Basic performance measurements of the intel optane
  DC persistent memory module}.
\newblock \bibinfo{journal}{\emph{arXiv preprint arXiv:1903.05714}}
  (\bibinfo{year}{2019}).
\newblock


\bibitem[Jun et~al\mbox{.}(2017)]%
        {jun2017hbm}
\bibfield{author}{\bibinfo{person}{Hongshin Jun}, \bibinfo{person}{Jinhee Cho},
  \bibinfo{person}{Kangseol Lee}, \bibinfo{person}{Ho-Young Son},
  \bibinfo{person}{Kwiwook Kim}, \bibinfo{person}{Hanho Jin}, {and}
  \bibinfo{person}{Keith Kim}.} \bibinfo{year}{2017}\natexlab{}.
\newblock \showarticletitle{Hbm (high bandwidth memory) dram technology and
  architecture}. In \bibinfo{booktitle}{\emph{2017 IEEE International Memory
  Workshop (IMW)}}. IEEE, \bibinfo{pages}{1--4}.
\newblock


\bibitem[Jun et~al\mbox{.}(2016)]%
        {jun2016high}
\bibfield{author}{\bibinfo{person}{Hongshin Jun}, \bibinfo{person}{Sangkyun
  Nam}, \bibinfo{person}{Hanho Jin}, \bibinfo{person}{Jong-Chern Lee},
  \bibinfo{person}{Yong~Jae Park}, {and} \bibinfo{person}{Jae~Jin Lee}.}
  \bibinfo{year}{2016}\natexlab{}.
\newblock \showarticletitle{High-bandwidth memory (HBM) test challenges and
  solutions}.
\newblock \bibinfo{journal}{\emph{IEEE Design \& Test}} \bibinfo{volume}{34},
  \bibinfo{number}{1} (\bibinfo{year}{2016}), \bibinfo{pages}{16--25}.
\newblock


\bibitem[Kateja et~al\mbox{.}(2017)]%
        {kateja2017viyojit}
\bibfield{author}{\bibinfo{person}{Rajat Kateja}, \bibinfo{person}{Anirudh
  Badam}, \bibinfo{person}{Sriram Govindan}, \bibinfo{person}{Bikash Sharma},
  {and} \bibinfo{person}{Greg Ganger}.} \bibinfo{year}{2017}\natexlab{}.
\newblock \showarticletitle{Viyojit: Decoupling battery and DRAM capacities for
  battery-backed DRAM}.
\newblock \bibinfo{journal}{\emph{ACM SIGARCH Computer Architecture News}}
  \bibinfo{volume}{45}, \bibinfo{number}{2} (\bibinfo{year}{2017}),
  \bibinfo{pages}{613--626}.
\newblock


\bibitem[Khan et~al\mbox{.}(2020)]%
        {khan2020persistent}
\bibfield{author}{\bibinfo{person}{Awais Khan}, \bibinfo{person}{Hyogi Sim},
  \bibinfo{person}{Sudharshan~S Vazhkudai}, \bibinfo{person}{Jinsuk Ma},
  \bibinfo{person}{Myeong-Hoon Oh}, {and} \bibinfo{person}{Youngjae Kim}.}
  \bibinfo{year}{2020}\natexlab{}.
\newblock \showarticletitle{Persistent memory object storage and indexing for
  scientific computing}. In \bibinfo{booktitle}{\emph{2020 IEEE/ACM Workshop on
  Memory Centric High Performance Computing (MCHPC)}}. IEEE,
  \bibinfo{pages}{1--9}.
\newblock


\bibitem[Kogge and Dally(2022)]%
        {kogge2022frontier}
\bibfield{author}{\bibinfo{person}{Peter~M Kogge} {and}
  \bibinfo{person}{William~J Dally}.} \bibinfo{year}{2022}\natexlab{}.
\newblock \showarticletitle{Frontier vs the Exascale Report: Why so long? and
  Are We Really There Yet?}. In \bibinfo{booktitle}{\emph{2022 IEEE/ACM
  International Workshop on Performance Modeling, Benchmarking and Simulation
  of High Performance Computer Systems (PMBS)}}. IEEE, \bibinfo{pages}{26--35}.
\newblock


\bibitem[Lee et~al\mbox{.}(2010)]%
        {lee2010phase}
\bibfield{author}{\bibinfo{person}{Benjamin~C Lee}, \bibinfo{person}{Engin
  Ipek}, \bibinfo{person}{Onur Mutlu}, {and} \bibinfo{person}{Doug Burger}.}
  \bibinfo{year}{2010}\natexlab{}.
\newblock \showarticletitle{Phase change memory architecture and the quest for
  scalability}.
\newblock \bibinfo{journal}{\emph{Commun. ACM}} \bibinfo{volume}{53},
  \bibinfo{number}{7} (\bibinfo{year}{2010}), \bibinfo{pages}{99--106}.
\newblock


\bibitem[Lee et~al\mbox{.}(2023)]%
        {lee2023elastic}
\bibfield{author}{\bibinfo{person}{Donghun Lee}, \bibinfo{person}{Thomas
  Willhalm}, \bibinfo{person}{Minseon Ahn}, \bibinfo{person}{Suprasad
  Mutalik~Desai}, \bibinfo{person}{Daniel Booss}, \bibinfo{person}{Navneet
  Singh}, \bibinfo{person}{Daniel Ritter}, \bibinfo{person}{Jungmin Kim}, {and}
  \bibinfo{person}{Oliver Rebholz}.} \bibinfo{year}{2023}\natexlab{}.
\newblock \showarticletitle{Elastic Use of Far Memory for In-Memory Database
  Management Systems}. In \bibinfo{booktitle}{\emph{Proceedings of the 19th
  International Workshop on Data Management on New Hardware}}.
  \bibinfo{pages}{35--43}.
\newblock


\bibitem[Liu et~al\mbox{.}(2003)]%
        {liu2003high}
\bibfield{author}{\bibinfo{person}{Jiuxing Liu}, \bibinfo{person}{Jiesheng Wu},
  \bibinfo{person}{Sushmitha~P Kini}, \bibinfo{person}{Pete Wyckoff}, {and}
  \bibinfo{person}{Dhabaleswar~K Panda}.} \bibinfo{year}{2003}\natexlab{}.
\newblock \showarticletitle{High performance RDMA-based MPI implementation over
  InfiniBand}. In \bibinfo{booktitle}{\emph{Proceedings of the 17th annual
  international conference on Supercomputing}}. \bibinfo{pages}{295--304}.
\newblock


\bibitem[Liu(2023)]%
        {liu2023fabric}
\bibfield{author}{\bibinfo{person}{Ming Liu}.} \bibinfo{year}{2023}\natexlab{}.
\newblock \showarticletitle{Fabric-Centric Computing}. In
  \bibinfo{booktitle}{\emph{Proceedings of the 19th Workshop on Hot Topics in
  Operating Systems}}. \bibinfo{pages}{118--126}.
\newblock


\bibitem[Lockwood et~al\mbox{.}(2023)]%
        {lockwood2023storage}
\bibfield{author}{\bibinfo{person}{Glenn~K Lockwood}, \bibinfo{person}{Damian
  Hazen}, \bibinfo{person}{Quincey Koziol}, \bibinfo{person}{R~Shane Canon},
  \bibinfo{person}{Katie Antypas}, \bibinfo{person}{Jan Balewski},
  \bibinfo{person}{Nicholas Balthaser}, \bibinfo{person}{Wahid Bhimji},
  \bibinfo{person}{James Botts}, \bibinfo{person}{Jeff Broughton},
  {et~al\mbox{.}}} \bibinfo{year}{2023}\natexlab{}.
\newblock \showarticletitle{Storage 2020: A vision for the future of hpc
  storage}.
\newblock  (\bibinfo{year}{2023}).
\newblock


\bibitem[Logan et~al\mbox{.}(2023)]%
        {logan2023evaluation}
\bibfield{author}{\bibinfo{person}{Luke Logan}, \bibinfo{person}{Jay Lofstead},
  \bibinfo{person}{Xian-He Sun}, {and} \bibinfo{person}{Anthony Kougkas}.}
  \bibinfo{year}{2023}\natexlab{}.
\newblock \showarticletitle{An Evaluation of DAOS for Simulation and Deep
  Learning HPC Workloads}. In \bibinfo{booktitle}{\emph{Proceedings of the 3rd
  Workshop on Challenges and Opportunities of Efficient and Performant Storage
  Systems}}. \bibinfo{pages}{9--16}.
\newblock


\bibitem[Malladi et~al\mbox{.}(2016)]%
        {malladi2016drampersist}
\bibfield{author}{\bibinfo{person}{Krishna~T Malladi}, \bibinfo{person}{Manu
  Awasthi}, {and} \bibinfo{person}{Hongzhong Zheng}.}
  \bibinfo{year}{2016}\natexlab{}.
\newblock \showarticletitle{DRAMPersist: Making DRAM Systems Persistent}. In
  \bibinfo{booktitle}{\emph{Proceedings of the Second International Symposium
  on Memory Systems}}. \bibinfo{pages}{94--95}.
\newblock


\bibitem[McCalpin(2007)]%
        {McCalpin2007}
\bibfield{author}{\bibinfo{person}{John~D. McCalpin}.}
  \bibinfo{year}{1991-2007}\natexlab{}.
\newblock \bibinfo{booktitle}{\emph{STREAM: Sustainable Memory Bandwidth in
  High Performance Computers}}.
\newblock \bibinfo{type}{{T}echnical {R}eport}.
  \bibinfo{institution}{University of Virginia},
  \bibinfo{address}{Charlottesville, Virginia}.
\newblock
\urldef\tempurl%
\url{http://www.cs.virginia.edu/stream/}
\showURL{%
\tempurl}
\newblock
\shownote{A continually updated technical report.
  http://www.cs.virginia.edu/stream/}.


\bibitem[McCalpin(1995a)]%
        {McCalpin1995}
\bibfield{author}{\bibinfo{person}{John~D. McCalpin}.}
  \bibinfo{year}{1995}\natexlab{a}.
\newblock \showarticletitle{Memory Bandwidth and Machine Balance in Current
  High Performance Computers}.
\newblock \bibinfo{journal}{\emph{IEEE Computer Society Technical Committee on
  Computer Architecture (TCCA) Newsletter}} (\bibinfo{date}{Dec.}
  \bibinfo{year}{1995}), \bibinfo{pages}{19--25}.
\newblock


\bibitem[McCalpin(1995b)]%
        {stream}
\bibfield{author}{\bibinfo{person}{John~D. McCalpin}.}
  \bibinfo{year}{1995}\natexlab{b}.
\newblock \bibinfo{title}{{STREAM Benchmark}}.
\newblock \bibinfo{howpublished}{\url{https://www.cs.virginia.edu/stream/}}.
\newblock
\newblock
\shownote{[Online]}.


\bibitem[McCalpin(1995c)]%
        {mccalpin1995stream}
\bibfield{author}{\bibinfo{person}{John~D McCalpin}.}
  \bibinfo{year}{1995}\natexlab{c}.
\newblock \showarticletitle{Stream benchmark}.
\newblock \bibinfo{journal}{\emph{Link: www. cs. virginia. edu/stream/ref.
  html\# what}} \bibinfo{volume}{22}, \bibinfo{number}{7}
  (\bibinfo{year}{1995}).
\newblock


\bibitem[McCalpin(2023)]%
        {McCalpin2023}
\bibfield{author}{\bibinfo{person}{John~D. McCalpin}.}
  \bibinfo{year}{2023}\natexlab{}.
\newblock \showarticletitle{Bandwidth Limits in the Intel Xeon Max (Sapphire
  Rapids with HBM) Processors}. In \bibinfo{booktitle}{\emph{ISC 2023 IXPUG
  Workshop}}. \bibinfo{pages}{1--24}.
\newblock


\bibitem[McKee(2004)]%
        {mckee2004reflections}
\bibfield{author}{\bibinfo{person}{Sally~A McKee}.}
  \bibinfo{year}{2004}\natexlab{}.
\newblock \showarticletitle{Reflections on the memory wall}. In
  \bibinfo{booktitle}{\emph{Proceedings of the 1st conference on Computing
  frontiers}}. \bibinfo{pages}{162}.
\newblock


\bibitem[Michelogiannakis et~al\mbox{.}(2023)]%
        {michelogiannakis2023efficient}
\bibfield{author}{\bibinfo{person}{George Michelogiannakis},
  \bibinfo{person}{Yehia Arafa}, \bibinfo{person}{Brandon Cook},
  \bibinfo{person}{Liang~Yuan Dai}, \bibinfo{person}{Abdel~Hameed Badawy},
  \bibinfo{person}{Madeleine Glick}, \bibinfo{person}{Yuyang Wang},
  \bibinfo{person}{Keren Bergman}, {and} \bibinfo{person}{John Shalf}.}
  \bibinfo{year}{2023}\natexlab{}.
\newblock \showarticletitle{Efficient Intra-Rack Resource Disaggregation for
  HPC Using Co-Packaged DWDM Photonics}.
\newblock \bibinfo{journal}{\emph{arXiv preprint arXiv:2301.03592}}
  (\bibinfo{year}{2023}).
\newblock


\bibitem[Michelogiannakis et~al\mbox{.}(2022)]%
        {michelogiannakis2022case}
\bibfield{author}{\bibinfo{person}{George Michelogiannakis},
  \bibinfo{person}{Benjamin Klenk}, \bibinfo{person}{Brandon Cook},
  \bibinfo{person}{Min~Yee Teh}, \bibinfo{person}{Madeleine Glick},
  \bibinfo{person}{Larry Dennison}, \bibinfo{person}{Keren Bergman}, {and}
  \bibinfo{person}{John Shalf}.} \bibinfo{year}{2022}\natexlab{}.
\newblock \showarticletitle{A case for intra-rack resource disaggregation in
  HPC}.
\newblock \bibinfo{journal}{\emph{ACM Transactions on Architecture and Code
  Optimization (TACO)}} \bibinfo{volume}{19}, \bibinfo{number}{2}
  (\bibinfo{year}{2022}), \bibinfo{pages}{1--26}.
\newblock


\bibitem[Mironov et~al\mbox{.}(2019)]%
        {mironov2019performance}
\bibfield{author}{\bibinfo{person}{Vladimir Mironov}, \bibinfo{person}{Igor
  Chernykh}, \bibinfo{person}{Igor Kulikov}, \bibinfo{person}{Alexander
  Moskovsky}, \bibinfo{person}{Evgeny Epifanovsky}, {and}
  \bibinfo{person}{Andrey Kudryavtsev}.} \bibinfo{year}{2019}\natexlab{}.
\newblock \showarticletitle{Performance evaluation of the intel optane dc
  memory with scientific benchmarks}. In \bibinfo{booktitle}{\emph{2019
  IEEE/ACM Workshop on Memory Centric High Performance Computing (MCHPC)}}.
  IEEE, \bibinfo{pages}{1--6}.
\newblock


\bibitem[Mutlu(2013)]%
        {mutlu2013memory}
\bibfield{author}{\bibinfo{person}{Onur Mutlu}.}
  \bibinfo{year}{2013}\natexlab{}.
\newblock \showarticletitle{Memory scaling: A systems architecture
  perspective}. In \bibinfo{booktitle}{\emph{2013 5th IEEE International Memory
  Workshop}}. IEEE, \bibinfo{pages}{21--25}.
\newblock


\bibitem[Narayanan and Hodson(2012)]%
        {narayanan2012whole}
\bibfield{author}{\bibinfo{person}{Dushyanth Narayanan} {and}
  \bibinfo{person}{Orion Hodson}.} \bibinfo{year}{2012}\natexlab{}.
\newblock \showarticletitle{Whole-system persistence}. In
  \bibinfo{booktitle}{\emph{Proceedings of the seventeenth international
  conference on Architectural Support for Programming Languages and Operating
  Systems}}. \bibinfo{pages}{401--410}.
\newblock


\bibitem[Nathan~Pham(2019)]%
        {tristian2021analyzing}
\bibfield{author}{\bibinfo{person}{Kevin~Huang Nathan~Pham}.}
  \bibinfo{year}{2019}\natexlab{}.
\newblock \bibinfo{title}{Analyzing the Performance of Intel Optane Persistent
  Memory 200 Series in Memory Mode with Lenovo ThinkSystem Servers}.
\newblock
\newblock


\bibitem[Newsroom(2022a)]%
        {samsung}
\bibfield{author}{\bibinfo{person}{Samsung Newsroom}.}
  \bibinfo{year}{2022}\natexlab{a}.
\newblock \bibinfo{booktitle}{\emph{Samsung Electronics Introduces Industry’s
  First 512GB CXL Memory Module}}.
\newblock
\urldef\tempurl%
\url{https://news.samsung.com/global/samsung-electronics-introduces-industrys-first-512gb-cxl-memory-module}
\showURL{%
\tempurl}


\bibitem[Newsroom(2022b)]%
        {hynix}
\bibfield{author}{\bibinfo{person}{SK~Hynix Newsroom}.}
  \bibinfo{year}{2022}\natexlab{b}.
\newblock \bibinfo{booktitle}{\emph{SK hynix Develops DDR5 DRAM CXLTM Memory to
  Expand the CXL Memory Ecosystem}}.
\newblock
\urldef\tempurl%
\url{https://news.skhynix.com/sk-hynix-develops-ddr5-dram-cxltm-memory-to-expand-the-cxl-memory-ecosystem/}
\showURL{%
\tempurl}


\bibitem[Park(2019)]%
        {negevhpc}
\bibfield{author}{\bibinfo{person}{Rotem~Industrial Park}.}
  \bibinfo{year}{2019}\natexlab{}.
\newblock \bibinfo{title}{{NegevHPC Project}}.
\newblock \bibinfo{howpublished}{\url{https://www.negevhpc.com}}.
\newblock
\newblock
\shownote{[Online]}.


\bibitem[Patil et~al\mbox{.}(2019)]%
        {patil2019performance}
\bibfield{author}{\bibinfo{person}{Onkar Patil}, \bibinfo{person}{Latchesar
  Ionkov}, \bibinfo{person}{Jason Lee}, \bibinfo{person}{Frank Mueller}, {and}
  \bibinfo{person}{Michael Lang}.} \bibinfo{year}{2019}\natexlab{}.
\newblock \showarticletitle{Performance characterization of a dram-nvm hybrid
  memory architecture for hpc applications using intel optane dc persistent
  memory modules}. In \bibinfo{booktitle}{\emph{Proceedings of the
  International Symposium on Memory Systems}}. \bibinfo{pages}{288--303}.
\newblock


\bibitem[Peng et~al\mbox{.}(2021)]%
        {peng2021holistic}
\bibfield{author}{\bibinfo{person}{Ivy Peng}, \bibinfo{person}{Ian Karlin},
  \bibinfo{person}{Maya Gokhale}, \bibinfo{person}{Kathleen Shoga},
  \bibinfo{person}{Matthew Legendre}, {and} \bibinfo{person}{Todd Gamblin}.}
  \bibinfo{year}{2021}\natexlab{}.
\newblock \showarticletitle{A holistic view of memory utilization on HPC
  systems: Current and future trends}. In \bibinfo{booktitle}{\emph{The
  International Symposium on Memory Systems}}. \bibinfo{pages}{1--11}.
\newblock


\bibitem[Peng et~al\mbox{.}(2020)]%
        {peng2020memory}
\bibfield{author}{\bibinfo{person}{Ivy Peng}, \bibinfo{person}{Roger Pearce},
  {and} \bibinfo{person}{Maya Gokhale}.} \bibinfo{year}{2020}\natexlab{}.
\newblock \showarticletitle{On the memory underutilization: Exploring
  disaggregated memory on hpc systems}. In \bibinfo{booktitle}{\emph{2020 IEEE
  32nd International Symposium on Computer Architecture and High Performance
  Computing (SBAC-PAD)}}. IEEE, \bibinfo{pages}{183--190}.
\newblock


\bibitem[Radulovic et~al\mbox{.}(2015)]%
        {radulovic2015another}
\bibfield{author}{\bibinfo{person}{Milan Radulovic}, \bibinfo{person}{Darko
  Zivanovic}, \bibinfo{person}{Daniel Ruiz}, \bibinfo{person}{Bronis~R de
  Supinski}, \bibinfo{person}{Sally~A McKee}, \bibinfo{person}{Petar
  Radojkovi{\'c}}, {and} \bibinfo{person}{Eduard Ayguad{\'e}}.}
  \bibinfo{year}{2015}\natexlab{}.
\newblock \showarticletitle{Another trip to the wall: How much will stacked
  dram benefit hpc?}. In \bibinfo{booktitle}{\emph{Proceedings of the 2015
  International Symposium on Memory Systems}}. \bibinfo{pages}{31--36}.
\newblock


\bibitem[Rai and Talawar(2023)]%
        {rai2023nonvolatile}
\bibfield{author}{\bibinfo{person}{Sadhana Rai} {and}
  \bibinfo{person}{Basavaraj Talawar}.} \bibinfo{year}{2023}\natexlab{}.
\newblock \showarticletitle{Nonvolatile Memory Technologies: Characteristics,
  Deployment, and Research Challenges}.
\newblock \bibinfo{journal}{\emph{Frontiers of Quality Electronic Design (QED)
  AI, IoT and Hardware Security}} (\bibinfo{year}{2023}),
  \bibinfo{pages}{137--173}.
\newblock


\bibitem[Reed et~al\mbox{.}(2022)]%
        {reed2022reinventing}
\bibfield{author}{\bibinfo{person}{Daniel Reed}, \bibinfo{person}{Dennis
  Gannon}, {and} \bibinfo{person}{Jack Dongarra}.}
  \bibinfo{year}{2022}\natexlab{}.
\newblock \showarticletitle{Reinventing high performance computing: challenges
  and opportunities}.
\newblock \bibinfo{journal}{\emph{arXiv preprint arXiv:2203.02544}}
  (\bibinfo{year}{2022}).
\newblock


\bibitem[Ruan et~al\mbox{.}(2023)]%
        {ruan2023persistent}
\bibfield{author}{\bibinfo{person}{Chaoyi Ruan}, \bibinfo{person}{Yingqiang
  Zhang}, \bibinfo{person}{Chao Bi}, \bibinfo{person}{Xiaosong Ma},
  \bibinfo{person}{Hao Chen}, \bibinfo{person}{Feifei Li},
  \bibinfo{person}{Xinjun Yang}, \bibinfo{person}{Cheng Li},
  \bibinfo{person}{Ashraf Aboulnaga}, {and} \bibinfo{person}{Yinlong Xu}.}
  \bibinfo{year}{2023}\natexlab{}.
\newblock \showarticletitle{Persistent Memory Disaggregation for Cloud-Native
  Relational Databases}. In \bibinfo{booktitle}{\emph{Proceedings of the 28th
  ACM International Conference on Architectural Support for Programming
  Languages and Operating Systems, Volume 3}}. \bibinfo{pages}{498--512}.
\newblock


\bibitem[Rudoff(2017)]%
        {rudoff2017persistent}
\bibfield{author}{\bibinfo{person}{Andy Rudoff}.}
  \bibinfo{year}{2017}\natexlab{}.
\newblock \showarticletitle{Persistent memory: The value to hpc and the
  challenges}. In \bibinfo{booktitle}{\emph{Proceedings of the Workshop on
  memory centric programming for hpc}}. \bibinfo{pages}{7--10}.
\newblock


\bibitem[Sainio et~al\mbox{.}(2016)]%
        {sainio2016nvdimm}
\bibfield{author}{\bibinfo{person}{Arthur Sainio} {et~al\mbox{.}}}
  \bibinfo{year}{2016}\natexlab{}.
\newblock \showarticletitle{NVDIMM: changes are here so what’s next}.
\newblock \bibinfo{journal}{\emph{Memory Computing Summit}}
  (\bibinfo{year}{2016}).
\newblock


\bibitem[Scargall(2020)]%
        {scargall2020programming}
\bibfield{author}{\bibinfo{person}{Steve Scargall}.}
  \bibinfo{year}{2020}\natexlab{}.
\newblock \showarticletitle{Programming Persistent Memory: A Comprehensive
  Guide for Developers}.
\newblock  (\bibinfo{year}{2020}).
\newblock


\bibitem[Scargall and Scargall(2020)]%
        {scargall2020libpmemobj}
\bibfield{author}{\bibinfo{person}{Steve Scargall} {and} \bibinfo{person}{Steve
  Scargall}.} \bibinfo{year}{2020}\natexlab{}.
\newblock \showarticletitle{libpmemobj: A Native Transactional Object Store}.
\newblock \bibinfo{journal}{\emph{Programming Persistent Memory: A
  Comprehensive Guide for Developers}} (\bibinfo{year}{2020}),
  \bibinfo{pages}{81--109}.
\newblock


\bibitem[Sharma et~al\mbox{.}(2023)]%
        {sharma2023introduction}
\bibfield{author}{\bibinfo{person}{Debendra~Das Sharma},
  \bibinfo{person}{Robert Blankenship}, {and} \bibinfo{person}{Daniel~S
  Berger}.} \bibinfo{year}{2023}\natexlab{}.
\newblock \showarticletitle{An Introduction to the Compute Express Link (CXL)
  Interconnect}.
\newblock \bibinfo{journal}{\emph{arXiv preprint arXiv:2306.11227}}
  (\bibinfo{year}{2023}).
\newblock


\bibitem[Shipman et~al\mbox{.}(2022)]%
        {shipman2022early}
\bibfield{author}{\bibinfo{person}{Galen~M Shipman}, \bibinfo{person}{Sriram
  Swaminarayan}, \bibinfo{person}{Gary Grider}, \bibinfo{person}{Jim Lujan},
  {and} \bibinfo{person}{R~Joseph Zerr}.} \bibinfo{year}{2022}\natexlab{}.
\newblock \showarticletitle{Early Performance Results on 4th Gen Intel (R) Xeon
  (R) Scalable Processors with DDR and Intel (R) Xeon (R) processors, codenamed
  Sapphire Rapids with HBM}.
\newblock \bibinfo{journal}{\emph{arXiv preprint arXiv:2211.05712}}
  (\bibinfo{year}{2022}).
\newblock


\bibitem[Technology(2022)]%
        {montage}
\bibfield{author}{\bibinfo{person}{Montage Technology}.}
  \bibinfo{year}{2022}\natexlab{}.
\newblock \bibinfo{booktitle}{\emph{Montage Technology Delivers the World’s
  First CXL™ Memory eXpander Controller}}.
\newblock
\urldef\tempurl%
\url{https://www.montage-tech.com/Press_Releases/20220506}
\showURL{%
\tempurl}


\bibitem[Tristian and Travis(2019)]%
        {tristian2019analyzing}
\bibfield{author}{\bibinfo{person}{TB Tristian} {and} \bibinfo{person}{L
  Travis}.} \bibinfo{year}{2019}\natexlab{}.
\newblock \bibinfo{title}{Analyzing the performance of Intel Optane DC
  persistent memory in app direct mode in Lenovo ThinkSystem servers}.
\newblock
\newblock


\bibitem[Wahlgren et~al\mbox{.}(2022)]%
        {wahlgren2022evaluating}
\bibfield{author}{\bibinfo{person}{Jacob Wahlgren}, \bibinfo{person}{Maya
  Gokhale}, {and} \bibinfo{person}{Ivy~B Peng}.}
  \bibinfo{year}{2022}\natexlab{}.
\newblock \showarticletitle{Evaluating Emerging CXL-enabled Memory Pooling for
  HPC Systems}. In \bibinfo{booktitle}{\emph{2022 IEEE/ACM Workshop on Memory
  Centric High Performance Computing (MCHPC)}}. IEEE, \bibinfo{pages}{11--20}.
\newblock


\bibitem[Wang et~al\mbox{.}(2023)]%
        {wang2023survey}
\bibfield{author}{\bibinfo{person}{Ying Wang}, \bibinfo{person}{Wen-Qing Jia},
  \bibinfo{person}{De-Jun Jiang}, {and} \bibinfo{person}{Jin Xiong}.}
  \bibinfo{year}{2023}\natexlab{}.
\newblock \showarticletitle{A Survey of Non-Volatile Main Memory File Systems}.
\newblock \bibinfo{journal}{\emph{Journal of Computer Science and Technology}}
  \bibinfo{volume}{38}, \bibinfo{number}{2} (\bibinfo{year}{2023}),
  \bibinfo{pages}{348--372}.
\newblock


\bibitem[Weiland et~al\mbox{.}(2019)]%
        {weiland2019early}
\bibfield{author}{\bibinfo{person}{Mich{\`e}le Weiland},
  \bibinfo{person}{Holger Brunst}, \bibinfo{person}{Tiago Quintino},
  \bibinfo{person}{Nick Johnson}, \bibinfo{person}{Olivier Iffrig},
  \bibinfo{person}{Simon Smart}, \bibinfo{person}{Christian Herold},
  \bibinfo{person}{Antonino Bonanni}, \bibinfo{person}{Adrian Jackson}, {and}
  \bibinfo{person}{Mark Parsons}.} \bibinfo{year}{2019}\natexlab{}.
\newblock \showarticletitle{An early evaluation of intel's optane dc persistent
  memory module and its impact on high-performance scientific applications}. In
  \bibinfo{booktitle}{\emph{Proceedings of the international conference for
  high performance computing, networking, storage and analysis}}.
  \bibinfo{pages}{1--19}.
\newblock


\end{thebibliography}
\end{document}